\documentclass[journal,twoside,web]{ieeecolor}
\usepackage{generic}
\usepackage{cite}
\usepackage{amsmath,amssymb,amsfonts}
\usepackage{graphicx}
\usepackage{algorithm}
\usepackage[noend]{algpseudocode}
\usepackage{hyperref}
\usepackage{textcomp}

\usepackage{tikz}
\usepackage{stfloats}
\usepackage{subcaption}

\usepackage{stmaryrd}

\usepackage{bm}

\newcommand{\calB}{\mathcal{B}}
\newcommand{\calC}{\mathcal{C}}

\newcommand{\calE}{\mathcal{E}}

\newcommand{\calI}{\mathcal{I}}
\newcommand{\calJ}{\mathcal{J}}

\newcommand{\calL}{\mathcal{L}}

\newcommand{\calN}{\mathcal{N}}
\newcommand{\calO}{\mathcal{O}}
\newcommand{\calP}{\mathcal{P}}
\newcommand{\calQ}{\mathcal{Q}}
\newcommand{\calR}{\mathcal{R}}

\newcommand{\calV}{\mathcal{V}}
\newcommand{\calW}{\mathcal{W}}

\newcommand{\cala}{\mathscr{a}}
\newcommand{\calc}{\mathscr{c}}
\newcommand{\cald}{\mathscr{d}}
\newcommand{\cale}{\mathscr{e}}

\newcommand{\rf}{{\mathrm {f}}}
\newcommand{\rs}{{\mathrm {s}}}
\newcommand{\rx}{{\mathrm {x}}}
\newcommand{\ry}{{\mathrm {y}}}
\newcommand{\rz}{{\mathrm {z}}}

\newcommand{\rA}{{\mathrm {A}}}
\newcommand{\rB}{{\mathrm {B}}}

\newcommand{\rF}{{\mathrm {F}}}

\newcommand{\vect}{{\mathrm {vec}}}
\newcommand{\dom}{{\mathrm {dom}}}
\newcommand{\rIm}{{\mathrm {Im}}}

\newcommand{\Eb}{\mathbb{E}}
\newcommand{\Pb}{\mathbb{P}}

\newcommand{\Rb}{\mathbb{R}}
\newcommand{\Sb}{\mathbb{S}}
\newcommand{\Zb}{\mathbb{Z}}

\newcommand{\diag}{{\mathrm {diag}}}
\newcommand{\bdiag}{{\mathrm {bdiag}}}
\newcommand{\tr}{{\mathrm {tr}}}

\newcommand{\argmin}{\operatornamewithlimits{argmin}}

\usepackage[dvipsnames]{xcolor}
\definecolor{myYellow}{RGB}{255,255,204}
\definecolor{myCyan}{RGB}{234,255,255}

\DeclareFontFamily{U}{mathc}{}
\DeclareFontShape{U}{mathc}{m}{it}%
{<->s*[1.03] mathc10}{}

\DeclareMathAlphabet{\mathscr}{U}{mathc}{m}{it}

\usepackage{makecell}
\usepackage{multirow}

\newcommand{\nablax}{\nabla_{\!\bar{x}}}
\newcommand{\nablaw}{\nabla_{\!\bar{w}}}
\newcommand{\nablaxx}{\nabla_{\!\bar{x} \bar{x}}}

\newcommand{\labeleditem}[3]{%
  \textbf{#1~#2}\ \textit{(#3):}%
}

\newcommand{\myemph}[1]{\textbf{#1}}

\def\BibTeX{{\rm B\kern-.05em{\sc i\kern-.025em b}\kern-.08em
    T\kern-.1667em\lower.7ex\hbox{E}\kern-.125emX}}
\begin{document}
\title{Distributed Covariance Steering via Non-Convex ADMM for Large-Scale Multi-Agent Systems}
\author{Augustinos D. Saravanos, 
Isin M. Balci, Arshiya Taj Abdul,
Efstathios Bakolas \\ and Evangelos A. Theodorou
\thanks{This work was supported by the ARO Award $\#$W911NF2010151. Augustinos Saravanos acknowledges support by the A. Onassis Foundation Scholarship. \textit{(Corresponding author: Augustinos D. Saravanos)}}
\thanks{Augustinos D. Saravanos was with the School of Electrical and Computer Engineering, Georgia Institute of Technology, Atlanta, GA 30332, USA, during this work. He is now with the Department of Aeronautics and Astronautics, Massachusetts Institute of Technology, Cambridge, MA 02139, USA
(e-mail: asaravanos@gatech.edu; asaravan@mit.edu).}
\thanks{Arshiya Taj Abdul is with the School of Electrical and Computer Engineering, Georgia Institute of Technology, Atlanta, GA 30332 USA 
(e-mail: aabdul6@gatech.edu).}
\thanks{Isin M. Balci and Efstathios Bakolas are with the Department of Aerospace Engineering and Engineering Mechanics, The University of Texas at Austin, Austin, TX 78712 USA (isinmertbalci@utexas.edu, bakolas@austin.utexas.edu).}
\thanks{Evangelos A. Theodorou is with the Daniel Guggenheim School of Aerospace Engineering, Georgia Institute of Technology, Atlanta, GA 30332 USA (e-mail: evangelos.theodorou@gatech.edu).}
}

\maketitle

\begin{abstract}
This paper studies the problem of steering large-scale multi-agent stochastic linear systems between Gaussian distributions under probabilistic collision avoidance constraints. We introduce a family of \textit{distributed covariance steering (DCS)} methods based on the Alternating Direction Method of Multipliers (ADMM), each offering different trade-offs between conservatism and computational efficiency. The first method, \textit{Full-Covariance-Consensus (FCC)-DCS}, enforces consensus over both the means and covariances of neighboring agents, yielding the least conservative safe solutions. The second approach, \textit{Partial-Covariance-Consensus (PCC)-DCS}, leverages the insight that safety can be maintained by exchanging only partial covariance information, reducing computational demands. The third method, \textit{Mean-Consensus (MC)-DCS}, provides the most scalable alternative by requiring consensus only on mean states. Furthermore, we establish novel convergence guarantees for distributed ADMM with iteratively linearized non-convex constraints, covering a broad class of consensus optimization problems, and show that the proposed DCS methods fall within this framework. Simulations in 2D and 3D multi-agent environments verify safety, illustrate the trade-offs between methods, and demonstrate scalability to thousands of agents.
\end{abstract}


\section{Introduction}
\label{sec:introduction}


\IEEEPARstart{T}HE increasing scale and complexity of multi-agent systems, ranging from self-driving cars \cite{malikopoulos2021optimal} and warehouse automation \cite{liu2020prediction} to UAV coordination \cite{kabore2021distributed} and swarm robotics \cite{cortes2017coordinated}, necessitate algorithms capable of ensuring reliable operation. Two fundamental challenges arise: (i) \textit{scalability}, requiring computational and communication efficiency as team sizes grow, and (ii) \textit{safety}, demanding probabilistic guarantees under uncertainty. This article addresses these challenges, by introducing a family of distributed methods for steering the state distributions of large-scale multi-agent stochastic systems to prescribed distributions, while ensuring collision avoidance.

        

Classical stochastic control approaches such as Linear Quadratic Gaussian (LQG) control indirectly minimize the state variance, often leading to overly aggressive behavior which might be undesirable in safety-critical multi-agent settings. Other common approaches such as stochastic model predictive control often rely on sampling-based approximations \cite{schildbach2014scenario}, fixed feedback gains \cite{arcari2023stochastic} or other conservative reformulations \cite{farina2016stochastic}, which can limit robustness and scalability. The idea of steering the distribution of a stochastic system from initial to target distributions offers an attractive alternative as it can be naturally associated with probabilistic guarantees under uncertainty. However, controlling the full density of distributions is known to be computationally intensive \cite{chen2021optimal}, rendering such approaches impractical for large-scale systems.

Covariance Steering (CS) has emerged as a powerful methodology for steering the state distribution of stochastic systems from a given initial distribution to a prescribed terminal one. 
Originally formulated for linear systems under Gaussian uncertainty \cite{chen2015optimal, bakolas2018finite, liu2025optimal}, CS methods have since been extended to nonlinear dynamics \cite{ridderhof2019nonlinear, ratheesh2025operator}, robust formulations \cite{kotsalis2021convex},
data-driven approaches \cite{pilipovsky2023data}, Gaussian mixture models (GMM) \cite{balci2024density}, general distribution steering \cite{wu2024distribution}, and various other settings. Successful applications are found in navigation \cite{yin2022trajectory}, manipulation \cite{shirai2023covariance}, aerospace systems \cite{kumagai2024sequential} and multi-agent control \cite{Saravanos-RSS-21}, among other domains.

Despite their promise for safety-critical systems, CS methods typically result in computationally intensive semidefinite programming (SDP) problems, which restricts their applicability to low-dimensional systems. To overcome this fundamental bottleneck, this article introduces a family of distributed CS methods with desirable trade-offs between conservatism and computational efficiency, that achieve scalability to large-scale multi-agent systems with safety guarantees.

\subsection{Related Work}

\myemph{Distribution Steering for Multi-Agent Systems.} A significant amount of the literature has studied the control of multi-agent systems through density control, where the collective behavior of a swarm is represented as a single distribution. Such approaches include mean-field formulations \cite{yonxin_chen2024density, rapakoulias2025steering}, Markov chain representations \cite{demir2015decentralized} and power moment-based methods \cite{wu2022group}. These approaches, however, differ fundamentally from the setting considered in this work, where each agent is itself modeled as a distribution whose mean and covariance must be steered to specific targets. The first distributed CS algorithm was introduced in \cite{Saravanos-RSS-21}, demonstrating scalability to dozens of agents; however, safety was enforced solely through constraints on the mean states. 
Similarly, the decentralized model predictive CS method in \cite{saravanos_2024distributedmpcs} and the hierarchical distribution steering framework in \cite{Saravanos-RSS-23} adopted formulations that relied on actively optimizing only the mean states for achieving safety. More recently, a centralized CS approach  was presented in \cite{bai2024_centralized_cs}, but its scalability is significantly limited to very few agents. Overall, existing works fall short in presenting decentralized methodologies that fully leverage distributional information to ensure safety, remain scalable to large systems and are supported by convergence guarantees.

\myemph{Distributed ADMM in Non-Convex Optimization.} Distributed optimization algorithms based on the Alternating Direction Method of Multipliers (ADMM) \cite{boyd2011distributed} have gained widespread popularity in autonomy, networked systems and other areas. Naturally, distributed multi-agent control methods leveraging ADMM have also been proposed, often achieving remarkable scalability \cite{saravanos2023distributed, shorinwa2024distributed, abdul2025scalable}. However, the convergence guarantees of such schemes typically hold only under convex settings. In contrast, the majority of multi-agent control problems in autonomy are inherently non-convex, e.g. due to collision avoidance constraints, so most distributed ADMM-based methods lack convergence guarantees in such settings.

Early convergence analyses of ADMM considered problems with non-convex objectives, but in the absence of constraints \cite{li2015global, guo2017convergence} or under convex ones \cite{hong2016convergence}. Linearized ADMM approaches have also been proposed, yet also without accounting for non-convex constraints \cite{liu2019linearized, lu2021linearized}. Several other works \cite{melo2017iteration, wang2019global, themelis2020douglas} have established  results for non-convex ADMM schemes, but rely on a restrictive assumption on the linear coupling constraints, typically not satisfied in distributed consensus optimization, as pointed out by Sun and Sun in \cite{sun2023two}. To accommodate that, the latter authors presented a two-level scheme in \cite{sun2023two} with an outer Augmented Lagrangian (AL) loop on top of the inner ADMM to ensure convergence under non-convex constraints. Several works such as \cite{sun2021two, tang2022fast} have followed a similar two-level setup, yet such schemes might require many iterations, limiting their applicability. In contrast to these approaches, we present a novel analysis for distributed ADMM with iterative linearization of the non-convex constraints which guarantees convergence to a stationary point. 

\subsection{Contributions}

This article introduces a family of \myemph{Distributed Covariance Steering (DCS)} methods based on ADMM that address these challenges. The contributions of this work are listed as follows:
\begin{enumerate}
    \item We present \myemph{Full-Covariance-Consensus (FCC)-DCS}, a distributed optimization approach which exploits both the mean and the full covariances of the states of the agents to effectively achieve safety.
    \item Next, we propose \myemph{Partial-Covariance-Consensus (PCC)-DCS}, a method which leverages that ensuring safety requires only partial covariance information, thus reducing computational and communication requirements. 
    \item Subsequently, we present \myemph{Mean-Consensus (MC)-DCS}, a further simplified approach which only requires a consensus among the mean states of the agents towards achieving even higher computational efficiency.
    \item We establish novel convergence guarantees for distributed ADMM with iteratively linearized non-convex constraints; a result of independent interest.
    As PCC-DCS and MC-DCS fall under this setup, their convergence to stationary points follows. We also discuss modifications for the convergence of FCC-DCS.
    \item We validate the proposed methods through simulations in 2D and 3D environments, highlighting their safety and scalability to systems with up to thousands of agents.
\end{enumerate}

\myemph{Paper Organization.} The rest of this article is organized as follows. Section \ref{sec: problem statement} formulates the multi-agent covariance steering problem. In Sections \ref{sec: fcc dcs}, \ref{sec: pcc dcs} and \ref{sec: mcc dcs}, we present the FCC-DCS, PCC-DCS and MC-DCS algorithms, respectively. Section \ref{sec: convergence analysis} provides the convergence analysis. In Section \ref{sec: simulation results}, we present simulation experiments that verify the effectiveness of the approaches. Finally, Section \ref{sec: conclusion} concludes this article.

\myemph{Notation.}
The space of positive (semi)definite matrices of dimension $n \times n$ is given by $\Sb_n^{++}$ ($\Sb_n^+$). 
%
The inner product between two vectors $x, y \in \Rb^n$ is denoted by $\langle x, y \rangle = x^\top y$, while the $\ell_2$-norm of $x$ is $\| x \|_2 = \sqrt{\langle x, x \rangle}$. 
Given a matrix $W \in \Sb_n^+$, we define the weighted semi-norm as $\| x \|_W = \sqrt{\langle x, W x \rangle}$.
The Frobenius inner product between two matrices $X, Y \in \Rb^{n \times m}$ is denoted with $\langle X, Y \rangle_{\rF} = \tr(X^\top Y)$, while the Frobenius norm of $X$ is $\| X \|_\rF = \sqrt{\langle X, X \rangle_\rF}$. 
Given a random variable (r.v.) $x$, we denote its mean with $\mu_x = \Eb[x]$ and its covariance with $\Sigma_x = \mathrm{Cov}[x]$. If a r.v. $x$ is Gaussian, then this is expressed as $x \sim \calN(\mu, \Sigma)$. The cumulative distribution function (CDF) of the standard Gaussian distribution is denoted by $\Phi(\cdot)$, while the CDF of the chi-squared distribution with $k$ d.o.f. is denoted by $F_{\chi_k^2}(\cdot)$. 
Further, given $a,b \in \Rb$, we denote the integer set $[a,b]~\cap~\Zb$ as $\llbracket a, b \rrbracket$. 
We say that a function $f: \Rb^n \rightarrow \Rb$, is $M$-partially strongly convex with $M \in \Sb_n^+$ if $f(x) - \frac{1}{2} \| x \|_M^2$ is convex.
%
%
Finally, we call a differentiable function $L$-partially smooth with 
$L \in \Sb_n^+$ if for any $x, y \in \dom f$, we have
$
\| \nabla f(y) - \nabla f(x)\|_2 \leq \| y - x \|_L^2.
$

\section{Problem Statement}
\label{sec: problem statement}
This section states the multi-agent covariance steering (MACS) problem considered in this article. Section \ref{subsec: problem setup} introduces the agent topology and local communication structure. Section \ref{subsec: dynamics cost constraints} details the dynamics, cost and constraints of the agents. The full MACS problem is formulated in Section \ref{subsec: problem formulation}.

\subsection{Agent Topology and Local Communication}
\label{subsec: problem setup}

Consider a set of $N$ agents $\calV =\{ 1, \dots, N \}$. Each agent $i \in \calV$ has a set of \textit{neighbors} $\calV_i \subseteq \calV$ (including $i$), typically comprising other agents in proximity to $i$. We adopt the following assumptions regarding the neighborhoods and local communication capabilities.

\refstepcounter{assumption}
\labeleditem{Assumption}{\theassumption}{Time-Invariant Neighborhoods}
The neighbor sets $\calV_i$, $i \in \calV$, remain fixed over time.

\refstepcounter{assumption}
\labeleditem{Assumption}{\theassumption}{Local Communication}
Each agent $i \in \calV$ can exchange information with all agents $j \in \calV_i$.

For convenience, we also define the \textit{neighbor-of} sets $\calW_i := \{ j \in \calV ~|~ i \in \calV_j \}$, which include all agents that consider $i$ as a neighbor. Note that $\calV_i$ and $\calW_i$ need not be equal.

\subsection{Dynamics, Cost and Constraints}
\label{subsec: dynamics cost constraints}

Each agent $i \in \calV$ is subject to the following stochastic discrete-time linear dynamics:
\begin{equation}
x_{k+1}^i = A_k^i x_k^i + B_k^i u_k^i + w_k^i,
\end{equation}
where $x_k^i \in \Rb^{n_i}$ is the state, $u_k^i \in \Rb^{m_i}$ is the control input, and the matrices $A_k^i \in \Rb^{n_i \times n_i}$ and $B_k^i \in \Rb^{n_i \times m_i}$ are known. The noise process $\{w_k^i\}_{k=0}^{T-1}$ over a time horizon $T$, is a sequence of independent and identically distributed zero-mean Gaussian r.v. $w_k^i \sim \mathcal{N}(0, W_k^i)$ with $W_k^i \in \Sb_{n_i}^+$ and $\Eb[w_k^i w_\kappa^{i \top}] = 0$ for any $k \neq \kappa$.
The initial state $x_0^i$ of each agent is also a Gaussian r.v. given by
\begin{equation}
x_0^i \sim \calN(\mu_\mathrm{s}^i, \Sigma_\mathrm{s}^i),
\end{equation}
with $\mu_\mathrm{s}^i \in \Rb^{n_i}$,  $\Sigma_\mathrm{s}^i \in \Sb_{n_i}^{++}$ and $\Eb[x_0^i w_k^{i \top}] = \Eb[w_k^i x_0^{i \top}] = 0$. The concatenated state, control and noise sequences over the horizon $T$ are defined as $x_i = [x_0^i; \dots; x_T^i] \in \Rb^{(T+1)n_i}$, $u_i = [u_0^i; \dots; u_{T-1}^i] \in \Rb^{T m_i}$ and $w_i = [w_0^i; \dots; w_{T-1}^i] \in \Rb^{T n_i}$. Hence, the dynamics over that horizon can be written in a compact form as
\begin{equation}
x_i = G_0^i x_0^i + G_u^i u_i + G_w^i w_i,
\label{eq: state compact}
\end{equation}
with the matrices $G_0^i$, $G_u^i$ and $G_w^i$ defined as in \cite{balci2024constrained}.

The aim of all agents is to minimize the collective objective
\begin{equation}
J = \sum_{i \in \calV} J_i(x_i, u_i),
\end{equation}
where each local cost function $J_i(x_i, u_i)$ is given by
\begin{equation}
J_i(x_i, u_i) = 
\Eb \bigg[ \sum_{k = 0}^T x_k^{i \top} Q_k^i x_k^i + \sum_{k = 0}^{T-1} u_k^{i \top} R_k^i u_k^i \bigg],
\label{eq: local cost definition}
\end{equation}
with $Q_k^i \in \Sb_{n_i}^+$ and $R_k^i \in \Sb_{m_i}^{++}$. 

For notational convenience, we refer to the state means and covariances as $\mu_k^i := \mu_{x_k^i}$ and $\Sigma_k^i := \Sigma_{x_k^i}$. The terminal distributions of the agents are constrained to satisfy:
\begin{align}
\mu_T^i = \mu_\rf^i, \quad 
\Sigma_T^i \preceq \Sigma_\rf^i, \quad \forall  i \in \calV,
\label{eq: terminal mean and cov constraints}
\end{align}
with $\mu_\mathrm{f}^i \in \Rb^{n_i}$ and $\Sigma_\mathrm{f}^i \in \Sb_{++}^{n_i}$.



In addition, we consider the following chance constraints for obstacle avoidance:
\begin{equation}
\Pb [x_k^i \notin \calR_o] \geq 1 - \epsilon, \quad
\forall 
k \in \llbracket 0, T \rrbracket, ~
o \in \calO, ~ 
i \in \calV,
\label{eq: obs avoid chance constraint - set version}   
\end{equation}
where $\calO$ is the set of obstacles, and $\calR_o$ is the region covered by each obstacle $o \in \calO$. Assuming spherical obstacles, these constraints can be further formulated as
\begin{equation}
\Pb [c_k^{io}(x_k^i) \leq 0] \geq 1 - \epsilon, \quad
\forall 
k \in \llbracket 0, T \rrbracket, ~
o \in \calO, ~ 
i \in \calV,
\label{eq: obs avoid chance constraint}   
\end{equation}
with 
$
c_k^{io}(x_k^i) = - \| p_k^i - p_o \|_2 + s_o,
$
where $p_k^i = P_i x_k^i \in \Rb^q$, $q \in \{2,3\}$ for 2D or 3D space, denotes the position of agent $i$, with matrix $P_i \in \Rb^{q \times n_i}$ defined accordingly, and $p_o \in \Rb^q$ and $s_o > 0$ are the center and radius of obstacle $o$.

Furthermore, we consider the inter-agent collision avoidance chance constraints:
\begin{equation}
\begin{aligned}
& \Pb [d_k^{ij}(x_k^i, x_k^j) \leq 0] \geq 1 - \epsilon,
\\
& ~~~~~~~~ \forall k \in \llbracket 0, T \rrbracket, ~
j \in \calV_i, ~
i \in \calV,
\end{aligned}
\label{eq: coll avoid chance constraint}  
\end{equation}
with 
$
d_k^{ij}(x_k^i, x_k^j) = - \| p_k^i - p_k^j \|_2 + s_{ij}, 
$
where $s_{ij} > 0$ is the minimum allowed distance between agents $i$ and $j$.

\refstepcounter{remark}
\labeleditem{Remark}{\theremark}{Additional Convex Constraints}
It is straightforward to incorporate additional constraints such as linear state or control chance constraints \cite{okamoto2019optimal}, bounds on the expected control effort \cite{balci2024constrained}, equality constraints on the state covariances \cite{rapakoulias2023discrete}, communication maintenance constraints, etc., since these typically admit convex reformulations. However, the primary focus of this article is on the more challenging case of non-convex constraints arising from collision avoidance, which are central to ensuring safety in multi-agent systems.

%

%
\subsection{Problem Formulation}
\label{subsec: problem formulation}

With the agent topology, dynamics, cost, and constraints defined, we now formally state the MACS problem.

\refstepcounter{problem}
\labeleditem{Problem}{\theproblem}{MACS Problem}
%
Find the optimal control sequences $u_i^*$, for all $i \in \calV$, that solve: 
\label{problem: macs}
\begin{align*}
& \qquad \qquad \qquad \min \sum_{i \in \calV} J_i(x_i, u_i)
\\[0.05cm]
\mathrm{s.t.} ~~
& x_{k+1}^i = A_k^i x_k^i + B_k^i u_k^i + w_k^i, 
\\
& \mu_0^i = \mu_\mathrm{s}^i, ~ \Sigma_0^i = \Sigma_\mathrm{s}^i, 
~ \mu_T^i = \mu_\mathrm{f}^i, ~ \Sigma_T^i \preceq \Sigma_\mathrm{f}^i, 
\\
& \Pb [c_k^{io}(x_k^i) \leq 0] \geq 1 - \epsilon,
~ 
\Pb [d_k^{ij}(x_k^i, x_k^j) \leq 0] \geq 1 - \epsilon,
\\
& \forall k \in \llbracket 0, T \rrbracket,
~ o \in \calO,
~ j \in \calV_i, 
~ i \in \calV.
\end{align*}

\section{Full-Covariance-Consensus \\ Distributed  Covariance Steering}
\label{sec: fcc dcs}

This section introduces the \myemph{Full-Covariance-Consensus (FCC)-DCS} approach for addressing the MACS problem. Section \ref{subsec: fcc dcs - problem transformation} presents a tractable transformation of the original problem. In Section \ref{subsec: fcc dcs - consensus opt}, we cast this reformulation as a consensus optimization. The derivation of FCC-DCS, as well as the final algorithm, are presented in Section \ref{subsec: fcc dcs - method}.  

\subsection{Problem Transformation}
\label{subsec: fcc dcs - problem transformation}

%
%
Let us consider affine disturbance history feedback control policies as in \cite{balci2024constrained}, for each agent $i \in \calV$,
%
%
\begin{equation}
u_k^i = v_k^i + \sum_{\kappa = k - \gamma}^k K_{k,\kappa}^i w_{\kappa}^i,
\label{eq: control policy}
\end{equation}
where $v_k^i \in \Rb^{m_i}$ are feed-forward control inputs, $K_{k,\kappa}^i \in \Rb^{m_i \times n_i}$ are feedback gains and $\gamma \in \llbracket 1, T \rrbracket$ is a truncation parameter that defines the length of the history of disturbances, with the convention that $w_{-1}^i = x_0^i - \mu_{\rs}^i$. As shown in \cite{balci2024constrained}, $\gamma$ equal to $2$ or $3$ works well in practice. Then, the control and state sequences are given by
%
%
\begin{align}
u_i & 
= v_i + K_i \hat{w}_i,
\label{eq: control compact}
\\[0.05cm]
x_i & = (\hat{G}_i + G_u^i K_i) \hat{w}_i + G_u^i v_i + G_0^i \mu_{\rs}^i,
\label{eq: state compact 2}
\end{align}
where $v_i \! = \! [v_i^0; \dots; v_i^{T-1}] \in \Rb^{T m_i}$,  
$K_i \in \Rb^{T m_i \times (T+1) n_i}$ is
\begin{equation}
[K_i]_{k,\kappa} = \begin{cases} K_{k,\kappa}^i & \text{if } k - \gamma \leq \kappa \leq k \\ 0 & \text{otherwise} \end{cases},
\end{equation}
$\hat{w}_i = [x_0^i - \mu_{\rs}^i; w_0^i; \dots; w_{T-1}^i] \in \Rb^{(T+1) n_i}$ 
and 
$\hat{G}_i = [G_0^i, G_w^i]$.
%
%
%
The mean $\mu_i := \mu_{x_i}$ and covariance $\Sigma_i := \Sigma_{x_i}$ of the state sequence $x_i$ are provided by
\begin{equation}
\mu_i = \theta_i (v_i), \quad  
\Sigma_i = \Theta_i (K_i) \Theta_i (K_i)^\top,
\end{equation}
where $\theta_i (v_i) \in \Rb^{(T+1) n_i}$ and $\Theta_i (K_i) \in \Rb^{(T+1) n_i \times (T+1) n_i}$ are affine functions given by
\begin{align}
\theta_i (v_i) & = G_0^i \mu_0^i + G_u^i v_i,
\\
\Theta_i (K_i) & = (\hat{G}_i + G_u^i  K_i) \mathrm{bdiag}(\Sigma_{\rs}^i, W_i)^{1/2},
\end{align}
with $W_i = \bdiag(W_0^i, \dots, W_T^i)$. 

It follows that each local cost function \eqref{eq: local cost definition} is then given by
\begin{align}
& \calJ_i(v_i, K_i) = 
\theta_i (v_i)^\top Q_i \theta_i (v_i)
+ v_i^\top R_i v_i
\label{eq: local cost definition 2}
\\
& + \tr \left[Q_i \Theta_i (K_i) \Theta_i (K_i) \! {}^\top \right] + \tr \left[ R_i K_i \mathrm{bdiag}(\Sigma_{\rs}^i, W_i) K_i^\top \right],
\nonumber
\end{align}
with $Q_i = \bdiag(Q_0^i, \dots, Q_T^i)$, $R_i = \bdiag(R_0^i, \dots, R_{T-1}^i)$.
In addition, the terminal mean and covariance constraints \eqref{eq: terminal mean and cov constraints}  can be expressed as the linear equality constraint:
\begin{equation}
\cala_i(v_i) := \Gamma_T^i \theta_i (v_i) - \mu_\rf^i = 0,
\label{eq: cala terminal mean constraint}
\end{equation}
and the linear matrix inequality (LMI) constraint:
\begin{equation}
\calB_i(K_i) := 
\begin{bmatrix}
\Sigma_\rf^i & \Gamma_T^i \Theta_i (K_i)
\\[0.05cm]
\Theta_i (K_i)^\top \Gamma_T^{i \top} & I_{(T+1) n_i}
\end{bmatrix}
\succeq 0,
\label{eq: calB for terminal mean constraint}
\end{equation}
respectively, where the matrix $\Gamma_k^i \in \Rb^{n_i \times (T+1) n_i}$ is defined such that $x_k^i = \Gamma_k^i x_i$, and the Schur complement is used
for reformulating $\Gamma_T^i \Theta_i (K_i) \Theta_i (K_i)^\top \Gamma_T^{i \top} \preceq \Sigma_\rf^i$ as \eqref{eq: calB for terminal mean constraint}.

\refstepcounter{remark}
\labeleditem{Remark}{\theremark}{Alternative Control Policy Parametrizations}
Several other policy parametrizations can be considered, based on state feedback or other auxiliary variables; for an overview, we refer the reader to \cite{balci2024constrained}. In this work, we adopt the disturbance feedback parametrization, as it offers a favorable balance between performance and computational tractability, and in addition, yields convex reformulations of linear chance constraints. Yet, the proposed algorithms can be extended to any available policy parametrization.

The most challenging constraints in Problem \ref{problem: macs} are the obstacle and inter-agent safety constraints due to their non-convex nature. In the following, we provide convex tractable reformulations for approximating these constraints. We start with reformulating the inter-agent collision avoidance ones as second-order conic (SOC) constraints as follows.

\refstepcounter{proposition}
\labeleditem{Proposition}{\theproposition}{Convex Approximation of Collision Avoidance Chance Constraints via Inner Linearization}
\label{prop: fcc dcs - linearized chance constraint}
%
The non-convex inter-agent collision avoidance chance constraint \eqref{eq: coll avoid chance constraint} is satisfied if the following SOC constraint holds:
\begin{equation}
\cald_{ij, k}^{\text{FCC}}(v_i, K_i, v_j, K_j) \leq 0,
\label{eq: fcc - coll avoid chance constraint}
\end{equation}
where
\begin{align}
\cald_{ij, k}^{\text{FCC}}& := 
\Phi^{-1}(1 - \epsilon)
\!
\left \lVert 
{\setlength{\arraycolsep}{1.5pt}
\begin{bmatrix}
P_i \Gamma_k^i \Theta_i (K_i) & 0 \\
0 & P_j \Gamma_k^j \Theta_j (K_j)
\end{bmatrix}
}^\top 
\!\!
\begin{bmatrix}
a_k^{ij}
\\[0.1cm]
a_k^{ij}
\end{bmatrix} \right \rVert_2
\nonumber
\\[0.05cm]
& ~~~~~~~~~~ - a_k^{ij \top} (P_i \Gamma_k^i \theta_i (v_i) - P_j \Gamma_k^j \theta_j (v_j)) - b_k^{ij},
\end{align}
with $a_k^{ij} = 2 (\hat{p}_k^i - \hat{p}_k^j)$, $b_k = - \| \hat{p}_k^i - \hat{p}_k^j \|_2^2 - s_{ij}^2$. The approximation points $\hat{p}_k^i$ and $\hat{p}_k^j$ are selected such that $\| \hat{p}_k^i - \hat{p}_k^j \|_2 = s_{ij}$.

%
%
\begin{proof}
Let us define the r.v. $q_k^{ij} = p_k^i - p_k^j$. Then, the chance constraint \eqref{eq: coll avoid chance constraint} can be rewritten as 
\begin{equation}
\Pb[\| q_k^{ij} \|_2^2 \geq s_{ij}^2] \geq 1 - \epsilon,
\label{eq: inter-agent chance constraint z}
\end{equation}
and it follows that $q_k \sim \calN(\mu_{q_k}, \Sigma_{q_k})$ with $\mu_{q_k} = \mu_{p_i^k} - \mu_{p_j^k}$ and $\Sigma_{q_k} = \Sigma_{p_i^k} + \Sigma_{p_j^k}$, where we temporarily drop the superscript $(ij)$ for notational convenience. By linearizing the inner part of the LHS of \eqref{eq: inter-agent chance constraint z} around a point $\hat{q}_k$ that satisfies $\| \hat{q}_k \|_2 = s_{ij}$, we obtain 
$
\| \hat{q}_k \|_2^2 + 2 \hat{q}_k^\top (q_k - \hat{q}_k) 
$
which yields the constraint
$
a_k^\top q_k + b_k \geq 0,
$
with
$a_k = 2 \hat{q}_k$ and $b_k = - \| \hat{q}_k \|_2^2 - s_{ij}^2$.
Note that this is a convex under-approximation of the constraint $\| q_k \|_2^2 \geq s_{ij}^2$,
thus $\mathbb{P}(a_k^\top q_k + b_k \geq 0 ) \leq \mathbb{P}(\| q_k \|_2^2 \geq s_{ij}^2)$. Consequently, the constraint
\begin{equation}
\Pb[a_k^\top q_k + b_k \geq 0] \geq 1 - \epsilon, 
\label{eq: inter-agent chance constraint z linearized}
\end{equation}
is a sufficient condition for constraint \eqref{eq: inter-agent chance constraint z} to hold.

Next, since $q_k$ is a multivariate Gaussian r.v., then $a_k^\top q_k + b_k$ is univariate Gaussian, thus constraint \eqref{eq: inter-agent chance constraint z linearized} is equivalent with 
\begin{equation}
\Phi \left( (a_k^\top \mu_{q_k} + b_k) / \sqrt{a_k^\top \Sigma_{q_k} a_k} \right) \geq 1 - \epsilon.
\end{equation}
%
%
%
Since $\Phi(\cdot)$ is a monotonically increasing function, we get
\begin{equation}
\! \! \! a_k^{ij \top} \! (\mu_{p_k^i} \! - \! \mu_{p_k^j}) + b_k^{ij} \geq \Phi^{-1}(1 \! - \! \epsilon) \sqrt{ a_k^{ij \top} \! (\Sigma_{p_k^i} \! + \! \Sigma_{p_k^j}) a_k^{ij} }
\!
\label{eq:inequality-avoidance}
\end{equation}
%
%
which yields the constraint \eqref{eq: fcc - coll avoid chance constraint}.
\end{proof}

Subsequently, we derive a similar SOC approximation for the obstacle avoidance chance constraints as well.

\refstepcounter{proposition}
\labeleditem{Proposition}{\theproposition}{Convex Approximation of Obstacle Avoidance Chance Constraints via Inner Linearization}
\label{prop: fcc dcs - linearized obs chance constraint}
The non-convex obstacle avoidance chance constraint \eqref{eq: obs avoid chance constraint} is satisfied if the following SOC constraint holds:
\begin{equation}
\calc_{io,k}^{\text{FCC}}(v_i, K_i) \leq 0,
\end{equation}
where
\begin{align}
\calc_{io,k}^{\text{FCC}} & := 
\Phi^{-1}(1-\epsilon) \left \lVert 
\left( P_i \Gamma_k^i \Theta_i (K_i) \right) ^\top 
a_k^{io}
 \right \rVert_2
\\[0.05cm]
& ~~~~~~~~~~ - a_k^{io \top} (P_i \Gamma_k^i \theta_i (v_i) - p_k^o) - b_k^{io},
\nonumber
\end{align}
with $a_k^{io} = 2 (\hat{p}_k^i - p_o)$, $b_k = - \| \hat{p}_k^i - p_o \|_2^2 - s_o^2$. The approximation point $\hat{p}_k^i$ is selected such that $\| \hat{p}_k^i - p_o \|_2 = s_o$.

\begin{proof}
Similar to Proposition \ref{prop: fcc dcs - linearized chance constraint} and is thus omitted.
\end{proof}

For notational convenience, we define the concatenated constraints $\calc_i^{\text{FCC}}(v_i, K_i) := \big[ \{ \calc_{io,k}^{\text{FCC}}(v_i, K_i) \}_{o \in \calO, k \in \llbracket 0,T \rrbracket} \big] \leq 0$ and $\cald_{ij}^{\text{FCC}}(v_i, K_i, v_j, K_j) := \big[ \{ \cald_{ij,k}^{\text{FCC}}(v_i, K_i, v_j, K_j) \}_{k \in \llbracket 0,T \rrbracket} \big] \leq 0$.
Therefore, a convex approximation of Problem \ref{problem: macs} can be formulated through the following optimization problem. We refer to this formulation as the Full-Covariance-Constrained (FCC) variation since both the obstacle and collision avoidance constraints exploit the full state covariance to enforce safety.

\refstepcounter{problem}
\labeleditem{Problem}{\theproblem}{MACS - Full-Covariance Constrained Reformulation}
\label{problem: multi-agent cs - fcc version - before consensus}
Find the optimal policies $\{v_i^*,  K_i^* \}_{i \in \calV}$ such that
\begin{align*}
& ~~~~~~~~ \min \sum_{i \in \calV} \calJ_i(v_i, K_i)
\\
\mathrm{s.t.} \quad 
& \cala_i(v_i) = 0,
~ \calB_i(K_i) \succeq 0,
~ \calc_i^{\text{FCC}}(v_i, K_i) \leq 0,
\\
& \cald_{ij}^{\text{FCC}}(v_i, K_i, v_j, K_j) \leq 0, ~~ \forall j \in \calV_i, ~ i \in \calV.
\end{align*}




\refstepcounter{remark}
\labeleditem{Remark}{\theremark}{Scalability Limitations of Centralized Approach}
Solving Problem \ref{problem: multi-agent cs - fcc version - before consensus} in a centralized manner yields an optimization with 
$N T (m_i + \gamma n_i m_i)$ variables, 
$N$ LMI constraints of dim. $(T+2) n_i$, 
$N T (|\calV_i| + |\calO|)$ SOC constraints of dim. $2 (T+1) n_i$ and 
$N n_i$ linear equality constraints  (see Table \ref{tab: distr cs - alg dims}).
As the number of agents $N$ increases, the dimension of the centralized problem renders it intractable for large-scale systems, motivating the need for distributed architectures.


\subsection{Consensus Optimization}
\label{subsec: fcc dcs - consensus opt}

Problem \ref{problem: multi-agent cs - fcc version - before consensus} cannot be directly solved in a decentralized manner due to the inter-agent constraints $\cald_{ij}^{\text{FCC}}(v_i, K_i, v_j, K_j) \leq 0$. To address this, for each agent $i \in \calV$, we introduce the \textit{copy variables} $v_j^{(i)}, K_j^{(i)}$, $j \in \calV_i$, which represent the decisions of agent $i$ regarding their neighbors $j \in \calV_i$. It follows that we can then define the augmented (local) decision variables: 
\begin{equation}
\tilde{v}_i = [\{ v_j^{(i)} \}_{j \in \calV_i}], \quad
\tilde{K}_i = [\{ K_j^{(i)} \}_{j \in \calV_i}].
\end{equation}
Hence, the inter-agent constraints can now be reformulated from the perspective of each agent $i \in \calV$ as 
\begin{equation}
\tilde{\cald}_i^{\text{FCC}}(\tilde{v}_i, \tilde{K}_i) :=
[\{ \cald_{ij}^{\text{FCC}}(v_i, K_i, v_j^{(i)}, K_j^{(i)})
\}_{j \in \calV_i}]
\leq 0.
\end{equation}
However, introducing these additional variables necessitates a consensus among those corresponding to the same agent. To achieve this, we introduce the global variables 
$z = [ \{ z_i \}_{i \in \calV} ]$, 
$Z = [ \{ Z_i \}_{i \in \calV} ]$, 
and impose the consensus constraints
\begin{align}
v_j^{(i)} = z_j, \quad K_j^{(i)} = Z_j, 
\quad \forall j \in \calV_i, ~ i \in \calV,
\end{align}
or written more compactly
\begin{equation}
\tilde{v}_i = \tilde{z}_i, \quad 
\tilde{K}_i = \tilde{Z}_i, \quad
\forall i \in \calV,
\end{equation}
with $\tilde{z}_i := [\{ z_j \}_{j \in \calV_i}]$ and $\tilde{Z}_i := [\{ Z_j \}_{j \in \calV_i}]$.

Therefore, Problem \ref{problem: multi-agent cs - fcc version - before consensus} can be equivalently reformulated as the following consensus optimization problem.

\refstepcounter{problem}
\labeleditem{Problem}{\theproblem}{MACS - Full-Covariance Consensus Reformulation}
Find the optimal policies $\{v_i^*,  K_i^* \}_{i \in \calV}$ such that
\label{problem: multi-agent cs - fcc version - consensus}
\begin{align*}
& ~~~~~~~~~~~~~ \min \sum_{i \in \calV} \calJ_i(v_i, K_i)
\\
\mathrm{s.t.} \quad 
& \cala_i(v_i) = 0,
~ \calB_i(K_i) \succeq 0,
~ \calc_i^{\text{FCC}}(v_i, K_i) \leq 0,
\\
& \tilde{\cald}_i^{\text{FCC}}(\tilde{v}_i, \tilde{K}_i) \leq 0,
~ \tilde{v}_i = \tilde{z}_i, 
~ \tilde{K}_i = \tilde{Z}_i, 
~ \forall i \in \calV.
\end{align*}

\subsection{Method}
\label{subsec: fcc dcs - method}

We proceed with deriving a distributed algorithm for solving \eqref{problem: multi-agent cs - fcc version - consensus}. To achieve that, we treat $\tilde{v} = [\{ \tilde{v}_i \}_{i \in \calV}], \tilde{K} = [\{ \tilde{K}_i \}_{i \in \calV}]$ as the first, and $z, Z$ as the second block of variables, following the two-block ADMM derivation \cite{boyd2011distributed}. The AL is given by
\begin{align}
& \calL_{\rho} = 
\sum_{i \in \calV} 
\calJ_i(v_i, K_i) 
+ \calI_{\cala_i, \calB_i, \calc_i^{\text{FCC}}, \tilde{\cald}_i^{\text{FCC}}}(\tilde{v}_i, \tilde{K}_i)
+ \langle y_i, \tilde{v}_i - \tilde{z}_i \rangle
\nonumber
\\
& ~~~ + \! \langle Y_i, \tilde{K}_i - \tilde{Z}_i \rangle_\rF 
\! + \! \frac{\rho_v}{2} \| \tilde{v}_i - \tilde{z}_i \|_2^2
\! + \! \frac{\rho_K}{2} \| \tilde{K}_i - \tilde{Z}_i \|_\rF^2,
\end{align}
where $y_i$ and $Y_i$ are the dual variables for the constraints $\tilde{v}_i = \tilde{z}_i$ and $\tilde{K}_i = \tilde{Z}_i$, and $\rho_v, \rho_K > 0$ are penalty parameters.

\myemph{Local primal updates.} The first block is derived through
\begin{equation}
\{\tilde{v}, \tilde{K}\}^{\ell+1} = 
\argmin_{\tilde{v}, \tilde{K}}
\calL_\rho(\tilde{v}, \tilde{K}, \{z, Z, y, Y \}^\ell)
\end{equation}
which yields the following parallelizable local subproblems
\begin{equation}
\begin{aligned}
& \{\tilde{v}_i, \tilde{K}_i \}^{\ell+1} = 
\argmin
\tilde{\calJ}_i^{\text{FCC}}(\tilde{v}_i, \tilde{K}_i)
\\
\mathrm{s.t.} \quad 
& \cala_i(v_i) = 0,
~ \calB_i(K_i) \succeq 0,
\\
& \calc_i^{\text{FCC}}(v_i, K_i) \leq 0, 
~ \tilde{\cald}_i^{\text{FCC}}(\tilde{v}_i, \tilde{K}_i) \leq 0,
\end{aligned}
\label{eq: fcc-dcs - local updates}
\end{equation}
with 
\begin{align*}
\tilde{\calJ}_i^{\text{FCC}} 
(\tilde{v}_i, \tilde{K}_i) 
& := \calJ_i(v_i, K_i)
+ \langle y_i^\ell, \tilde{v}_i \rangle
+ \langle Y_i^\ell, \tilde{K}_i \rangle
\\
& ~~~~~~~~~ 
+ \frac{\rho_v}{2} \| \tilde{v}_i - \tilde{z}_i^\ell \|_2^2
+ \frac{\rho_K}{2} \| \tilde{K}_i - \tilde{Z}_i^\ell \|_\rF^2.
\nonumber
\end{align*}

\refstepcounter{remark}
\labeleditem{Remark}{\theremark}{Successive Convex Approximations for Local Subproblems}
\label{remark: successive convexification}
The constraint functions $\calc_i^{\text{FCC}}(v_i, K_i)$ and $\tilde{\cald}_i^{\text{FCC}}(\tilde{v}_i, \tilde{K}_i)$ are recomputed at each ADMM round based on the current iterates, to yield more accurate convex approximations of the original non-convex constraints in \eqref{eq: obs avoid chance constraint} and \eqref{eq: coll avoid chance constraint}.

\begin{algorithm}[t]
\caption{Full-Covariance-Consensus DCS (FCC-DCS)}\label{FCC-DCS Algorithm}
\begin{algorithmic}[1] 
\State \textbf{Initialize:} 
$\tilde{v}_i, \tilde{K}_i, z_i, Z_i, y_i, Y_i \leftarrow 0$. 
\While{not converged \textbf{and} $\ell \leq \ell_{\text{max}}$}
\State $\calc_{i, \text{lin}}^{\text{PCC}}, \tilde{\cald}_{i, \text{lin}}^{\text{PCC}} \leftarrow$ Get convexified constraints. 
\State $\!$$\tilde{v}_i, \tilde{K}_i \leftarrow$ Solve \eqref{eq: fcc-dcs - local updates} in parallel $ \forall \ i \in \calV$.
\State $\!$\textit{Each agent $i \! \in \! \calV$ receives $v_i^{(j)} \! \! \!, K_i^{(j)}$ from all $j \! \in \! \calW_i $\!$ \backslash  \{  i \}$.} 
\State $\!$$z_i, Z_i \leftarrow$ Update with \eqref{eq: fcc-dcs - global updates} in parallel $ \forall \ i \in \calV$.
\State $\!$\textit{Each agent $i \in \calV$ receives $z_j, Z_j$ from all $j \in \calV_i \backslash \{i\}$.} 
\State $\!$$y_i, Y_i \leftarrow$ Update with \eqref{eq: fcc-dcs - dual updates} in parallel $ \forall \ i \in \calV$.
\EndWhile
\end{algorithmic}
\label{alg: fcc-dcs}
\end{algorithm}

\myemph{Global primal updates.} The second block is derived as 
\begin{equation}
\{z, Z\}^{\ell+1} = \argmin_{z, Z}
\calL_\rho( \{ \tilde{v}, \tilde{K} \}^{\ell+1}, z, Z, \{ y, Y \}^\ell)
\end{equation}
which results into the parallelizable updates
\begin{equation}
z_i^{\ell+1} \! = \! \frac{1}{|\calW_i|} \! \sum_{j \in \calW_i} \! v_i^{(j), \ell+1}, 
~~
Z_i^{\ell+1} \! = \! \frac{1}{|\calW_i|} \! \sum_{j \in \calW_i} \! K_i^{(j), \ell+1}.
\label{eq: fcc-dcs - global updates}
\end{equation}

\myemph{Dual updates.} Finally, the dual variables are updated through the following dual ascent steps
\begin{subequations}
\begin{align}
y_i^{\ell+1} & = y_i^\ell + \rho_v (\tilde{v}_i^{\ell+1} - \tilde{z}_i^{\ell+1}), 
\\
Y_i^{\ell+1} & = Y_i^\ell + \rho_K (\tilde{K}_i^{\ell+1} - \tilde{Z}_i^{\ell+1}).
\end{align}
\label{eq: fcc-dcs - dual updates}%
\end{subequations}
\indent \myemph{Algorithm.} The FCC-DCS algorithm is detailed in Alg. \ref{alg: fcc-dcs}. 
During each ADMM round, the local variables $\tilde{v}_i, \tilde{K}_i$ are first updated via solving subproblems \eqref{eq: fcc-dcs - local updates}. Then, each agent $i$ receives the copy variables $v_i^{(j)}, K_i^{(j)}$ from all $j \in \calW_i \backslash \{i\}$, so that the global variables $z_i, Z_i$ are updated with \eqref{eq: fcc-dcs - global updates}. Next, each agent $i$ receives $z_j, Z_j$ from all $j \in \calV_i \backslash \{i\}$, and the dual updates \eqref{eq: fcc-dcs - dual updates} take place. The ADMM loop repeats until a termination criterion is met.

\refstepcounter{remark}
\labeleditem{Remark}{\theremark}{Decentralized Structure of FCC-DCS}
\label{remark: fcc-dcs decentralized}
The FCC-DCS algorithm is fully decentralized since all computations can be parallelized across the agents and only local communication is required.



\refstepcounter{remark}
\labeleditem{Remark}{\theremark}{Computational Benefits of FCC-DCS}
The FCC-DCS method is substantially more computationally efficient than centralized CS, as the local subproblems \eqref{eq: fcc-dcs - local updates} involve only 
$|\calV_i| T (m_i + \gamma n_i m_i)$ variables, 
a single LMI constraint of dim. $(T+2) n_i$, 
$T (|\calV_i| + |\calO|)$ SOC constraints of dim. $2 (T+1) n_i$ and 
$n_i$ linear equality constraints  (see Table \ref{tab: distr cs - alg dims}), and are solved in parallel.
Furthermore, the amount of required ADMM rounds $H$ to achieve an acceptable accuracy typically ranges from tens to hundreds in practice \cite{boyd2011distributed}. Therefore, given that for large-scale systems $|\calV_i| \ll M$, FCC-DCS offers a substantial computational improvement. 

\section{Partial-Covariance-Consensus \\ Distributed Covariance Steering}
\label{sec: pcc dcs}

This section presents the \myemph{Partial-Covariance-Consensus (PCC)-DCS} method which further reduces the computational burden of solving the MACS problem. In Section \ref{subsec: fcc dcs - problem transformation}, we present a reformulation that substantially reduces the number of variables and computationally demanding constraints, and in Section \ref{subsec: fcc dcs - consensus opt}, we cast this new problem again as a consensus optimization. Section \ref{subsec: fcc dcs - method} presents the derivation and final algorithm for PCC-DCS.  

\subsection{Problem Transformation}

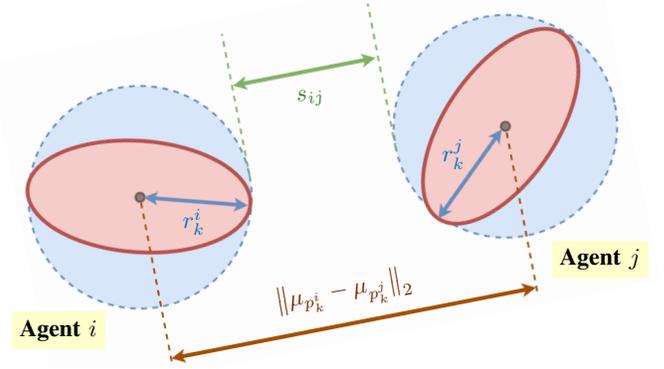
\begin{figure}[t]
\centering
\begin{tikzpicture}
    \node[anchor=south west,inner sep=0] at (0,0){    \includegraphics[width=0.5\textwidth, trim={0cm 0.5cm 0cm 0.0cm
         },clip]{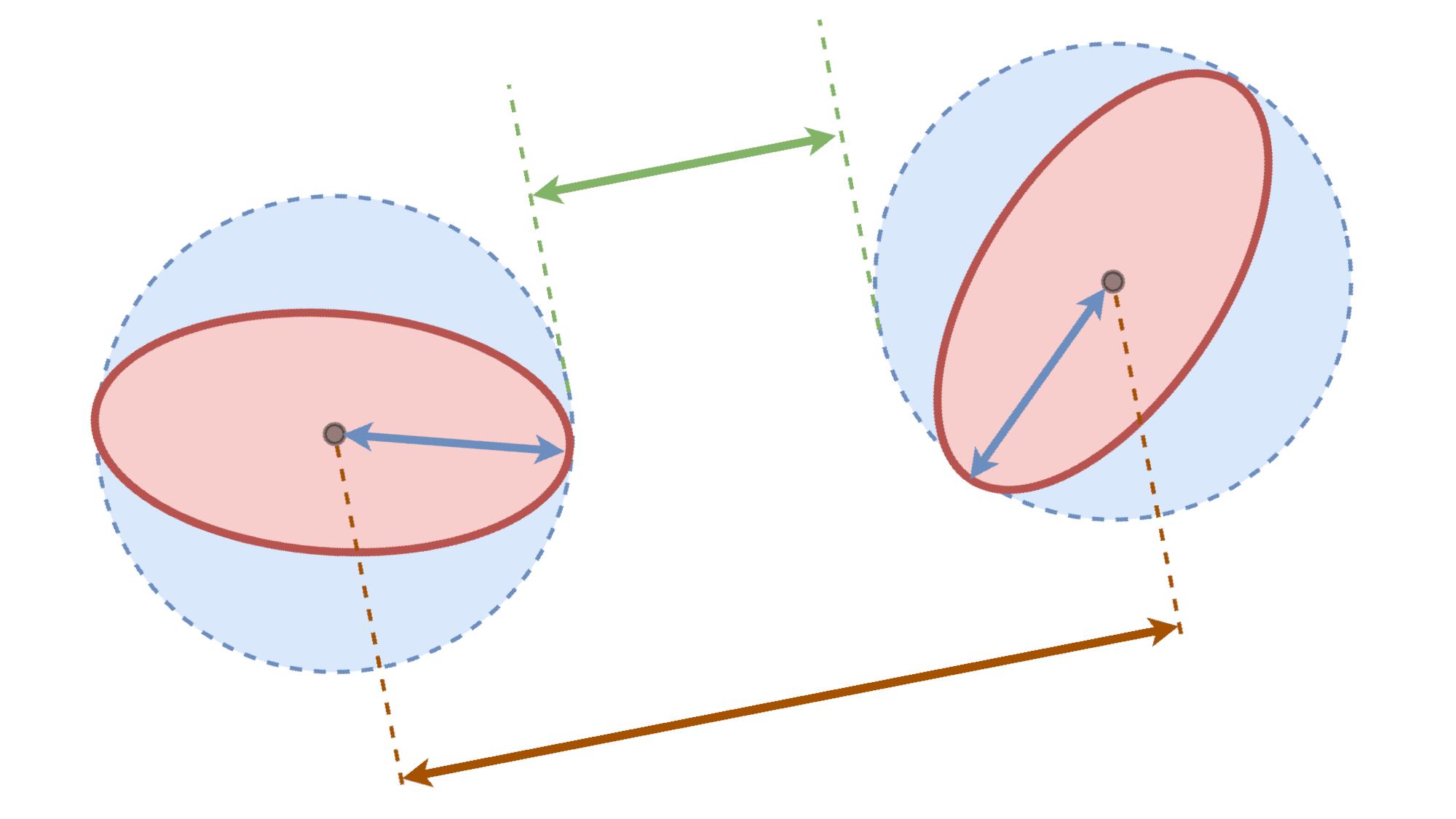}};
    \node[align=center, fill = myYellow, scale = 0.9] (c) at (1.0, 0.5) {\textbf{Agent} $i$};
    \node[align=center, fill = myYellow, scale = 0.9] (c) at (8.2, 1.45) {\textbf{Agent} $j$};
    \node[align=center, color = OliveGreen, scale = 0.9, rotate = 10] (c) at (4.35, 3.6) {$s_{ij}$};
    \node[align=center, color = Brown, scale = 0.9, rotate = 10] (c) at (4.8, 1.0) {$\big\| \mu_{p_k^i} - \mu_{p_k^j} \big\|_2$};
    \node[align=center, color = NavyBlue, scale = 0.9, rotate = 0] (c) at (2.8, 1.95) {$r_k^i$};
    \node[align=center, color = NavyBlue, scale = 0.9, rotate = 0] (c) at (6.25, 2.85) {$r_k^j$};
\end{tikzpicture}
\caption{Illustration of inter-agent constraint components via confidence ball separation in the PCC-DCS method.}
\vspace{-0.5cm}
\label{fig: pcc-dcs constraints separation}
\end{figure}

The key insight underlying the PCC-DCS approach is that to enforce the probabilistic safety constraints, it is not necessary to leverage---and therefore establish consensus upon---the full covariance information of the agents, but only the part associated with the major axis of their confidence ellipsoids. This is formalized through the following proposition in terms of the inter-agent collision avoidance constraints (Fig. \ref{fig: pcc-dcs constraints separation}).

\refstepcounter{proposition}
\labeleditem{Proposition}{\theproposition}{Sufficient Conditions for Collision Avoidance  via Confidence Ball Separation}
\label{prop: pcc-dcs main prop}
The non-convex chance constraint \eqref{eq: coll avoid chance constraint} is satisfied if the following constraints hold:
\begin{subequations}
\begin{align}
\phi_k^{ij}(\mu_k^i, \mu_k^j, r_k^i, r_k^j) \! & := \!
\| \mu_{p_k^i} \! - \! \mu_{p_k^j} \|_2 \! - \! r_k^i \! - \! r_k^j \! - \! s_{ij} \geq 0,
\label{eq: pcc-dcs proposition - system - mean}
\\
\alpha_k^i(\Sigma_k^i, r_k^i) & := \sqrt{\beta_i \lambda_{\max} (\Sigma_{p_k^i})} - r_k^i \leq 0,
\label{eq: pcc-dcs proposition - system - cov i}
\\
\alpha_k^j(\Sigma_k^j, r_k^j) & := \sqrt{\beta_j \lambda_{\max} (\Sigma_{p_k^j})} - r_k^j \leq 0,
\label{eq: pcc-dcs proposition - system - cov j}
\end{align}
\label{eq: pcc-dcs proposition - system}%
\end{subequations}
where $r_k^i$, $r_k^j > 0$ are auxiliary variables,
$\beta_i = F_{\chi_q^2}^{-1}(1 - \epsilon_i)$, 
$\beta_j = F_{\chi_q^2}^{-1}(1 - \epsilon_j)$
and
$\epsilon_i + \epsilon_j \leq \epsilon$.

\begin{proof} Given a multivariate Gaussian variable $x \sim \calN(\mu, \Sigma)$ with $\mu \in \Rb^n$, $\Sigma \in \Sb_n^{++}$, the confidence ellipsoid $\calC_{1 - \epsilon}(x)$ such that $\Pb[x \in \calC_{1 - \epsilon}(x)] = 1 - \epsilon$, is given by $\calC_{1 - \epsilon}(x) := \{ x:~ (x -\mu)^\top \Sigma^{-1} (x -\mu) \leq \beta \}$ with $\beta = F_{\chi_n^2}^{-1}(1 - \epsilon)$. 
Let us denote the confidence ellipsoids of $p_k^i$ and $p_j^k$ as 
$\calC_i := \calC_{1 - \epsilon_i}(p_k^i)$ 
and 
$\calC_j := \calC_{1 - \epsilon_j}(p_k^j)$, respectively.
If $p_k^i \in \calC_i$ and $p_k^j \in \calC_j$, 
then by definition we have $\Pb[p_k^i \in \calC_i] = 1 - \epsilon_i$ and $\Pb[p_k^j \in \calC_j] = 1 - \epsilon_j$. Now, let us also define the ball over-approximations of these ellipsoids as:
\begin{subequations}
\begin{align}
\!\! \hat{\calC}_i \! & := \! \{ p_k^i \! : \! 
\lambda_{\max}(\Sigma_{p_k^i})^{-1}
(p_k^i -\mu_{p_k^i})^{\! \top} \!
(p_k^i -\mu_{p_k^i}) \leq \beta_i \},
\\[0.05cm]
\!\! \hat{\calC}_j \! & := \! \{ p_k^j \! : \!
\lambda_{\max}(\Sigma_{p_k^j})^{-1}
(p_k^j -\mu_{p_k^j})^{\! \top} \! 
(p_k^j -\mu_{p_k^j}) \leq \beta_j \},
\end{align}
\end{subequations}
with
$\beta_i = F_{\chi_q^2}^{-1}(1 - \epsilon_i)$, 
$\beta_j = F_{\chi_q^2}^{-1}(1 - \epsilon_j)$.
Since $\calC_i \subseteq \hat{\calC}_i$ and $\calC_j \subseteq \hat{\calC}_j$, then $\Pb[p_k^i \in \hat{\calC}_i] \geq 1 - \epsilon_i$ and $\Pb[p_k^j \in \hat{\calC}_j] \geq 1 - \epsilon_j$.

Next, we will show that a sufficient condition for constraint \eqref{eq: coll avoid chance constraint}, i.e., $\Pb [\| p_k^i - p_k^j \|_2 \geq s_{ij}] \geq 1 - \epsilon$, is the following one:
\begin{equation}
\min_{p_k^i \in \hat{\calC}_i, p_k^j \in \hat{\calC}_j} 
\| p_k^i - p_k^j \|_2 \geq s_{ij},
\label{eq: pcc-dcs proposition proof - useful condition}
\end{equation}
with $\epsilon_i + \epsilon_j \leq \epsilon$. Using $P(A \, \cap \, B) \geq 1 - P(A^\text{c}) - P(B^\text{c})$,
\begin{equation}
\Pb [
p_k^i \in \hat{\calC}_i
~\cap~  
p_k^j \in \hat{\calC}_j
]
\geq 1 - \epsilon_i - \epsilon_j \geq 1 - \epsilon.
\end{equation}
Further, if the condition \eqref{eq: pcc-dcs proposition proof - useful condition} holds, then for any $p_k^i \in \hat{\calC}_i$, $p_k^j \in \hat{\calC}_j$, we have $\| p_k^i - p_k^j \|_2 \geq s_{ij}$. Therefore, in the event $(p_k^i \in \hat{\calC}_i) \cap (p_k^j \in \hat{\calC}_j)$, the inequality $\| p_k^i - p_k^j \|_2 \geq s_{ij}$ always holds. As a result, \eqref{eq: pcc-dcs proposition proof - useful condition} implies that
\begin{equation}
\Pb [\| p_k^i - p_k^j \|_2 \geq s_{ij}] \geq 
\Pb [
p_k^i \in \hat{\calC}_i
\cap  
p_k^j \in \hat{\calC}_j
]
\geq
1 - \epsilon.
\end{equation}
Subsequently, the condition \eqref{eq: pcc-dcs proposition proof - useful condition} holds if the constraints \eqref{eq: pcc-dcs proposition - system} hold,
%
%
since each $\hat{\calC}_i$ has center $\mu_{p_k^i}$ and radius $(\beta_i \lambda_{\max} (\Sigma_{p_k^i}))^{\frac{1}{2}}$.
Consequently, we have shown that system \eqref{eq: pcc-dcs proposition - system} is a sufficient condition for \eqref{eq: pcc-dcs proposition proof - useful condition}, which in turn suffices for constraint \eqref{eq: coll avoid chance constraint}. 
\end{proof}

\refstepcounter{proposition}
\labeleditem{Proposition}{\theproposition}{Sufficient Conditions for Obstacle Avoidance via Confidence Ball Separation}
\label{prop: pcc-dcs main prop obs}
The non-convex chance constraint \eqref{eq: obs avoid chance constraint} is satisfied if the following constraints hold:
\begin{subequations}
\begin{align}
\psi_k^{io}(\mu_k^i, r_k^i) & := \| \mu_{p_k^i} - p_o \|_2 - r_k^i - s_o \geq 0,
\label{eq: pcc-dcs proposition - system - mean - obs}
\\
\alpha_k^i(\Sigma_k^i, r_k^i) & \leq 0,
\end{align}
\end{subequations}
where $\alpha_k^i(\Sigma_k^i, r_k^i)$ is defined as in Proposition \ref{prop: pcc-dcs main prop}.

\begin{proof}
With a similar argument as in the proof of Proposition \ref{prop: pcc-dcs main prop}, we can show that a sufficient condition for the constraint \eqref{eq: obs avoid chance constraint} to hold, i.e., $\! \Pb [\| p_k^i \! - \! p_o \|_2 \! \geq \! s_o] \! \geq \! 1 \! - \! \epsilon$, is:
\begin{equation}
\min_{p_k^i \in \hat{\calC}_i} 
\| p_k^i - p_o \|_2 \geq d_o,
\label{eq: pcc-dcs proposition proof - useful condition obstacle}
\end{equation}
where $\hat{\calC}_i$ refers to the ball over-approximation of the confidence ellipsoid of $p_k^i$ with probability $1 - \epsilon_i > 1 - \epsilon$, as in Proposition \ref{prop: pcc-dcs main prop}. 
Subsequently, the condition \eqref{eq: pcc-dcs proposition proof - useful condition obstacle} is satisfied if in addition to the constraint \eqref{eq: pcc-dcs proposition - system - cov i}, 
the constraint \eqref{eq: pcc-dcs proposition - system - mean - obs} holds.
%

Consequently, we have shown that the constraints \eqref{eq: pcc-dcs proposition - system - cov i} and \eqref{eq: pcc-dcs proposition - system - mean - obs} are a sufficient condition for \eqref{eq: pcc-dcs proposition proof - useful condition obstacle}, which in turn is sufficient for the constraint \eqref{eq: obs avoid chance constraint} to be satisfied. 
\end{proof}

Although Propositions \ref{prop: pcc-dcs main prop} and \ref{prop: pcc-dcs main prop obs} provide conditions under which the original inter-agent collision and obstacle avoidance chance constraints are satisfied, the resulting constraints \eqref{eq: pcc-dcs proposition - system - mean} and \eqref{eq: pcc-dcs proposition - system - mean - obs} are still non-convex w.r.t. $\mu_{p_k^i}, \mu_{p_k^j}$ and the constraints \eqref{eq: pcc-dcs proposition - system - cov i} and \eqref{eq: pcc-dcs proposition - system - cov j} are non-convex w.r.t. $\Sigma_{p_k^i}, \Sigma_{p_k^j}$. 
Considering the control policies \eqref{eq: control policy}, we will now reformulate these constraints w.r.t. $v_i$ and $K_i$.  Before that, let us define the concatenated variables $r_i = [r_0^i; \dots; r_K^i]$ for each agent $i \in \calV$. 
%

\refstepcounter{proposition}
\labeleditem{Proposition}{\theproposition}{Reformulation of Constraints in Propositions \ref{prop: pcc-dcs main prop} and \ref{prop: pcc-dcs main prop obs} w.r.t. Decision Variables}
\label{prop: pcc-dcs main prop - convex}
The constraints 
$\psi_k^{io}(\mu_k^i, r_k^i) \geq 0$, 
$\phi_k^{ij}(\mu_k^i, \mu_k^j, r_k^i, r_k^j) \geq 0$ and
$\alpha_k^i(\Sigma_k^i, r_k^i) \leq 0$
%
%
can be equivalently reformulated 
as follows, respectively:
\begin{subequations}
\begin{align}
& \! \calc_{io, k}^{\text{PCC}}(v_i, r_i) := 
- \| P_i \Gamma_k^i \theta_i (v_i) - p_o \|_2
+ r_k^i + s_o
\leq 0,
\label{eq: pcc-dcs obs constraint reformulated}
\\[0.05cm]
& \! \cald_{ij, k}^{\text{PCC}} (v_i, r_i, v_j, r_j) := 
- \| P_i \Gamma_k^i \theta_i (v_i) - P_j \Gamma_k^j \theta_j (v_j)  \|_2
\label{eq: pcc-dcs coll constraint reformulated}
\\
& ~~~~~~~~~~~~~~~~~~~~~~~~~~~~~~~~~~~~~~~~~~~~ + r_k^i + r_k^j + s_{ij}
\leq 0,
\nonumber
\\[0.05cm]
& \! \calE_{i,k}^{\text{PCC}}(K_i, r_i) \! := \!
{\setlength{\arraycolsep}{1.2pt}
\begin{bmatrix}
r_i^k I_q &  P_i \Gamma_k^i \Theta_i (K_i)
\\[0.05cm]
\Theta_i (K_i)^\top \Gamma_k^{i \top} \! P_i^\top & r_k^i / \beta_i I_{(T+1) n_i}    
\end{bmatrix}
}
\! \! \succeq \! 0.
\label{eq: pcc-dcs lmi}
\end{align}
\end{subequations}

\begin{proof}
The constraints 
$\phi_k^{ij}(\mu_k^i, \mu_k^j, r_k^i, r_k^j) \geq 0$ 
and
$\psi_k^{io}(\mu_k^i, r_k^i) \geq 0$
can be rewritten as \eqref{eq: pcc-dcs obs constraint reformulated} and \eqref{eq: pcc-dcs coll constraint reformulated}, by simply substituting $\mu_{p_k^i} = P_i \Gamma_k^i \theta_i(v_i)$ and $\mu_{p_k^j} = P_j \Gamma_k^j \theta_j(v_j)$. The constraint $\alpha_k^i(\Sigma_k^i, r_k^i) \leq 0$ can be expressed as
\begin{equation}
\sqrt{ \beta_i \lambda_{\max} 
\left( 
P_i \Gamma_k^i \Theta_i (K_i) 
( P_i \Gamma_k^i \Theta_i (K_i) )^\top
\right)
}
\leq r_k^i,
\end{equation}
which is a convex constraint since it can be written in terms of the spectral norm of $P_i \Gamma_k^i \Theta_i (K_i)$ as follows 
\begin{equation}
\| P_i \Gamma_k^i \Theta_i (K_i) \|_2 
\leq r_k^i / \sqrt{\beta_i},
\end{equation}
or equivalently as the semidefinite constraint:
\begin{equation}
P_i \Gamma_k^i \Theta_i (K_i) 
( P_i \Gamma_k^i \Theta_i (K_i) )^\top \preceq (r_k^i)^2 / \beta_i I_{q}.
\end{equation}
Using Schur's complement, we arrive to the LMI in \eqref{eq: pcc-dcs lmi}.
%
\end{proof}

\refstepcounter{corollary}
\textbf{Corollary~\thecorollary:}
Combining Propositions \ref{prop: pcc-dcs main prop}, \ref{prop: pcc-dcs main prop obs} and \ref{prop: pcc-dcs main prop - convex}, it follows that 
the constraint \eqref{eq: obs avoid chance constraint} is satisfied if
$\cald_{ij, k}^{\text{PCC}} (v_i, r_i, v_j, r_j) \leq 0$, $\calE_{i,k}^{\text{PCC}}(K_i, r_i) \succeq 0$ and $\calE_{j,k}^{\text{PCC}}(K_j, r_j) \succeq 0$,
and the constraint \eqref{eq: coll avoid chance constraint} is satisfied if 
$\calc_{io, k}^{\text{PCC}}(v_i, r_i) \leq 0$ and $\calE_{i,k}^{\text{PCC}}(K_i, r_i) \succeq 0$.

Note that although the constraints $\calE_{i,k}^{\text{PCC}}(K_i, r_i) \succeq 0$ are convex, the constraints $\calc_{io, k}^{\text{PCC}}(v_i, r_i) \leq 0$ and 
$\cald_{ij, k}^{\text{PCC}} (v_i, r_i, v_j, r_j) \leq 0$ are still non-convex. As shown later, we accommodate for that through a successive linearization strategy.

\begin{table*}
\centering
\caption{Overview of Variable and Constraint Dimensions of Different Approaches}
\vspace{-0.1cm}
\setlength{\tabcolsep}{3pt}
\begin{tabular}{|c|c|c|c|cc|}
\hline
 & 
\multirow{3}{*}{\textbf{Centralized CS}} & 
\multirow{3}{*}{\makecell{\textbf{FCC-DCS} \\ (per local problem)}} &
\multirow{3}{*}{\makecell{\textbf{PCC-DCS} \\ (per local problem)}} &
\multicolumn{2}{c|}{\rule{0pt}{0.9\normalbaselineskip}\textbf{MC-DCS}}
\\
& & & & \multirow{2}{*}{\makecell{Cov. part \\ (solved once)}} & 
\multirow{2}{*}{\makecell{Mean part \\ (per local problem)}} 
\\
& & & & & 
\\[0.05cm]
\hline
\multirow{2}{*}{Num. of variables} & 
\multirow{2}{*}{$N T (m_i + \gamma n_i m_i)$}  & 
\multirow{2}{*}{$|\calV_i| T (m_i + \gamma n_i m_i)$} &
\multirow{2}{*}{$|\calV_i| T (m_i + 1) + \gamma T n_i m_i$} &
\multirow{2}{*}{$\gamma T n_i m_i$} & 
\multirow{2}{*}{$|\calV_i| T m_i$}
\\[0.3cm]
LMI constraints & 
\makecell{$N$ constraints \\ of dim. $(T+2) n_i$} & 
\makecell{$1$ constraint \\ of dim. $(T+2) n_i$} &
\makecell{$1$ constraint \\ of dim. $(T+2) n_i$} & 
\makecell{$1$ constraint \\ of dim. $(T+2) n_i$} & 
-
\\[0.3cm]
SOC constraints & 
\makecell{$N T (|\calV_i| + |\calO|)$ constraints \\ of dim. $(T+2) n_i$} & 
\makecell{$T (|\calV_i| + |\calO|)$ constraints \\ of dim. $(T+2) n_i$} & 
\makecell{$T$ constraints \\ of dim. $(T+2) n_i$} & 
- & 
- 
\\[0.3cm]
Linear ineq. constraints & 
- & 
- & 
$T (|\calV_i| + |\calO|)$ & 
- & 
$T (|\calV_i| + |\calO|)$
\\[0.2cm]
Linear eq. constraints & 
$N n_i$ & 
$n_i$ & 
$n_i$ & 
- & 
$n_i$
\\[0.1cm]
\hline
\end{tabular}
\label{tab: distr cs - alg dims}
\end{table*}

For notational convenience, let us define the concatenated constraints 
$\calc_i^{\text{PCC}}(v_i, r_i) := \big[ \{ \calc_{io,k}^{\text{PCC}}(v_i, r_i) \}_{o \in \calO, k \in \llbracket 0,T \rrbracket} \big]$, 
$\cald_{ij}^{\text{PCC}}(v_i, r_i, v_j, r_j) := \big[ \{ \cald_{ij,k}^{\text{PCC}}(v_i, r_i, v_j, r_j) \}_{k \in \llbracket 0,T \rrbracket} \big]$ 
and 
$\calE_i^{\text{PCC}}(K_i, r_i) := \big[ \{ \calE_{i,k}^{\text{PCC}}(K_i, r_i) \}_{k \in \llbracket 0, T \rrbracket} \big]$. 
Therefore, we arrive to the following new problem.

\refstepcounter{problem}
\labeleditem{Problem}{\theproblem}{MACS - Partial-Covariance Constrained Reformulation}
\label{problem: multi-agent cs - pcc version}
Find the optimal $\{v_i^*,  K_i^*, r_i^* \}_{i \in \calV}$ such that
\begin{align*}
& ~~~ \min \sum_{i \in \calV} \calJ_i(v_i, K_i)
\\
\mathrm{s.t.} \quad 
& \cala_i(v_i) = 0, ~ \calB_i(K_i) \succeq 0, 
\\
& \calc_i^{\text{PCC}}(v_i, r_i) \leq 0,~
\cald_{ij}^{\text{PCC}}(v_i, r_i, v_j, r_j) \leq 0,  
\\
& \calE_i^{\text{PCC}}(K_i, r_i) \succeq 0, ~ \forall j \in \calV_i, ~ i \in \calV.
\end{align*}

\subsection{Consensus Optimization}

Similar to Problem \ref{problem: multi-agent cs - fcc version - before consensus}, Problem \ref{problem: multi-agent cs - pcc version} cannot be directly solved in a decentralized manner due of the coupling constraints $\cald_{ij}^{\text{PCC}}(v_i, r_i, v_j, r_j) \leq 0$ between neighboring agents. Yet, in contrast to FCC-DCS which requires a consensus on the feed-forward controls and the feedback gains, this formulation will require introducing copy variables only for the feed-forward controls $v_j$ and the auxiliary variables $r_j$ of neighbor agents. 

In this context, we introduce the copy variables $v_j^{(i)}, r_j^{(i)}$, $j \in \calV_i$, from the perspective of each agent $i$, which gives rise to the augmented local decision variables 
$\tilde{v}_i = [\{ v_j^{(i)} \}_{j \in \calV_i}]$ and 
$\tilde{r}_i = [\{ r_j^{(i)} \}_{j \in \calV_i}]$. 
Therefore, the inter-agent constraints can be expressed from the point of view of each $i \in \calV$ as
\begin{equation}
\tilde{\cald}_i^{\text{PCC}}(\tilde{v}_i, \tilde{r}_i) :=
[\{ \cald_{ij}^{\text{PCC}}(v_i, r_i, v_j^{(i)}, r_j^{(i)})
\}_{j \in \calV_i}]
\leq 0.
\end{equation}
As in FCC-DCS, the presence of these copy variables also mandates introducing the global variables $z = [ \{ z_i \}_{i \in \calV} ]$, 
$\zeta = [ \{ \zeta_i \}_{i \in \calV} ]$ and the consensus constraints
$\tilde{v}_i = \tilde{z}_i$, 
$\tilde{r}_i = \tilde{\zeta}_i$,
$\forall i \in \calV$,
with $\tilde{z}_i := [\{ z_j \}_{j \in \calV_i}]$ and $\tilde{\zeta}_i := [\{ \zeta_j \}_{j \in \calV_i}]$.

\refstepcounter{remark}
\labeleditem{Remark}{\theremark}{More computationally efficient SOC constraint}
Despite the significant reduction in the amount of variables, a potential computational drawback could be the additional LMI constraints $\calE_i^{\text{PCC}}(K_i, r_i) \succeq 0$. To accommodate that, we replace the spectral norm with the Frobenius norm and obtain the more conservative SOC constraint:
%
%
\begin{equation}
\cale_{i,k}^{\text{PCC}}(K_i, r_i) :
= \| P_i \Gamma_k^i \Theta_i (K_i) \|_\rF - r_k^i / \sqrt{\beta_i} \leq 0.
\end{equation}
Note that although typically replacing a spectral norm constraint with a Frobenius norm one, introduces conservatism for high-dimensional matrices, this effect is mitigated in our case since $P_i \Gamma_k^i \Theta_i (K_i) 
( P_i \Gamma_k^i \Theta_i (K_i) )^\top \in \Rb^{q \times q}$ with $q=2$ or $q=3$ for 2D or 3D spaces, respectively.

Therefore, we arrive to the consensus optimization problem.
\refstepcounter{problem}
\labeleditem{Problem}{\theproblem}{MACS - Partial-Covariance Consensus Version}
For each $i \in \calV$, find the optimal $\{v_i^*, r_i^*, K_i^* \}$ such that
\label{problem: multi-agent cs - pcc version - consensus}
\begin{align*}
& ~~~~~~~~~~~~ \min \sum_{i \in \calV} \calJ_i(v_i, K_i)
\\
\mathrm{s.t.} \quad 
& \cala_i(v_i) = 0, ~ \calB_i(K_i) \succeq 0, ~ 
\calc_i^{\text{PCC}}(v_i, r_i) \leq 0, 
\\ 
& \tilde{\cald}_i^{\text{PCC}}(\tilde{v}_i, \tilde{r}_i) \leq 0, ~ 
\cale_i^{\text{PCC}}(K_i, r_i) \leq 0, 
\\
& \tilde{v}_i = \tilde{z}_i, ~
\tilde{r}_i = \tilde{\zeta}_i, ~
\forall i \in \calV.
\end{align*}

\subsection{Method}

Subsequently, we present a distributed algorithm for solving Problem \ref{problem: multi-agent cs - pcc version - consensus}, following a two-block ADMM derivation as in Section \ref{subsec: fcc dcs - method}. The first block of variables is 
$\tilde{v} = \{ \tilde{v}_i \}_{i \in \calV}, \tilde{r} = \{ \tilde{r}_i \}_{i \in \calV}, K = [\{ K_i \}_{i \in \calV}]$, 
and the second block is $z, \zeta$. 
At every ADMM round, the non-convex constraints in the local problems are iteratively linearized around the previous iterates.

\refstepcounter{proposition}
\textbf{Proposition~\theproposition:}
The constraints $\calc_{io, k}^{\text{PCC}}(v_i, r_i) \leq 0$ and $\cald_{ij, k}^{\text{PCC}} (v_i, r_i, v_j, r_j) \leq 0$ are satisfied if the following linearized inequalities hold:
\begin{subequations}
\begin{align}
& \! \calc_{io, k, \text{lin}}^{\text{PCC}}(v_i, r_i) \! := \! 
- a_k^{io \top} \!
( P_i \Gamma_k^i \theta_i (v_i) \! - \! p_o)
\! + \! r_k^i \! + \! s_o
\leq 0, \!
\\ 
& \! \cald_{ij, k, \text{lin}}^{\text{PCC}} (v_i, r_i, v_j, r_j) \! := \! 
- a_k^{ij \top} \! 
( P_i \Gamma_k^i \theta_i (v_i) \! - \! P_j \Gamma_k^j \theta_j (v_j)) \! \!
\nonumber
\\
& ~~~~~~~~~~~~~~~~~~~~~~~~~~~~~~~ \! + r_k^i + r_k^j + s_{ij}
\leq 0,
\end{align}
\label{eq: pcc-dcs linearized constraints}%
\end{subequations}
respectively, where 
$a_k^{io} = (\hat{p}_k^i - p_o) / \| \hat{p}_k^i - p_o \|_2$, 
$a_k^{ij} = (\hat{p}_k^i - \hat{p}_k^j) / \| \hat{p}_k^i - \hat{p}_k^j \|_2$
and $\hat{p}_k^i$, $\hat{p}_k^j$ are the approximation points. 

\begin{proof}
For brevity, we show the derivation of the inter-agent constraint; the obstacle avoidance constraint follows similarly.
If we define $q_k = \mu_{p_k^i} - \mu_{p_k^j}$, then the first-order Taylor approximation of $\| q_k \|_2$ around $\hat{q}_k = \hat{p}_k^i - \hat{p}_k^j$, yields
\begin{equation}
\| \hat{q}_k \|_2 + \hat{q}_k^\top (q_k - \hat{q}_k) / \| \hat{q}_k \|_2 
= 
\hat{q}_k^\top q_k /
\| \hat{q}_k \|_2
\end{equation}
which yields the constraint
\begin{equation}
a_k^{ij \top} (\mu_{p_k^i} - \mu_{p_k^j})
\geq r_k^i + r_k^j + s_{ij},
\end{equation}
where $a_k^{ij} = (\hat{p}_k^i - \hat{p}_k^j) / \| \hat{p}_k^i - \hat{p}_k^j \|_2$. 
Note that this is a convex under-approximation of \eqref{eq: pcc-dcs proposition - system - mean} from the convexity of norms. 
\end{proof}

We also define the concatenated expressions 
$\calc_{i, \text{lin}}^{\text{PCC}}(v_i, r_i)$,   
$\cald_{ij, \text{lin}}^{\text{PCC}} (v_i, r_i, v_j, r_j)$ and 
$\tilde{\cald}_{i, \text{lin}}^{\text{PCC}}(\tilde{v}_i, \tilde{r}_i)$ accordingly. The AL for Problem \ref{problem: multi-agent cs - pcc version - consensus} is formulated as
\begin{align}
& \calL_\rho = 
\sum_{i \in \calV} 
\calJ_i(v_i, K_i) 
+ \calI_{\cala_i, \calB_i, \calc_i^{\text{PCC}}, \tilde{\cald}_i^{\text{PCC}}, \cale_i^{\text{PCC}}}
(\tilde{v}_i, \tilde{r}_i, K_i)
\\[-0.05cm]
& ~~~~~
+ \! \langle y_i, \tilde{v}_i \! - \! \tilde{z}_i \rangle
\! + \! \langle \xi_i, \tilde{r}_i \! - \! \tilde{\zeta}_i \rangle
\! + \! \frac{\rho_v}{2} \| \tilde{v}_i \! - \! \tilde{z}_i \|_2^2
\! + \! \frac{\rho_r}{2} \| \tilde{r}_i \! - \! \tilde{\zeta}_i \|_2^2,
\nonumber
\end{align}
where $y_i$ and $\xi_i$ are the dual variables for the constraints $\tilde{v}_i = \tilde{z}_i$ and $\tilde{r}_i = \tilde{\zeta}_i$, and $\rho_v, \rho_r > 0$ are penalty parameters.
Then, the algorithm updates are derived as follows.

\myemph{Local primal updates.}
The first block yields the following updates for the local variables
\begin{equation}
\begin{aligned}
& \{ \tilde{v}_i, \tilde{r}_i, K_i \}^{\ell+1} = \argmin
\tilde{\calJ}_i^{\text{PCC}}(\tilde{v}_i, \tilde{r}_i, K_i)
\\
\mathrm{s.t.} \quad 
& \cala_i(v_i) = 0, ~ \calB_i(K_i) \succeq 0, ~ 
\calc_{i, \text{lin}}^{\text{PCC}}(v_i, r_i) \leq 0, 
\\ 
& \tilde{\cald}_{i, \text{lin}}^{\text{PCC}}(\tilde{v}_i, \tilde{r}_i) \leq 0, ~ 
\cale_i^{\text{PCC}}(K_i, r_i) \leq 0,
\end{aligned}
\label{eq: pcc-dcs - local updates}
\end{equation}
with
\begin{align}
\tilde{\calJ}_i^{\text{PCC}} 
(\tilde{v}_i, \tilde{r}_i, K_i) 
& := \calJ_i(v_i, K_i)
+ \langle y_i^\ell, \tilde{v}_i \rangle
+ \langle \xi_i^\ell, \tilde{r}_i \rangle
\\
& ~~~~~~~~~ 
+ \frac{\rho_v}{2} \| \tilde{v}_i - \tilde{z}_i^\ell \|_2^2
+ \frac{\rho_r}{2} \| \tilde{r}_i - \tilde{\zeta}_i^\ell \|_\rF^2,
\nonumber
\end{align}
where the constraints $\calc_{i, \text{lin}}^{\text{PCC}}(v_i, r_i)$ and $\cald_{ij, \text{lin}}^{\text{PCC}}(v_i, r_i, v_j^{(i)}, r_j^{(i)})$ are linearized using $\mu_i^{\ell}$ and $\mu_j^\ell$ as approximation points.

\myemph{Global primal updates.}
The global variables $z$ and $\zeta$ are updated through
\begin{equation}
z_i^{\ell+1} \! = \! \frac{1}{|\calW_i|} \! \sum_{j \in \calW_i} \! v_i^{(j), \ell+1}, 
\quad
\zeta_i^{\ell+1} \! = \! \frac{1}{|\calW_i|} \! \sum_{j \in \calW_i} \! r_i^{(j), \ell+1}.
\label{eq: pcc-dcs - global updates}
\end{equation}

\myemph{Dual updates.} Finally, the dual variables are updated as
\begin{subequations}
\begin{align}
y_i^{\ell+1} & = y_i^\ell + \rho_v (\tilde{v}_i^{\ell+1} - \tilde{z}_i^{\ell+1}), 
\\
\xi_i^{\ell+1} & = \xi_i^\ell + \rho_r (\tilde{r}_i^{\ell+1} - \tilde{\zeta}_i^{\ell+1}).
\end{align}
\label{eq: pcc-dcs - dual updates}%
\end{subequations}

\indent \myemph{Algorithm.} The PCC-DCS algorithm is described in Alg. \ref{PCC-DCS Algorithm}. 
During each ADMM round, the variables $\tilde{v}_i, \tilde{r}_i, K_i$ are updated first by solving the local problems \eqref{eq: pcc-dcs - local updates}. Then, each agent $i$ receives $v_i^{(j)}, r_i^{(j)}$ from all $j \in \calW_i \backslash \{i\}$ and the global updates \eqref{eq: pcc-dcs - global updates} take place so that the variables $z_i, \zeta_i$ are updated. Finally, every agent $i$ receives $z_j, \zeta_j$ from all $j \in \calV_i \backslash \{i\}$, so that the dual updates \eqref{eq: pcc-dcs - dual updates} are performed. 

\refstepcounter{remark}
\labeleditem{Remark}{\theremark}{Decentralized Structure of PCC-DCS}
\label{remark: pcc-dcs decentralized}
Similar to FCC-DCS, the PCC-DCS algorithm is fully decentralized.

\refstepcounter{remark}
\labeleditem{Remark}{\theremark}{Computational Benefits of PCC-DCS}
The PCC-DCS method offers a substantial computational advantage compared to FCC-DCS, as the number of variables in each local subproblem is reduced to $|\calV_i| T (m_i + 1) + \gamma T n_i m_i$. Further, each local problem involves a single LMI constraint and $T$ SOC constraints of dim. $(T+2)n_i$ (see Table \ref{tab: distr cs - alg dims}).


\section{Mean-Consensus \\ Distributed Covariance Steering}
\label{sec: mcc dcs}

Towards further improving computational efficiency, we also propose an approach that restricts inter-agent coupling, and therefore the need for consensus, only on the mean states. This is achieved by modifying the obstacle and inter-agent collision avoidance constraints \eqref{eq: pcc-dcs proposition - system - mean - obs} and \eqref{eq: pcc-dcs proposition - system - mean}, where the auxiliary variables $r_k^i$ are replaced with fixed parameters $\hat{r}_i$ for all $k \in \llbracket 1, T \rrbracket$. The resulting constraints take the form:
\begin{subequations}
\begin{align}
\hat{\psi}_k^{io}(\mu_k^i) & := \| \mu_{p_k^i} - p_o \|_2 - \hat{r}_i - s_o \geq 0,
\label{eq: mc-dcs - obs}
\\
\hat{\phi}_k^{ij}(\mu_k^i, \mu_k^j) & :=
\| \mu_{p_k^i} - \mu_{p_k^j} \|_2 - \hat{r}_i - \hat{r}_j - s_{ij} \geq 0.
\label{eq: mc-dcs - inter-agent}
\end{align}
\label{eq: mc-dcs - constraints}%
\end{subequations}
Then, following similar derivations as in Section \ref{sec: pcc dcs}, we obtain the equivalent constraints:
\begin{subequations}
\begin{align}
& \calc_{io, k}^{\text{MC}}(v_i) :=  
- \| P_i \Gamma_k^i \theta_i (v_i) - p_o \|_2
+ \hat{r}_i + s_o
\leq 0,
\\
& \cald_{ij, k}^{\text{MC}} (v_i, v_j) := 
- \| P_i \Gamma_k^i \theta_i (v_i) \! - \! P_j \Gamma_k^j \theta_j (v_j) \|_2
\\
& ~~~~~~~~~~~~~~~~~~~~~~~~~~~~~~~~~~~~ + \hat{r}_i + \hat{r}_j + s_{ij}
\leq 0,
\nonumber
\end{align}
\end{subequations}
which leads to the following problem formulation.

\begin{algorithm}[t]
\caption{Partial-Covariance-Consensus DCS (PCC-DCS)}\label{PCC-DCS Algorithm}
\begin{algorithmic}[1] 
\State \textbf{Initialize:} 
$\tilde{v}_i \leftarrow 0$,
$\tilde{r}_i \leftarrow [ \{ r_j' \}_{j \in \calV_i} ]$,
$z_i \leftarrow \tilde{v}_i$, 
$\zeta_i \leftarrow \tilde{r}_i$.
\While{not converged \textbf{and} $\ell \leq \ell_{\text{max}}$}
\State $\calc_{i, \text{lin}}^{\text{PCC}}, \tilde{\cald}_{i, \text{lin}}^{\text{PCC}} \leftarrow$ Get linearized constraints with \eqref{eq: pcc-dcs linearized constraints}.
\State $\tilde{v}_i, \tilde{r}_i, K_i \leftarrow$ Solve \eqref{eq: pcc-dcs - local updates} in parallel $ \forall \ i \in \calV$.
\State \textit{Each agent $i \in \calV$ receives $v_i^j, \! K_i^j$ from all $j \in \calW_i \backslash \{i\}$.} 
\State $z_i, \zeta_i \leftarrow$ Update with \eqref{eq: pcc-dcs - global updates} in parallel $ \forall \ i \in \calV$.
\State \textit{Each agent $i \in \calV$ receives $z_j, \zeta_j$ from all $j \in \calV_i \backslash \{i\}$.} 
\State $y_i, \xi_i \leftarrow$ Update with \eqref{eq: pcc-dcs - dual updates} in parallel $ \forall \ i \in \calV$.
\EndWhile
\end{algorithmic}
\end{algorithm}

\refstepcounter{problem}
\labeleditem{Problem}{\theproblem}{MACS - Mean-Constrained Reformulation}
\label{problem: multi-agent cs - mc version}
For each $i \in \calV$, find the optimal $\{v_i^*,  K_i^*\}$ such that
\begin{align*}
& ~~~~~ \min \sum_{i \in \calV} \calJ_i(v_i, K_i)
\\
\mathrm{s.t.} \quad 
& \cala_i(v_i) = 0, ~ 
\calB_i(K_i) \succeq 0, ~ 
\calc_i^{\text{MC}}(v_i) \leq 0,
\\ 
& 
\cald_{ij}^{\text{MC}}(v_i, v_j) \leq 0, ~ \forall j \in \calV_i, ~ i \in \calV.
\end{align*}

In this formulation, the inter-agent coupling that hinders decentralization involves only the feed-forward control variables. Consequently, it suffices to maintain only the augmented local variables $\tilde{v}_i$ and enforce consensus with the global variables $z = [\{ z_i \}_{i \in \calV}]$ through
$
\tilde{v}_i = \tilde{z}_i, ~
\forall i \in \calV,
$
with $\tilde{z}_i := [\{ z_j \}_{j \in \calV_i}]$. The resulting consensus optimization becomes as follows.

\refstepcounter{problem}
\labeleditem{Problem}{\theproblem}{MACS - Mean Consensus Version}
For each $i \in \calV$, find the optimal $\{v_i^*,  K_i^*\}$ such that
\label{problem: multi-agent cs - mc version - consensus}
\begin{align*}
& ~~~~~ \min \sum_{i \in \calV} \calJ_i(v_i, K_i)
\\
\mathrm{s.t.} \quad 
& \cala_i(v_i) = 0, ~ 
\calB_i(K_i) \succeq 0, ~ 
\calc_i^{\text{MC}}(v_i) \leq 0,
\\ 
& 
\tilde{\cald}_i^{\text{MC}}(\tilde{v}_i) \leq 0, ~
\tilde{v}_i = \tilde{z}_i, ~
i \in \calV.
\end{align*}

In addition, each cost function $\calJ_i(v_i, K_i)$ decomposes into mean- and covariance-dependent components as follows:
\begin{equation}
\calJ_i(v_i, K_i) = 
\calJ_i^{\text{mean}}(v_i) 
+ \calJ_i^{\text{cov}}(K_i),
\end{equation}
with $\calJ_i^{\text{mean}}(v_i) := \theta_i (v_i)^\top Q_i \theta_i (v_i)
+ v_i^\top R_i v_i$ and
\begin{align}
\calJ_i^{\text{cov}}(K_i) & :=
\tr \left[Q_i \Theta_i (K_i) \Theta_i (K_i)^\top \right]
\\
& ~~~~~~  + \tr \left[ R_i K_i (G_0^i \Sigma_0^i G_0^{i \top} + G_w^i W_i G_w^{i \top}) K_i^\top \right].
\nonumber
\end{align}
Notably, $\calJ_i^{\text{mean}}(v_i)$ depends only on the feed-forward control inputs $v_i$, while $\calJ_i^{\text{cov}}(K_i)$ depends only on the feedback gains $K_i$. This structure enables a complete decoupling of the problem into two parts. The mean part is given by:
\begin{subequations}
\begin{align}
& ~~~ \min \sum_{i \in \calV} \calJ_i^{\text{mean}}(v_i)
\\
\mathrm{s.t.} \quad 
& \cala_i(v_i) = 0, ~ 
\calc_i^{\text{MC}}(v_i) \leq 0,
\\ 
& 
\tilde{\cald}_i^{\text{MC}}(\tilde{v}_i) \leq 0, ~
\tilde{v}_i = \tilde{z}_i, ~
i \in \calV.
\end{align}
\label{problem: multi-agent cs - mc version - mean part - consensus}%
\end{subequations}
The covariance part can be further decoupled fully across all agents, and for each $i \in \calV$ reduces to:
\begin{equation}
\min \calJ_i^{\text{cov}}(K_i) \quad \mathrm{s.t.} \quad
\calB(K_i) \succeq 0,
\label{eq: mc-dcs - single-agent - cov problems}
\end{equation}
which are only required to be solved once for each agent. 

To solve the consensus-constrained mean part \eqref{problem: multi-agent cs - mc version - consensus}, we derive a distributed algorithm through the two-block ADMM derivation with $\tilde{v} = \{ \tilde{v}_i \}_{i \in \calV}$ as the first block of variables and $z$ as the second one. The non-convex constraints are iteratively linearized as in PCC-DCS.
The updates are as follows.

\begin{algorithm}[t]
\caption{Mean-Consensus DCS (MC-DCS)}\label{MC-DCS Algorithm}
\begin{algorithmic}[1] 
\State \textbf{Initialize:} 
$\tilde{v}_i \leftarrow [ \{ v_j' \}_{j \in \calV_i} ]$,
$z_i \leftarrow \tilde{v}_i$, 
$y_i \leftarrow 0$. 
\While{not converged \textbf{and} $\ell \leq \ell_{\text{max}}$}
\State $\calc_{i, \text{lin}}^{\text{MC}}, \tilde{\cald}_{i, \text{lin}}^{\text{MC}} \leftarrow$ Get linearized constraints.
\State $\tilde{v}_i \leftarrow$ Solve \eqref{eq: mc-dcs - local updates} in parallel $ \forall \ i \in \calV$.
\State \textit{Each agent $i \in \calV$ receives $v_i^j$ from all $j \in \calW_i \backslash \{i\}$.} 
\State $z_i \leftarrow$ Update with \eqref{eq: mc-dcs - global updates} in parallel $ \forall \ i \in \calV$.
\State \textit{Each agent $i \in \calV$ receives $z_j$ from all $j \in \calV_i \backslash \{i\}$.} 
\State $y_i \leftarrow$ Update with \eqref{eq: mc-dcs - dual updates} in parallel $ \forall \ i \in \calV$.
\EndWhile
\end{algorithmic}
\end{algorithm}

\myemph{Local primal updates.} The local variables $\tilde{v}_i$ are updated through solving the following quadratic programs:
\begin{equation}
\begin{aligned}
& ~~~~~~~ \tilde{v}_i^{\ell+1} = \argmin_{\tilde{v}_i}
\tilde{\calJ}_i^{\text{MC}}(\tilde{v}_i)
\\
\mathrm{s.t.} \quad 
& \cala_i(v_i) = 0, ~
\calc_{i, \text{lin}}^{\text{MC}}(v_i) \leq 0, ~
\tilde{\cald}_{i, \text{lin}}^{\text{MC}}(\tilde{v}_i) \leq 0,
\end{aligned}
\label{eq: mc-dcs - local updates}
\end{equation}
with 
$\tilde{\calJ}_i^{\text{MC}}(\tilde{v}_i) := 
\calJ_i^{\text{mean}}(v_i)
+ \langle y_i^\ell, \tilde{v}_i \rangle
+ \frac{\rho_v}{2} \| \tilde{v}_i - \tilde{z}_i^{\ell} \|_2^2$, and
\begin{subequations}
\begin{align}
& \! \! \! \calc_{io, k, \text{lin}}^{\text{MC}}(v_i) :=  
- a_k^{io \top} \!
( P_i \Gamma_k^i \theta_i (v_i) - p_o)
+ \hat{r}_i + s_o
\leq 0,
\label{eq: mc-dcs - linearized obs constraint}
\\[0.05cm]
& \! \! \! \cald_{ij, k, \text{lin}}^{\text{MC}} (v_i, v_j) := 
- a_k^{ij \top} \! 
( P_i \Gamma_k^i \theta_i (v_i) \! - \! P_j \Gamma_k^j \theta_j (v_j)) \!
\label{eq: mc-dcs - linearized inter-agent constraint}
\\
& ~~~~~~~~~~~~~~~~~~~~~~~~~~ + \hat{r}_i + \hat{r}_j + s_{ij}
\leq 0,
\nonumber
\end{align}
\end{subequations}
where the linearized constraints $\calc_{i, \text{lin}}^{\text{MC}}(v_i, r_i)$ and $\tilde{\cald}_{i, \text{lin}}^{\text{MC}}(\tilde{v}_i, \tilde{r}_i)$ are obtained using $\mu_i^{\ell}$ and $\mu_j^\ell$ as the approximation points.

\myemph{Global primal updates.} The global updates for $z$ are 
\begin{equation}
z_i^{\ell+1} = \frac{1}{|\calW_i|} \sum_{j \in \calW_i} v_i^{(j), \ell+1}.
\label{eq: mc-dcs - global updates}
\end{equation}

\myemph{Dual updates.} Finally, the dual variables are updated with
\begin{equation}
y_i^{\ell+1} = y_i^\ell + \rho_v (\tilde{v}_i^{\ell + 1} - \tilde{z}_i^{\ell + 1}).
\label{eq: mc-dcs - dual updates}
\end{equation}

\myemph{Algorithm.} 
The MC-DCS algorithm is presented in Alg. \ref{MC-DCS Algorithm}. Initially, each agent solves in parallel the single-agent covariance problem \eqref{eq: mc-dcs - single-agent - cov problems} to obtain $K_i$. 
Then in each ADMM loop, the local variables $\tilde{v}_i$ are updated by solving problems \eqref{eq: mc-dcs - local updates}. Subsequently, each agent $i$ receives $v_i^{(j)}$ from all $j \in \calW_i \backslash \{i\}$ and the global variables $z_i$ are updated through \eqref{eq: mc-dcs - global updates}. Lastly, each agent $i$ receives $z_j$ from all $j \in \calV_i \backslash \{i\}$ and the dual variables $y_i$ are updated with \eqref{eq: mc-dcs - dual updates}. 

\refstepcounter{remark}
\labeleditem{Remark}{\theremark}{Decentralized Structure of MC-DCS}
\label{remark: mc-dcs decentralized}
Similar to Remarks \ref{remark: fcc-dcs decentralized} and \ref{remark: pcc-dcs decentralized}, the MC-DCS method is also a fully decentralized algorithm.

\refstepcounter{remark}
\labeleditem{Remark}{\theremark}{Computational Benefits of MC-DCS}
The MC-DCS method exhibits remarkable computational advantages, even over PCC-DCS. The local subproblems \eqref{eq: mc-dcs - local updates} are quadratic programs involving $|\calV_i| T m_i$ variables, $T(|\calV_i| + |\calO|)$ linear inequality constraints, and $n_i$ equality constraints. As a result, they are solved significantly faster than the subproblems in FCC-DCS and PCC-DCS, which involve LMI and SOC constraints. Moreover, the single-agent SDPs for the covariance part \eqref{eq: mc-dcs - single-agent - cov problems} are fully decoupled and only solved once.


\section{Convergence Analysis}
\label{sec: convergence analysis}

This section presents a novel convergence analysis for distributed ADMM methods with iteratively linearized non-convex constraints. As PCC-DCS and MC-DCS fall under this setup, their convergence is guaranteed.
Section \ref{sec: convergence - general} introduces a general consensus optimization 
problem formulation and assumptions. Section \ref{sec: convergence - lemmas} establishes intermediate lemmas that lead to the sufficient descent of a Lyapunov function, which is then used in Section \ref{sec: convergence - main} to prove the main theorem, establishing convergence to KKT points.
We further discuss modifications for the convergence of FCC-DCS.

\subsection{General Problem Formulation and Assumptions}
\label{sec: convergence - general}

Let us consider the following general consensus optimization problem formulation. 
For each $i \in \calV$, we denote the local variables subject to consensus with 
$\bar{x}_i \in \Rb^{\bar{n}_i}$, 
the local variables not subject to consensus with
$\bar{w}_i \in \Rb^{\bar{n}_i'}$, 
and the global variable with $\bar{z} \in \Rb^{\bar{m}}$. The functions $f_i(\bar{x}_i, \bar{w}_i)$ are the local objectives, $g_i(\bar{x}_i, \bar{w}_i) \leq 0$ and $h_i(\bar{x}_i) \leq 0$ denote local convex and non-convex constraints, respectively, and the matrices $\bar{C}_i \in \Rb^{\bar{n}_i \times \bar{m}}$ define the consensus structure.

\refstepcounter{problem}
\labeleditem{Problem}{\theproblem}{General Consensus Optimization Problem}
\label{problem: compact}
For each $i \in \calV$, find the optimal $\bar{x}_i^*, \bar{w}_i^*$ such that
\begin{align*}
& ~~~~~~~~~~~~~~~~~~~~ \min \sum_{i \in \calV} f_i(\bar{x}_i, \bar{w}_i)
\\
& \mathrm{s.t.} \quad g_i(\bar{x}_i, \bar{w}_i) \leq 0, ~
h_i(\bar{x}_i) \leq 0, ~
\bar{x}_i = \bar{C}_i \bar{z}, \quad
\forall i \in \calV.
\end{align*}

Table \ref{tab: compact notation convergence} shows that both Problems \ref{problem: multi-agent cs - pcc version - consensus} (PCC-DCS) and \ref{problem: multi-agent cs - mc version - consensus} (MC-DCS) are captured through Problem \ref{problem: compact}. We define the concatenated variables 
$\bar{x} = [\{ \bar{x}_i \}_{i \in \calV}] \in \Rb^{\bar{n}}$, 
%
%
$\bar{w} = [\{ \bar{w}_i \}_{i \in \calV}] \in \Rb^{\bar{n}'}$,  
%
%
and functions 
$f(\bar{x}, \bar{w}) = \sum_{i \in \calV} f_i(\bar{x}_i, \bar{w}_i) : \Rb^{\bar{n} + \bar{n}'} \rightarrow \Rb$,
$g(\bar{x}, \bar{w}) = [\{ g_i(\bar{x}_i, \bar{w}_i) \}_{i \in \calV}] : \Rb^{\bar{n} + \bar{n}'} \rightarrow \Rb^{\bar{p}}$, 
and 
$h(\bar{x}) = [\{ h_i(\bar{x}_i) \}_{i \in \calV}] : \Rb^{\bar{n}} \rightarrow \Rb^{\bar{q}}$. 
We also consider the (re-ordering) partition $\bar{x} = [\bar{x}_\rA; \bar{x}_\rB]$, where $\bar{x}_\rA \in \mathbb{R}^{\bar{n}_\rA}$ contains the variables appearing nonlinearly in the objective or constraints, and $\bar{x}_\rB \in \mathbb{R}^{\bar{n}_\rB}$ contains variables appearing only linearly. The consensus structure respects this partition with $\bar{x}_\rA = \bar{C}_\rA \bar{z}_\rA$ and $\bar{x}_\rB = \bar{C}_\rB \bar{z}_\rB$, where $\bar{z} = [\bar{z}_\rA; \bar{z}_\rB]$ and $\bar{C} = \mathrm{bdiag}(\bar{C}_\rA, \bar{C}_\rB) \in \Rb^{\bar{n} \times \bar{m}}$.

In the context of PCC/MC-DCS, the variables $\bar{x}_A$ correspond to the feed-forward controls $\tilde{v}$, $\bar{x}_B$ correspond to the auxiliary variables $\tilde{r}$ for PCC-DCS and are empty for MC-DCS, and finally, $\bar{w}$ correspond to the feedback gains $K$.


Next, we outline the following assumptions, which are straightforward to verify for both PCC-DCS and MC-DCS.

\refstepcounter{assumption}
\textbf{Assumption~\theassumption:}
\label{assumption: strongly convex f}
The function $f$ is convex and differentiable. In addition, $f$ is $M$-partially strongly convex with $M = \mathrm{bdiag}(M_x, 0_{{\bar{n}_\rB} \times \bar{n}_{\rB}}, 0_{{\bar{n}'} \times \bar{n}'})$, 
$M_x = \mu_x I_{{\bar{n}_\rA}}$ and 
$\mu_x > 0$.

\refstepcounter{assumption}
\textbf{Assumption~\theassumption:}
\label{assumption: convex g}
The functions $g_j$, $j \in \llbracket 1, \bar{p} \rrbracket$, are convex and differentiable.

\refstepcounter{assumption}
\textbf{Assumption\;\theassumption:}
\label{assumption: non-convex h}
The (non-convex) functions $h_j$, $j \in \llbracket 1, \bar{q} \rrbracket$, are concave and $L_j$-partially smooth with $L_j = \mathrm{bdiag}(l_j I_{\bar{n}_\rA}, 0_{\bar{n}_\rB \times \bar{n}_\rB})$ and $l_j > 0$.

\refstepcounter{remark}
\labeleditem{Remark}{\theremark}{Local Smoothness in Feasible Regions of PCC- and MC-DCS}
For PCC-DCS and MC-DCS, the non-convex constraints involve norms and are not globally smooth due to the singularity at zero. However, concavity ensures that starting from a feasible initialization, all subsequent iterates remain feasible for the original non-convex constraints, and therefore the norm arguments will always be non-zero. In this feasible region, the $L_j$-partial smoothness required by Assumption \ref{assumption: non-convex h} holds, so the convergence analysis applies.

\refstepcounter{assumption}
\textbf{Assumption~\theassumption:}
\label{assumption: C full column rank}
The matrix $\bar{C}$ is full column rank, since each global variable is associated with at least one local variable.  


\begin{table}
\centering
\caption{Compact Notation for Convergence Analysis}
\vspace{-0.15cm}
\label{table}
\setlength{\tabcolsep}{3pt}
\renewcommand{\arraystretch}{1.5} 
\begin{tabular}{|c|c|c|}
\hline
\textbf{Compact Notation} & 
\textbf{PCC-DCS} &
~~~~\textbf{MC-DCS}~~~~
\\
\hline
Local variables $\{\bar{x}_i, \bar{w}_i\}$ & 
$\{[ \tilde{v}_i; \tilde{r}_i], \vect(K_i)] \}$ &
$\{ \tilde{v}_i, \vect(K_i)\}$ 
\\
Global variables $\bar{z}$ & 
$[z; \zeta]$ & 
$z$ 
\\
Dual variables $\bar{y}_i$  & 
$[y_i; \xi_i]$ & 
$y_i$ 
\\
Objective function $f_i(\bar{x}_i, \bar{w}_i)$ & 
$\calJ_i(v_i, K_i)$ & 
$\calJ_i(v_i, K_i)$
\\
\multirow{2}{*}{\shortstack{Consensus constraints\\$\! \bar{x}_i = \bar{C}_i \bar{z} \!$}}
  & 
\multirow{2}{*}{$[\tilde{v}_i; \tilde{r}_i] \! = \! 
[\tilde{z}_i; \tilde{\zeta}_i]$} & 
\multirow{2}{*}{$\tilde{v}_i = 
\tilde{z}_i$} 
\\[0.35cm]
\multirow{2}{*}{\shortstack{Local convex\\constraints $g_i(\bar{x}_i, \bar{w}_i) \! \leq \! 0$}} & 
\multirow{2}{*}{\shortstack{$\cala_i(v_i) \! = \! 0,
\calB_i(K_i) \! \succeq \! 0$\\$\cale_i^{\text{PCC}} (K_i, r_i) \! \leq \! 0$}}
 & 
\multirow{2}{*}{\shortstack{$\cala_i(v_i) \! = \! 0$\\$\calB_i(K_i) \! \succeq \! 0$}} 
\\[0.45cm]
\multirow{2}{*}{\shortstack{Local non-convex\\constraints $h_i(\bar{x}_i) \! \leq \! 0$}} & 
\multirow{2}{*}{\shortstack{$\calc_i^{\text{PCC}}(v_i, r_i) \! \leq \! 0$,\\
$\tilde{\cald}_i^{\text{PCC}}(\tilde{v}_i, \tilde{r}_i) \! \leq \! 0$}} & 
\multirow{2}{*}{\shortstack{$\calc_i^{\text{MC}}(v_i) \! \leq \! 0$,\\
$\tilde{\cald}_i^{\text{MC}}(\tilde{v}_i) \! \leq \! 0$}}
\\[0.45cm]
\hline
\end{tabular}
\label{tab: compact notation convergence}
\end{table}

Subsequently, considering the distributed ADMM algorithms with iterative linearization of the non-convex constraints, as in Sections \ref{sec: pcc dcs} and \ref{sec: mcc dcs}, the local subproblems \eqref{eq: pcc-dcs - local updates} and \eqref{eq: mc-dcs - local updates} at iteration $\ell+1$ can be written more compactly as
\begin{align}
&\!\!\!\! \{\bar{x}_i^{\ell+1}, \bar{w}_i^{\ell+1} \} = \argmin_{\bar{x}_i, \bar{w}_i} f_i(\bar{x}_i, \bar{w}_i)  
+ \frac{\rho}{2} \| \bar{x}_i - \bar{C}_i \bar{z}^\ell + \bar{y}_i^\ell/\rho \|_2^2
\nonumber
\\
&\!\!\!\! ~ \mathrm{s.t.} \quad g_i(\bar{x}_i, \bar{w}_i) \leq 0, ~
h_i(\bar{x}_i^\ell) + \nabla h_i(\bar{x}_i^\ell) (\bar{x}_i - \bar{x}_i^\ell) \leq 0, \!
\label{eq: compact local}
\end{align}
where $\bar{y}_i$ denote the dual variables for the consensus constraints. Similarly, the global updates can then be written as
$\bar{z}^{\ell+1} = (\bar{C}^\top \bar{C})^{-1} \bar{C}^\top \bar{x}^{\ell+1}$
since $\bar{C}^\top \bar{C}$ is invertible as $\bar{C}$ is full column rank. By denoting with $\bar{y} \in \Rb^n$, the concatenation of all $\bar{y}_i$, $i \in \calV$, while respecting the ordering of $\bar{x}$, the dual updates are then expressed as
$\bar{y}^{\ell+1} = \bar{y}^\ell + \rho (\bar{x}^{\ell+1} - \bar{C} \bar{z}^{\ell+1})$.

The KKT conditions for Problem \ref{problem: compact} are given as follows.

\refstepcounter{definition}
\labeleditem{Definition}{\thedefinition}{KKT Conditions of General Consensus Problem}
A point $(\bar{x}^*, \bar{w}^*, \bar{z}^*, \bar{y}^*, \eta^*, \lambda^*)$ is a stationary point of Problem \ref{problem: compact} if and only if
\begin{subequations}
\begin{align}
& \! \! \! \! \! \!
\nablax f(\bar{x}^* \!, \bar{w}^*) 
\! + \! \nablax g(\bar{x}^* \!, \bar{w}^*)\!{}^\top \eta^*
\! + \! \nabla h(\bar{x}^*)\!{}^\top \lambda^*
\! + \bar{y}^* \! = 0,
\! \! \!
\label{eq: KKT compact - optimality x}
\\
& \! \! \! \! \! \!
\nablaw f(\bar{x}^* \!, \bar{w}^*) 
\! + \! \nablaw g(\bar{x}^* \!, \bar{w}^*)^\top \eta^* = 0,
\label{eq: KKT compact - optimality w}
\\
& \! \! \! \!
- \bar{C}^\top \bar{y}^* = 0,
\label{eq: KKT compact - optimality z}
\\
& \! \! \! \!
\eta_j^* g_j(\bar{x}^* \!, \bar{w}^*) = 0, \quad
\forall j \in \llbracket 1, \bar{p} \rrbracket,
\label{eq: KKT compact - slackness convex constraints}
\\
& \! \! \! \!
\lambda_j^* h_j(\bar{x}^*) = 0, \quad
\forall j \in \llbracket 1, \bar{q} \rrbracket,
\label{eq: KKT compact - slackness concave constraints}
\\
& \! \! \! \!
g(\bar{x}^* \!, \bar{w}^*) \leq 0, ~ 
h(\bar{x}^*) \leq 0, ~
\bar{x}^* = \bar{C} \bar{z}^*,
\label{eq: KKT compact - feasibility}
\\
& \! \! \! \!
\eta^* \geq 0, ~
\lambda^* \geq 0,
\end{align}
\label{eq: KKT compact}%
\end{subequations}
where $\eta \in \Rb^{\bar{p}}$ and $\lambda \in \Rb^{\bar{q}}$ are the Lagrange multipliers for the constraints $g(\bar{x}, \bar{w}) \leq 0$ and $h(\bar{x}) \leq 0$, respectively.

The KKT conditions for the local subproblems \eqref{eq: compact local} can be written in a concatenated form for all $i \in \calV$, as follows. 

\refstepcounter{definition}
\labeleditem{Definition}{\thedefinition}{KKT Conditions of Local Subproblems}
A point $(\bar{x}^{\ell+1}, \bar{w}^{\ell+1}, \sigma^{\ell+1}, \nu^{\ell+1})$ is a stationary point of the local subproblems \eqref{eq: compact local} if and only if
\begin{subequations}
\begin{align}
& \! \! 
\nablax f(\bar{x}^{\ell+1}, \bar{w}^{\ell+1}) 
+ y^\ell + \rho (\bar{x}^{\ell+1} - \bar{C} \bar{z}^\ell) 
\label{eq: KKT local - optimality x}
\\
& ~~~~~~~~~~~ 
+ \nablax g(\bar{x}^{\ell+1}, \bar{w}^{\ell+1})^\top \sigma^{\ell+1}
+ \nabla h(\bar{x}^\ell)^\top \nu^{\ell+1} = 0, 
\nonumber
\\
& \! \! 
\nablaw f(\bar{x}^{\ell+1}, \bar{w}^{\ell+1}) 
+ \nablaw g(\bar{x}^{\ell+1}, \bar{w}^{\ell+1})^\top \sigma^{\ell+1} = 0,
\label{eq: KKT local - optimality w}
\\
& \!
\sigma_j^{\ell+1} g_j(\bar{x}^{\ell+1}, \bar{w}^{\ell+1}) = 0, ~
\forall j \in \llbracket 1, \bar{p} \rrbracket, 
\label{eq: KKT local - slackness convex}
\\
& \! 
\nu_j^{\ell+1} [ h_j(\bar{x}^\ell) \! + \! \nabla h_j(\bar{x}^\ell)^{\top \!} (\bar{x}^{\ell+1} \! - \! \bar{x}^\ell) ] \! = \! 0, ~ 
\forall j \! \in \! \llbracket 1, \bar{q} \rrbracket,
\label{eq: KKT local - slackness concave linearized}
\\
& \!
g(\bar{x}^{\ell+1}, \bar{w}^{\ell+1}) \leq 0, ~
h(\bar{x}^\ell) \! + \! \nabla h(\bar{x}^\ell)^\top (\bar{x}^{\ell+1} \! - \! \bar{x}^\ell) \leq 0,
\label{eq: KKT local - feasibility}
\\
& \!
\sigma^{\ell+1} \geq 0, ~
\nu^{\ell+1} \geq 0,
\end{align}
\label{eq: KKT local}%
\end{subequations}
where $\sigma$ and $\nu$ are the Lagrange multipliers for the constraints $g(\bar{x}, \bar{w}) \leq 0$ and $h(\bar{x}^\ell) + \nabla h(\bar{x}^\ell)^\top (\bar{x} - \bar{x}^\ell) \leq 0$, respectively.

Let us define the function $V^\ell$ given by
\begin{align}
V^\ell 
& = 
\| \bar{y}^\ell \! - \! \bar{y}^* \|_{\frac{1}{\rho} I}^2
+ 
\| \bar{r}^\ell \|_{T_\nu}^2
+
\| \bar{C} (\bar{z}^\ell \! - \! \bar{z}^*) \|_{\rho I + T_\nu}^2 
,
\end{align}
with $\bar{r}^\ell = \bar{x}^\ell - \bar{C} \bar{z}^\ell$. We will prove that $V^\ell$ is a Lyapunov function with the convergence points satisfying the KKT conditions \eqref{eq: KKT compact}. We further consider the following assumption.

\refstepcounter{assumption}
\textbf{Assumption~\theassumption:}
\label{assumption: boundedness and SOOC type assumption}
The sum $\sum_{j = 1}^{\bar{q}} \lambda_j^* L_j$ is upper bounded by a constant matrix $T_\lambda = \mathrm{bdiag}(\tau_\lambda I_{\bar{n}_\rA}, 0_{\bar{n}_\rB \times \bar{n}_\rB})$ with $\tau_\lambda > 0$, i.e., 
$\sum_{j = 1}^{\bar{q}} \lambda_j^* L_j \preceq T_{\lambda}$.
Similarly, the sum 
$\sum_{j = 1}^{\bar{q}} \nu_j^\ell L_j \preceq T_\nu$,
with $T_\nu = \mathrm{bdiag}(\tau_\nu I_{\bar{n}_\rA}, 0_{\bar{n}_\rB \times \bar{n}_\rB})$ and $\tau_\nu > 0$, for any iteration $\ell$. In addition, $\mu_x \geq \tau_\lambda$ and $\mu_x \geq \tau_\nu$.

\refstepcounter{remark}
\labeleditem{Remark}{\theremark}{Interpretation of Assumption \ref{assumption: boundedness and SOOC type assumption}}
Assumption \ref{assumption: boundedness and SOOC type assumption} requires the  boundedness of the Lagrange multipliers, a mild regularity condition in constrained optimization \cite{nocedal2006numerical}. 
%
%
When $f,g$ and $h$, are twice differentiable, the conditions $\mu_x \geq \tau_\lambda$ and $\mu_x \geq \tau_\nu$ would imply that the Hessian of the Lagrangians w.r.t. $\bar{x}$,
$
\nablaxx f (\bar{x}^*, \bar{w}^*) + 
\nablaxx g (\bar{x}^*, \bar{w}^*)^\top \eta^* +
\nablaxx h (\bar{x}^*)^\top \lambda^* 
\succeq 0
$
and
$
\nablaxx f (\bar{x}^{\ell}, \bar{w}^{\ell}) + 
\nablaxx g (\bar{x}^{\ell}, \bar{w}^{\ell}) \sigma^{\ell} +
\nablaxx h(\bar{x}^{\ell})^\top \nu^{\ell}
\succeq 0
$
,
aligning with the second-order necessary optimality condition, which is widely used in the analysis of non-convex optimization methods \cite{nocedal2006numerical, boggs1995sequential}. 
As later shown in Lemma \ref{lemma 3}, these conditions guarantee the sufficient descent of $V^\ell$ at each iteration. 

%
%

\subsection{Intermediate Lemmas} Let us establish the following necessary lemmas.
\label{sec: convergence - lemmas}

\refstepcounter{lemma}
\textbf{Lemma~\thelemma:}
\label{lemma 1}
Under Assumption \ref{assumption: non-convex h}, the following relationships hold for each $j \in \llbracket 1, \bar{q} \rrbracket$, at every iteration $\ell$:
\begin{align}
- \nu_j^{\ell+1} \langle \nabla h_j(\bar{x}^\ell), \bar{x}^{\ell+1} - \bar{x}^* \rangle 
& \leq
\frac{\nu_j^{\ell+1}}{2} \| \bar{x}^\ell - \bar{x}^* \|_{L_j}^2,
\label{R1}
\tag{R1}
\\
\lambda_j^* \langle \nabla h_j(\bar{x}^*), \bar{x}^{\ell+1} - \bar{x}^{*} \rangle
& \leq 
\frac{\lambda_j^*}{2} \| \bar{x}^{\ell+1} - \bar{x}^* \|_{L_j}^2.
\label{R2}
\tag{R2}
\end{align}
\begin{proof}
Let us denote the left-hand side (LHS) of \eqref{R1} with $\bar{A}_1$. Then, we have
\begin{equation}
\begin{aligned}
\bar{A}_1
& = 
- \nu_j^{\ell+1} \langle \nabla h_j(\bar{x}^\ell), \bar{x}^{\ell+1} - \bar{x}^\ell + \bar{x}^\ell - \bar{x}^* \rangle
\\
& = 
\nu_j^{\ell+1} ( h_j(\bar{x}^\ell)
+ \langle \nabla h_j(\bar{x}^\ell), \bar{x}^* - \bar{x}^\ell \rangle ),
\end{aligned}
\label{eq: r1 - LHS - eq1}
\end{equation}
using the slackness condition \eqref{eq: KKT local - slackness concave linearized}. Subsequently, using the fact that each function $-h_j$ is also $L_j$-smooth, we have
\begin{equation}
h_j(\bar{x}^\ell)
+ \langle \nabla h_j(\bar{x}^\ell), \bar{x}^* - \bar{x}^\ell \rangle 
\leq
h_j(\bar{x}^*)
+ \frac{1}{2} \| \bar{x}^\ell - \bar{x}^* \|_{L_j}^2.
\label{eq: h l-smooth}
\end{equation}
Combining \eqref{eq: r1 - LHS - eq1}, \eqref{eq: h l-smooth}, $\nu_j^{\ell+1} \geq 0$ and $h_j(\bar{x}^*) \leq 0$, we obtain
\begin{equation}
\bar{A}_1 \leq \nu_j^{\ell+1} 
\big( 
h_j(\bar{x}^*)
+ \frac{1}{2} \| \bar{x}^\ell - \bar{x}^* \|_{L_j}^2
\big)
\leq \frac{\nu_j^{\ell+1}}{2} \| \bar{x}^\ell - \bar{x}^* \|_{L_j}^2
,
\nonumber
\end{equation}
which proves \eqref{R1}. 

The LHS of \eqref{R2}, denoted with $\bar{A}_2$, can be written as
\begin{equation}
\bar{A}_2 = \lambda_j^* [ h_j(\bar{x}^*) + \nabla h_j(\bar{x}^*)^\top (\bar{x}^{\ell+1} - \bar{x}^{*})],
\end{equation}
using the slackness condition \eqref{eq: KKT compact - slackness concave constraints}. Then, using again the $L_{j}$- smoothness of $-h_j$, we have
$
h_j(\bar{x}^*)
+ \langle \nabla h_j(\bar{x}^*), \bar{x}^{\ell+1}- \bar{x}^* \rangle \leq
h_j(\bar{x}^{\ell+1})
+ \frac{1}{2} \| \bar{x}^{\ell+1} - \bar{x}^* \|_{L_j}^2$,
so since $\lambda_j^* \geq 0$, we obtain
\begin{equation*}
\bar{A}_2 \leq \lambda_j^* \big( h_j(\bar{x}^{\ell+1})
+ \frac{1}{2} \| \bar{x}^{\ell+1} - \bar{x}^* \|_{L_j}^2 \big)
\leq \frac{\lambda_j^*}{2} \| \bar{x}^{\ell+1} \! - \bar{x}^* \|_{L_j}^2,
\end{equation*}
where we also used the fact that $h_j(\bar{x}^{\ell+1}) \leq h_j(\bar{x}^\ell) + \langle \nabla h_j(\bar{x}^\ell), \bar{x}^{\ell+1} - \bar{x}^\ell \rangle \leq 0$ from the concavity of $h_j$.
\end{proof}

\refstepcounter{lemma}
\textbf{Lemma~\thelemma:}
\label{lemma 2}
Under Assumption \ref{assumption: C full column rank}, the following relationships hold at every iteration $\ell$:
\begin{align}
&
\langle \bar{C} (\bar{z}^{\ell+1} - \bar{z}^\ell), \bar{x}^{\ell+1} - \bar{x}^{*} \rangle
= 
\frac{1}{2}
\big(
\| \bar{C} (\bar{z}^{\ell+1} - \bar{z}^\ell) \|_2^2,
\label{R3}
\tag{R3}
\\
& ~~~~~~~~~~~~~~~~~~~~~ + \| \bar{C} (\bar{z}^{\ell+1} - \bar{z}^*) \|_2^2
- \| \bar{C} (\bar{z}^\ell - \bar{z}^*) \|_2^2 \big),
\nonumber
\\
& \langle \bar{y}^{\ell+1} - \bar{y}^*, \bar{x}^{\ell+1} - \bar{x}^* \rangle 
=
\frac{1}{2 \rho} 
\big( \| \bar{y}^{\ell+1} - \bar{y}^* \|_2^2
\label{R4}
\tag{R4}
\\
& ~~~~~~~~~~~~~~~~~~~~~~  - \| \bar{y}^\ell - \bar{y}^* \|_2^2 \big)
+ \frac{\rho}{2} \| \bar{x}^{\ell+1} - \bar{C} \bar{z}^{\ell+1} \|_2^2.
\nonumber
\end{align}
\begin{proof}
We begin with proving \eqref{R3}. Let $\calQ(\bar{C}), \calP(\bar{C}) \in \Rb^{\bar{n} \times \bar{n}}$ be the orthogonal projection matrices onto $\rIm(\bar{C})$ and $\rIm(\bar{C})^\bot$, respectively. Since $\bar{C}$ has full column rank, then $\calQ(\bar{C}) = \bar{C} (\bar{C}^\top \bar{C})^{-1} \bar{C}^\top$ and $\calP(\bar{C}) = I - \calQ(\bar{C})$.
The LHS of \eqref{R3}, denoted with $\bar{A}_3$, can be written as
\begin{align*}
\bar{A}_3 = \langle \bar{C} (\bar{z}^{\ell+1} - \bar{z}^\ell), \calP(\bar{C}) \bar{x}^{\ell+1} + \calQ(\bar{C}) \bar{x}^{\ell+1} - \bar{x}^* \rangle.
\end{align*}
Through $\bar{C}^\top \calP(\bar{C}) = 0$, $\calQ(\bar{C}) \bar{x}^{\ell+1} = \bar{C} \bar{z}^{\ell+1}$ and $\bar{x}^* = \bar{C} \bar{z}^*$,
\begin{align*}
\bar{A}_3 = \langle \bar{C} (\bar{z}^{\ell+1} - \bar{z}^\ell), \bar{C} (\bar{z}^{\ell+1} - \bar{z}^{*}) \rangle.
\end{align*}
Then, using $2 \langle a, b \rangle = \| a \|_2^2 + \| b \|_2^2 - \| a -b \|_2^2$, we obtain
%
%
\eqref{R3}.

The LHS of \eqref{R4}, denoted with $\bar{A}_4$, can be written as
\begin{align}
\bar{A}_4 & = \langle \bar{y}^{\ell+1} - \bar{y}^*, \bar{x}^{\ell+1} - \bar{C} \bar{z}^* \rangle
= \langle \bar{y}^{\ell+1} - \bar{y}^*, \bar{x}^{\ell+1} \rangle
\nonumber
\\
& = \langle \bar{y}^{\ell+1} - \bar{y}^*, \calP(\bar{C}) \bar{x}^{\ell+1} + \calQ(\bar{C}) \bar{x}^{\ell+1} \rangle
\label{eq: r4 proof - eq1}
\\
& = \langle \bar{y}^{\ell+1} - \bar{y}^*, \bar{x}^{\ell+1} - \bar{C} \bar{z}^{\ell+1} \rangle,
\nonumber
\end{align}
since $\bar{C}^\top \bar{y}^* = \bar{C}^\top \bar{y}^{\ell+1} = 0$ and $\calP(\bar{C}) \bar{x}^{\ell+1} = \bar{x}^{\ell+1} - \bar{C} \bar{z}^{\ell+1}$. In addition, we have
\begin{align}
2 \langle \bar{y}^{\ell+1} - \bar{y}^*, \bar{y}^{\ell+1} - \bar{y}^\ell \rangle
& = \| \bar{y}^{\ell+1} - \bar{y}^* \|_2^2
\label{eq: r4 proof - eq2}
\\
& \quad + \| \bar{y}^{\ell+1} - \bar{y}^\ell \|_2^2
- \| \bar{y}^\ell - \bar{y}^* \|_2^2.
\nonumber
\end{align}
Substituting $\bar{y}^{\ell+1} - \bar{y}^\ell = \rho (\bar{x}^{\ell+1} - \bar{C} \bar{z}^{\ell+1})$ into \eqref{eq: r4 proof - eq2}, we get 
\begin{align}
& \! 2 \rho \langle \bar{y}^{\ell+1} - \bar{y}^*,  \bar{x}^{\ell+1} - \bar{C} \bar{z}^{\ell+1}
\rangle
= \| \bar{y}^{\ell+1} - \bar{y}^* \|_2^2
\label{eq: r4 proof - eq3}
\\
& \qquad \qquad \qquad \qquad \quad + \rho^2 \| \bar{x}^{\ell+1} - \bar{C} \bar{z}^{\ell+1} \|_2^2
- \| \bar{y}^\ell - \bar{y}^* \|_2^2.
\nonumber
\end{align}
Then, using \eqref{eq: r4 proof - eq3} in \eqref{eq: r4 proof - eq1} yields \eqref{R4}.
\end{proof}

\refstepcounter{lemma}
\labeleditem{Lemma}{\thelemma}{Sufficient Descent}
\label{lemma 3}
Under Assumptions \ref{assumption: strongly convex f}-\ref{assumption: boundedness and SOOC type assumption}, the following inequality holds at every iteration $\ell$:
\begin{align}
V^{\ell+1} - V^\ell
\leq  
& - c_1
\| \bar{r}^{\ell+1} \|_2^2
- c_2
\| \bar{C} (\bar{z}^{\ell+1} - \bar{z}^\ell) \|_2^2,
\label{eq: lemma 3 ineq}
\end{align}
with $c_1, c_2 > 0$.
%
%

\begin{proof}
%
%
Taking the inner product of \eqref{eq: KKT local - optimality x} with $\bar{x}^{\ell+1} - \bar{x}^*$, and replacing $\bar{y}^{\ell+1} = \bar{y}^\ell + \rho (\bar{x}^{\ell+1} - \bar{C} \bar{z}^{\ell+1})$, gives
\begin{align}
& \langle \nablax f(\bar{x}^{\ell+1}, \bar{w}^{\ell+1}) 
+ \bar{y}^{\ell+1} + \rho \bar{C} (\bar{z}^{\ell+1} \! - \! \bar{z}^\ell), \bar{x}^{\ell+1} \! - \! \bar{x}^* \rangle
\label{lemma 3 proof - local KKT inner product x}
\\
& ~ = - \langle \nablax g(\bar{x}^{\ell+1}, \bar{w}^{\ell+1})^\top \sigma^{\ell+1} + \nabla h(\bar{x}^\ell)^\top \nu^{\ell+1}, \bar{x}^{\ell+1} \! - \! \bar{x}^* \rangle. 
\nonumber
\end{align}
In addition, the inner product of \eqref{eq: KKT local - optimality w} with $\bar{w}^{\ell+1} - \bar{w}^*$, gives
\begin{align}
& \langle \nablaw f(\bar{x}^{\ell+1}, \bar{w}^{\ell+1}), \bar{w}^{\ell+1} \! - \! \bar{w}^* \rangle
\label{lemma 3 proof - local KKT inner product w}
\\
& \quad = - \langle \nablaw g(\bar{x}^{\ell+1}, \bar{w}^{\ell+1})^\top \sigma^{\ell+1}, \bar{w}^{\ell+1} \! - \! \bar{w}^* \rangle. 
\nonumber
\end{align}
Next, we observe that 
\begin{align}
& - \langle \nablax g(\bar{x}^{\ell+1}, \bar{w}^{\ell+1})^\top \sigma^{\ell+1},  \bar{x}^{\ell+1} - \bar{x}^* \rangle
\label{lemma 3 proof - g convexity at local}
\\
& - \langle \nablaw g(\bar{x}^{\ell+1}, \bar{w}^{\ell+1})^\top \sigma^{\ell+1},  \bar{w}^{\ell+1} - \bar{w}^* \rangle
\nonumber
\\
& \leq 
\sigma^{\ell+1\!} {}^\top \! (g(\bar{x}^*, \bar{w}^*) - g(\bar{x}^{\ell+1}, \bar{w}^{\ell+1}) )
\leq \sigma^{\ell+1\!} {}^\top \! g(\bar{x}^*, \bar{w}^*)
\leq 0,
\nonumber
\end{align}
where the first step uses the convexity of $g$,
i.e., 
$
g(\bar{x}^*, \bar{w}^*) 
\geq 
g(\bar{x}^{\ell+1}, \bar{w}^{\ell+1}) 
+ \nablax g (\bar{x}^{\ell+1}, \bar{w}^{\ell+1})
(\bar{x}^* - \bar{x}^{\ell+1})
+ \nablaw g (\bar{x}^{\ell+1}, \bar{w}^{\ell+1})
(\bar{w}^* - \bar{w}^{\ell+1})
$,
the second one that $\sigma^{\ell+1} {}^\top g(\bar{x}^{\ell+1}, \bar{w}^{\ell+1}) = 0$, and the third one that $\sigma^{\ell+1} \geq 0$ and $g(\bar{x}^*, \bar{w}^*) \leq 0$. 

Similarly, the inner product of \eqref{eq: KKT compact - optimality x} with $\bar{x}^* - \bar{x}^{\ell+1}$, yields
\begin{align}
& \langle \nablax f(\bar{x}^*, \bar{w}^*) 
+ \bar{y}^*, \bar{x}^* - \bar{x}^{\ell+1} \rangle =
\label{lemma 3 proof - full KKT inner product x}
\\
& ~~~~~~~~~~~ - \langle \nablax g(\bar{x}^*, \bar{w}^*)^\top \eta^* + \nabla h(\bar{x}^*)^\top \lambda^*, 
\bar{x}^* - \bar{x}^{\ell+1} \rangle,
\nonumber
\end{align}
and the inner product of of \eqref{eq: KKT compact - optimality w} with $\bar{w}^* - \bar{w}^{\ell+1}$, gives
\begin{align}
& \langle \nablaw f(\bar{x}^*, \bar{w}^*), \bar{w}^* - \bar{w}^{\ell+1} \rangle =
\label{lemma 3 proof - full KKT inner product w}
\\
& ~~~~~~~~~~~ - \langle \nablaw g(\bar{x}^*, \bar{w}^*)^\top \eta^*, 
\bar{w}^* - \bar{w}^{\ell+1} \rangle,
\nonumber
\end{align}
We also observe that 
\begin{align}
& - \langle \nablax g(\bar{x}^*, \bar{w}^*)^\top \eta^*,  \bar{x}^* - \bar{x}^{\ell+1} \rangle
\label{lemma 3 proof - g convexity at full}
\\
& - \langle \nablaw g(\bar{x}^*, \bar{w}^*)^\top \eta^*,  \bar{w}^* - \bar{w}^{\ell+1} \rangle
\nonumber
\\
& \leq 
 \eta^{* \top} \! (g(\bar{x}^{\ell+1}, \bar{w}^{\ell+1}) - g(\bar{x}^*, \bar{w}^*) )
\leq \eta^{* \top} \! g(\bar{x}^{\ell+1}, \bar{w}^{\ell+1})
\leq 0,
\nonumber
\end{align}
where the first step uses  
$g(\bar{x}^{\ell+1}, \bar{w}^{\ell+1}) \geq g(\bar{x}^*, \bar{w}^*) 
+ \nablax g (\bar{x}^*, \bar{w}^*) (\bar{x}^{\ell+1} - \bar{x}^*)
+ \nablaw g (\bar{x}^*, \bar{w}^*) (\bar{w}^{\ell+1} - \bar{w}^*)$,
the second one that $ \eta^{* \top} g(\bar{x}^*, \bar{w}^*) = 0$, and the third one that $\eta^* \geq 0$ and $g(\bar{x}^{\ell+1}, \bar{w}^{\ell+1}) \leq 0$.

Adding together 
\eqref{lemma 3 proof - local KKT inner product x} and \eqref{lemma 3 proof - local KKT inner product w}, 
subtracting  
\eqref{lemma 3 proof - full KKT inner product x} and \eqref{lemma 3 proof - full KKT inner product w}, 
and leveraging 
\eqref{lemma 3 proof - g convexity at local}, 
\eqref{lemma 3 proof - g convexity at full}, 
\eqref{R1} and 
\eqref{R2} 
gives
%
%
%
%
\begin{align}
& \langle \nablax f(\bar{x}^{\ell+1}, \bar{w}^{\ell+1}) - \nablax f(\bar{x}^*, \bar{w}^*), \bar{x}^{\ell+1} - \bar{x}^* \rangle
\nonumber
\\[0.1cm]
& + \langle \nablaw f(\bar{x}^{\ell+1}, \bar{w}^{\ell+1}) - \nablaw f(\bar{x}^*, \bar{w}^*), 
\bar{w}^{\ell+1} - \bar{w}^* \rangle
\label{lemma 3 proof - eq 5}
\\[0.1cm]
& + \langle \bar{y}^{\ell+1} - \bar{y}^* + \rho \bar{C} (\bar{z}^{\ell+1} - \bar{z}^\ell) , \bar{x}^{\ell+1} - \bar{x}^* \rangle
\nonumber
\\
& \leq \sum_{j = 1}^{\bar{q}} \frac{\nu_j^{\ell+1}}{2} \| \bar{x}^\ell - \bar{x}^* \|_{L_j}^2 + \frac{\lambda_j^*}{2} \| \bar{x}^{\ell+1} - \bar{x}^* \|_{L_j}^2 .
\nonumber
\end{align}
Next, from the $M$-partially strong convexity of $f$, we have
\begin{align}
& \langle \nablax f(\bar{x}^{\ell+1}, \bar{w}^{\ell+1}) - \nablax f(\bar{x}^*, \bar{w}^*),  \bar{x}^{\ell+1} - \bar{x}^* \rangle
\label{eq: f strong convex ineq}
\\
& \! \! + \! \langle \nablaw f(\bar{x}^{\ell+1} \!, \!
\bar{w}^{\ell+1}) \! - \! \nablaw f(\bar{x}^* \!, \! \bar{w}^*),  \bar{w}^{\ell+1} \! \! - \! \bar{w}^* \rangle
\nonumber
\\
& \geq 
\| \bar{x}^{\ell+1} - \bar{x}^{*} \|_{M_x}^2.
\nonumber
\end{align}
Substituting \eqref{eq: f strong convex ineq}, \eqref{R3} and \eqref{R4} into \eqref{lemma 3 proof - eq 5}, we obtain
\begin{align}
& \frac{1}{\rho} 
\big[ \| \bar{y}^{\ell+1} - \bar{y}^* \|_2^2
- \| \bar{y}^\ell - \bar{y}^* \|_2^2 \big]
+ \rho \| \bar{x}^{\ell+1} - \bar{C} \bar{z}^{\ell+1} \|_2^2
\\
& + \rho
\big[
\| \bar{C} (\bar{z}^{\ell+1} - \bar{z}^\ell) \|_2^2
+ \| \bar{C} (\bar{z}^{\ell+1} - \bar{z}^*) \|_2^2
- \| \bar{C} (\bar{z}^\ell - \bar{z}^*) \|_2^2 \big]
\nonumber
\\
& \leq \! - \| \bar{x}^{\ell+1} - \bar{x}^* \|_{2 M_x - T_\lambda}^2 
\! + \| \bar{x}^\ell - \bar{x}^* \|_{T_\nu}^2.
\nonumber
\end{align}
Next, note that for any weighted semi-norm $\| \cdot \|_{\bar{\Omega}}$ with $\bar{\Omega} = \mathrm{bdiag}(\bar{\omega} I_{{\bar{n}_\rA}}, 0_{{\bar{n}_\rB} \times \bar{n}_{\rB}})$ and $\bar{\omega} > 0$, we have
\begin{align}
\| \bar{x}^{\ell} - \bar{x}^* \|_{\bar{\Omega}}^2 & =
\| \calP(\bar{C}) \bar{x}^{\ell} + \calQ(\bar{C}) \bar{x}^{\ell} - \bar{x}^* \|_{\bar{\Omega}}^2 
\label{eq: lemma 3 proof - tricky decomposition}
\\
& = 
\| \calP(\bar{C}) \bar{x}^{\ell} + \bar{C} (\bar{z}^{\ell} - \bar{z}^*) \|_{\bar{\Omega}}^2 
\nonumber
\\
& = \| \calP(\bar{C}) \bar{x}^{\ell} \|_{\bar{\Omega}}^2 + \| \bar{C} (\bar{z}^{\ell} - \bar{z}^*) \|_{\bar{\Omega}}^2 \nonumber
\\
& =  \| \bar{x}^{\ell} - \bar{C} \bar{z}^{\ell} \|_{\bar{\Omega}}^2 + \| \bar{C} (\bar{z}^{\ell} - \bar{z}^*) \|_{\bar{\Omega}}^2 \nonumber
\end{align}
using that $\bar{x}^* = \bar{C} \bar{z}^*$, $\calQ(\bar{C}) \bar{x}^{\ell} = \bar{C} \bar{z}^{\ell}$, $\calP(\bar{C}) \bar{\Omega} \bar{C} = 0$, and $\bar{x}^{\ell} - \bar{C} \bar{z}^{\ell} = \calP(\bar{C}) \bar{x}^{\ell}$.
The fact that $\calP(\bar{C}) \bar{\Omega} \bar{C} = 0$ follows from $\calP(\bar{C}) \bar{\Omega} \bar{C} = 
\mathrm{bdiag}(\bar{\omega} \calP(\bar{C}_A) \bar{C}_A, 0)$ and $\calP(\bar{C}_A) \bar{C}_A = 0$. 
%
%
As a result, the inequality becomes
\begin{align}
& \frac{1}{\rho} 
\big[ \| \bar{y}^{\ell+1} - \bar{y}^* \|^2
- \| \bar{y}^\ell - \bar{y}^* \|^2 \big]
+ \big[ \| \bar{r}^{\ell+1} \|_{T_\nu}^2 - \| \bar{r}^{\ell} \|_{T_\nu}^2 \big]
\nonumber
\\
& +
\big[
\| \bar{C} (\bar{z}^{\ell+1} - \bar{z}^*) \|_{T_\nu + \rho I}^2
- \| \bar{C} (\bar{z}^\ell - \bar{z}^*) \|_{T_\nu + \rho I}^2 \big]
\\
& \leq  
- 
\| \bar{r}^{\ell+1} \|_{2 M_x - T_\lambda - T_\nu + \rho I}^2
-
\| \bar{C} (\bar{z}^{\ell+1} \! - \! \bar{z}^*) \|_{2 M_x - T_\lambda - T_\nu}^2
\nonumber
\\
& - \rho
\| \bar{C} (\bar{z}^{\ell+1} - \bar{z}^\ell) \|^2,
\nonumber
\end{align}
%
%
and then, given that $2 M_x - T_\lambda - T_\nu \succeq 0$, we obtain
\begin{align*}
& \frac{1}{\rho} 
\big[ \| \bar{y}^{\ell+1} - \bar{y}^* \|^2
- \| \bar{y}^\ell - \bar{y}^* \|^2 \big]
+ (\| \bar{r}^{\ell+1} \|_{T_\nu}^2 - \| \bar{r}^\ell \|_{T_\nu}^2)
\\
& + 
\left(
\| \bar{C} (\bar{z}^{\ell+1} - \bar{z}^*) \|_{T_\nu + \rho I}^2
- \| \bar{C} (\bar{z}^\ell - \bar{z}^*) \|_{T_\nu + \rho I}^2 \right)
\leq
\\
& \!  
\! - 
\| \bar{r}^{\ell+1} \|_{2 M_x - T_\lambda - T_\nu + \rho I}^2
\!
- \! \rho
\| \bar{C} (\bar{z}^{\ell+1} \! \! - \! \bar{z}^\ell) \|^2,
\end{align*}
which implies \eqref{eq: lemma 3 ineq}
%
%
as $\rho >  0$, $2 M_x \! - \! T_\lambda \! - T_\nu \! + \! \rho I \succ \! 0$.
\end{proof}

%

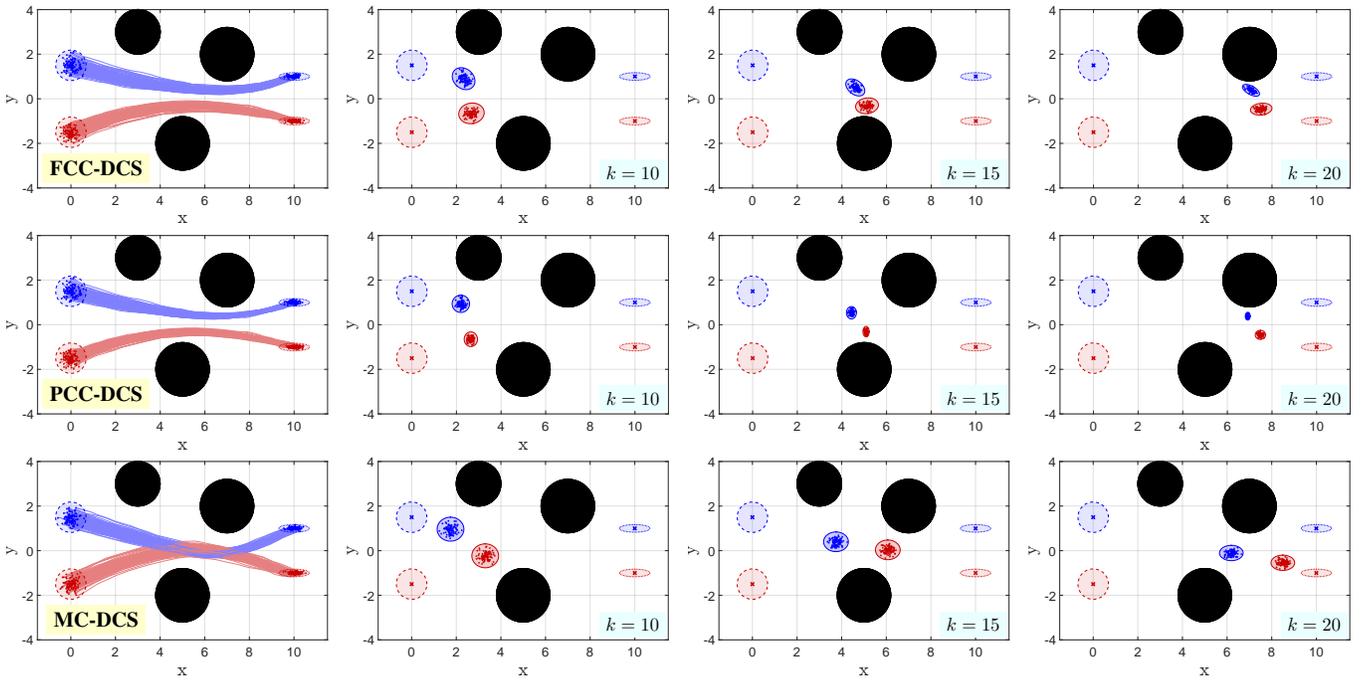
\begin{figure*}[!t]
\centering
\hfil
\subfloat{
\begin{tikzpicture}
    \node[anchor=south west,inner sep=0] at (0,0){
    \includegraphics[width=0.236\textwidth, trim={0cm 0cm 0cm 0cm},clip]{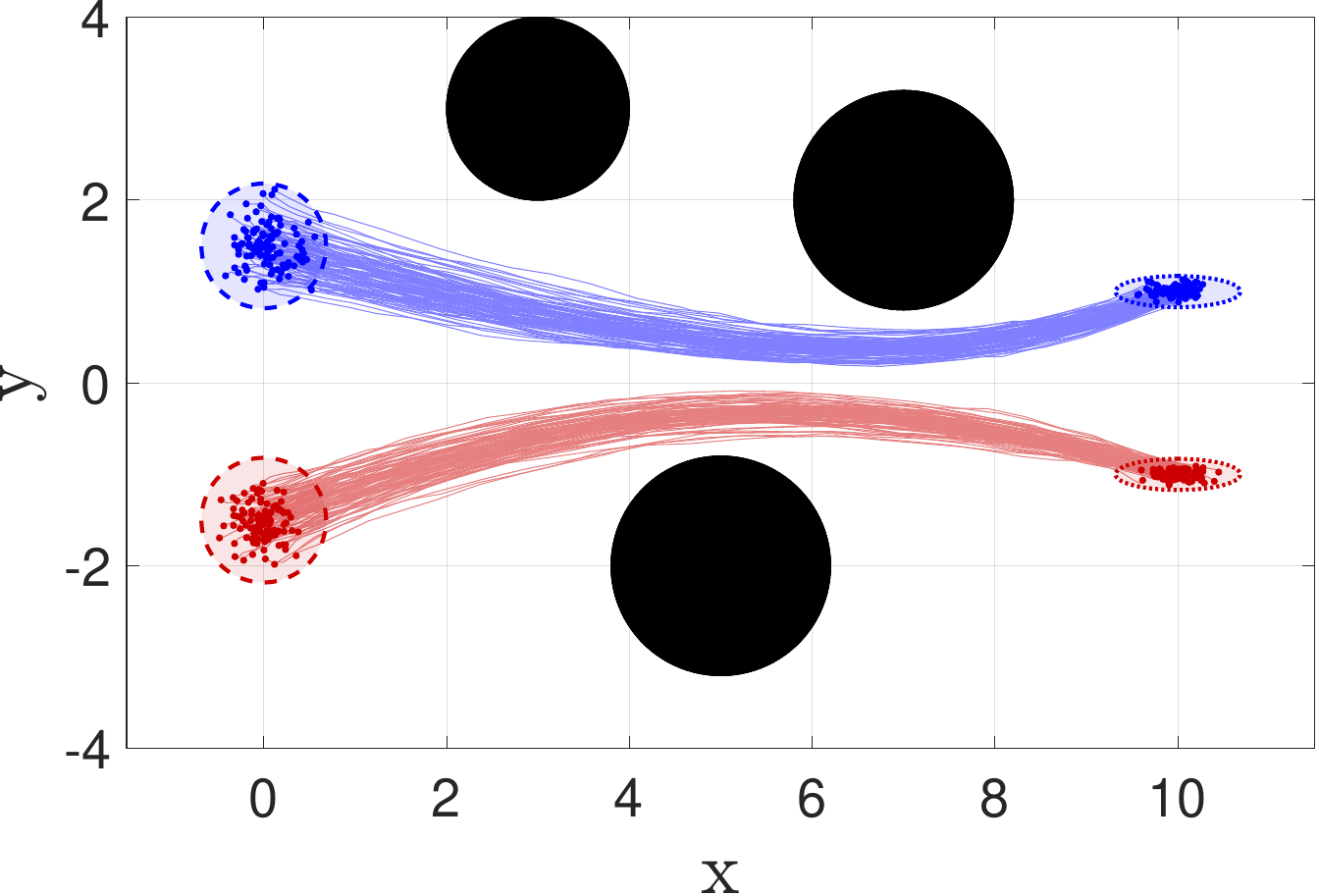}};
    \node[align=center, fill = myYellow, scale = 0.8] (c) at (1.2, 0.75) {\textbf{FCC-DCS}};
\end{tikzpicture}
}
\hfil
\subfloat{
\begin{tikzpicture}
    \node[anchor=south west,inner sep=0] at (0,0){    \includegraphics[width=0.236\textwidth, trim={0cm 0cm 0cm 0cm},clip]{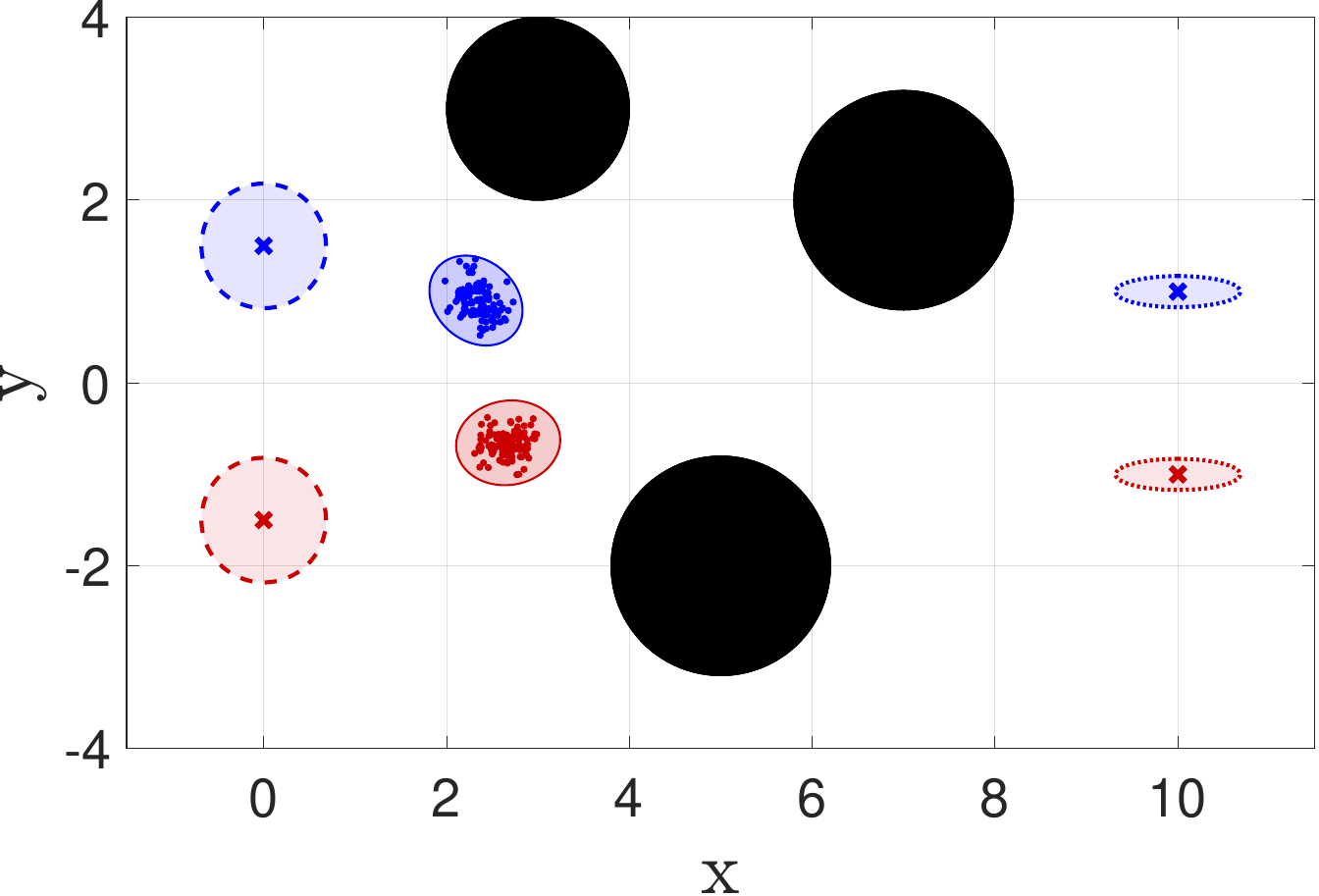}};
    \node[align=center, fill = myCyan, scale = 0.7] (c) at (3.8, 0.69) {$k=10$};
\end{tikzpicture}
}
\hfil
\subfloat{
\begin{tikzpicture}
    \node[anchor=south west,inner sep=0] at (0,0){
    \includegraphics[width=0.236\textwidth, trim={0cm 0cm 0cm 0cm},clip]{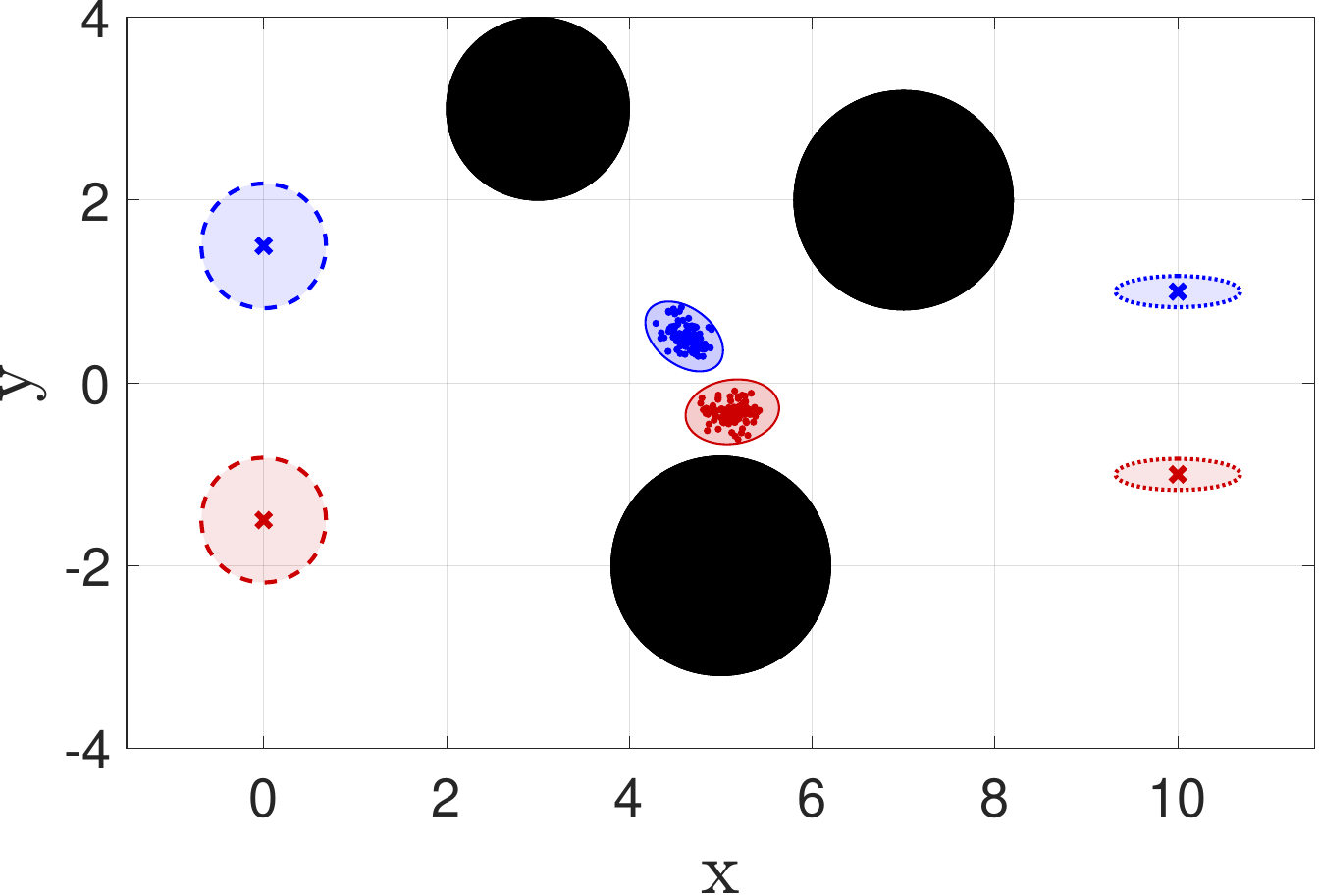}};
    \node[align=center, fill = myCyan, scale = 0.7] (c) at (3.8, 0.69) {$k=15$};
\end{tikzpicture}
}
\hfil
\subfloat{
\begin{tikzpicture}
    \node[anchor=south west,inner sep=0] at (0,0){
    \includegraphics[width=0.236\textwidth, trim={0cm 0cm 0cm 0cm},clip]{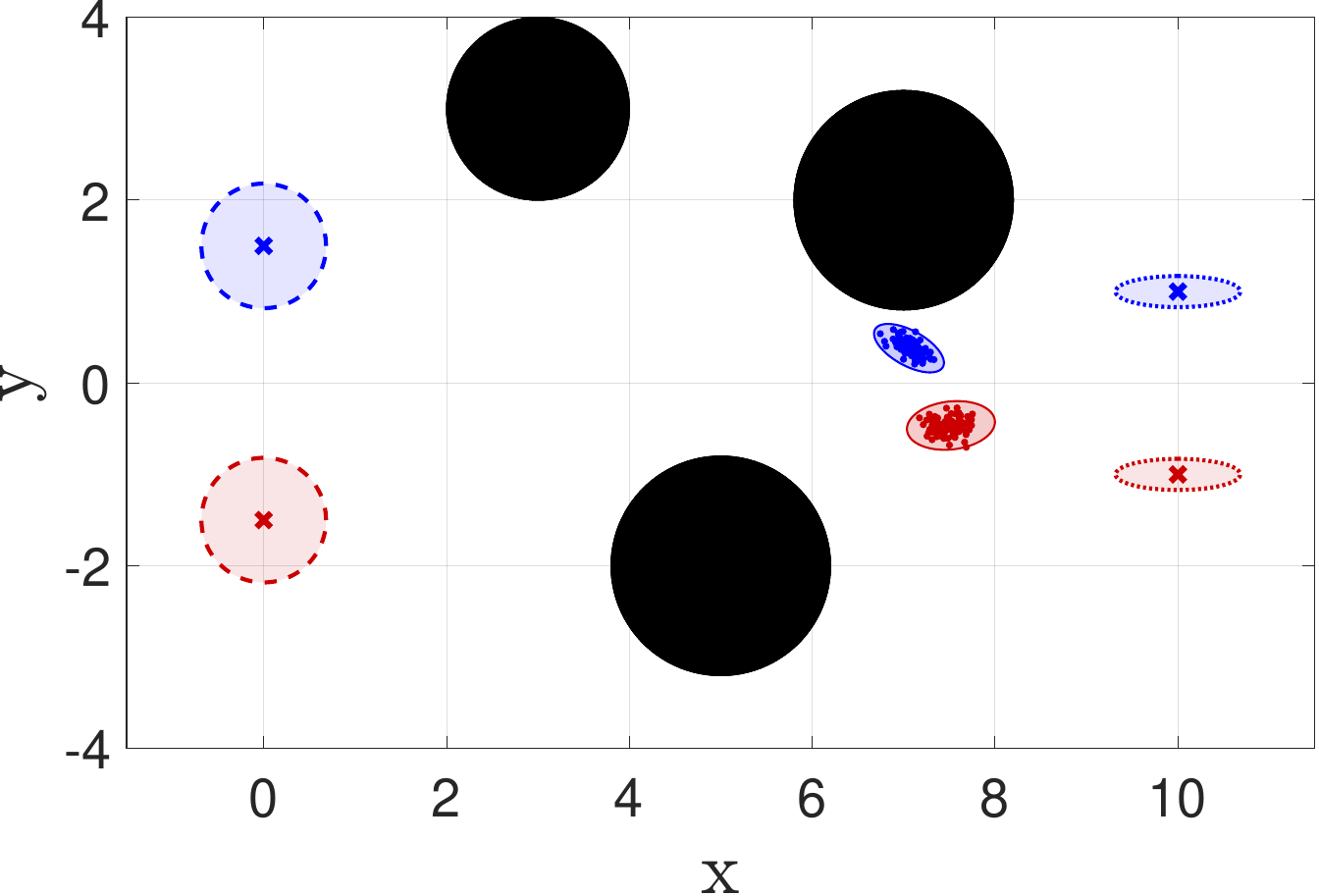}};
    \node[align=center, fill = myCyan, scale = 0.7] (c) at (3.8, 0.69) {$k=20$};
\end{tikzpicture}
}
\hfil
\\[0.1cm]
\hfil
\subfloat{
\begin{tikzpicture}
    \node[anchor=south west,inner sep=0] at (0,0){    \includegraphics[width=0.236\textwidth, trim={0cm 0cm 0cm 0cm},clip]{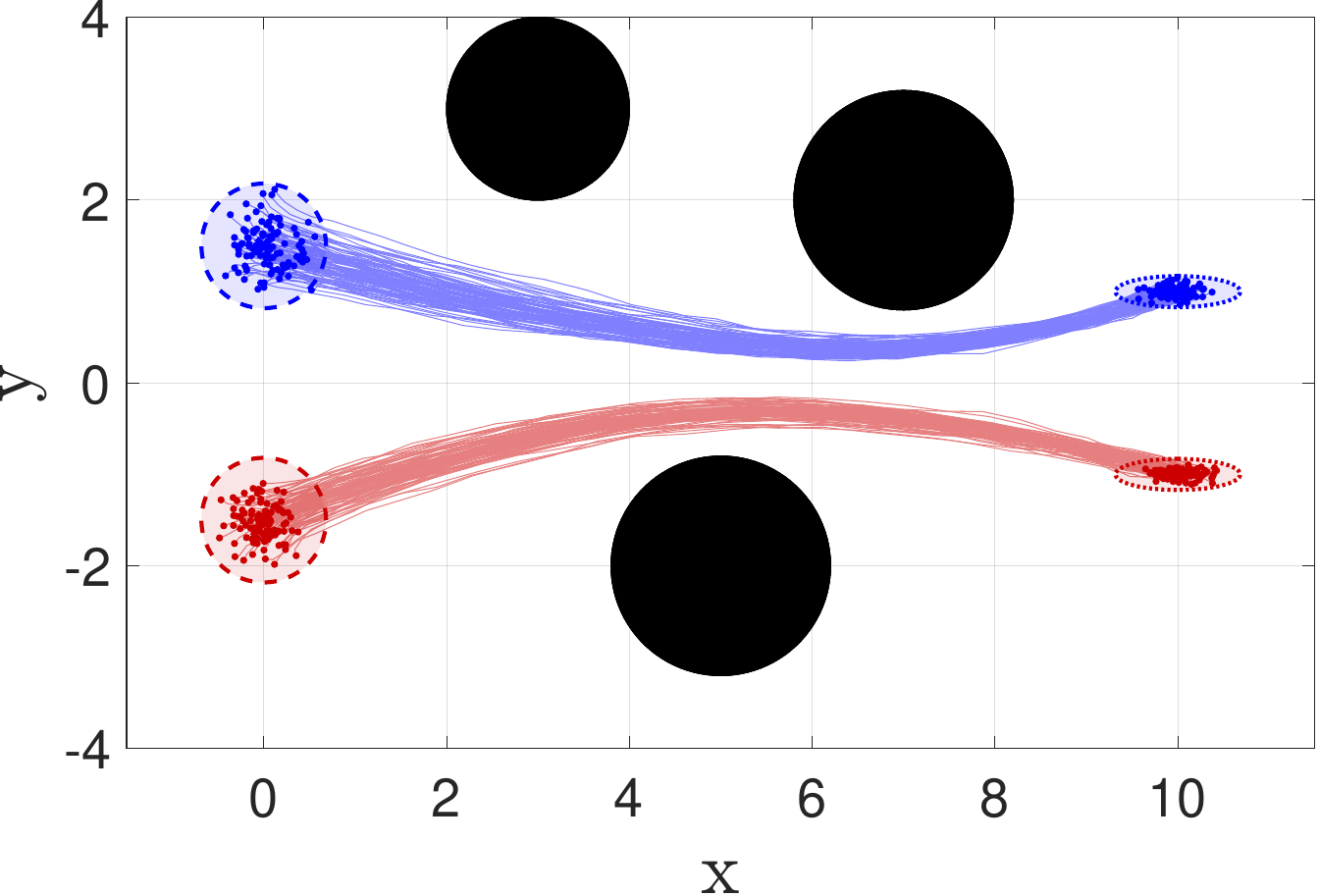}};
    \node[align=center, fill = myYellow, scale = 0.8] (c) at (1.2, 0.75) {\textbf{PCC-DCS}};
\end{tikzpicture}
}
\hfil
\subfloat{
\begin{tikzpicture}
    \node[anchor=south west,inner sep=0] at (0,0){
    \includegraphics[width=0.236\textwidth, trim={0cm 0cm 0cm 0cm},clip]{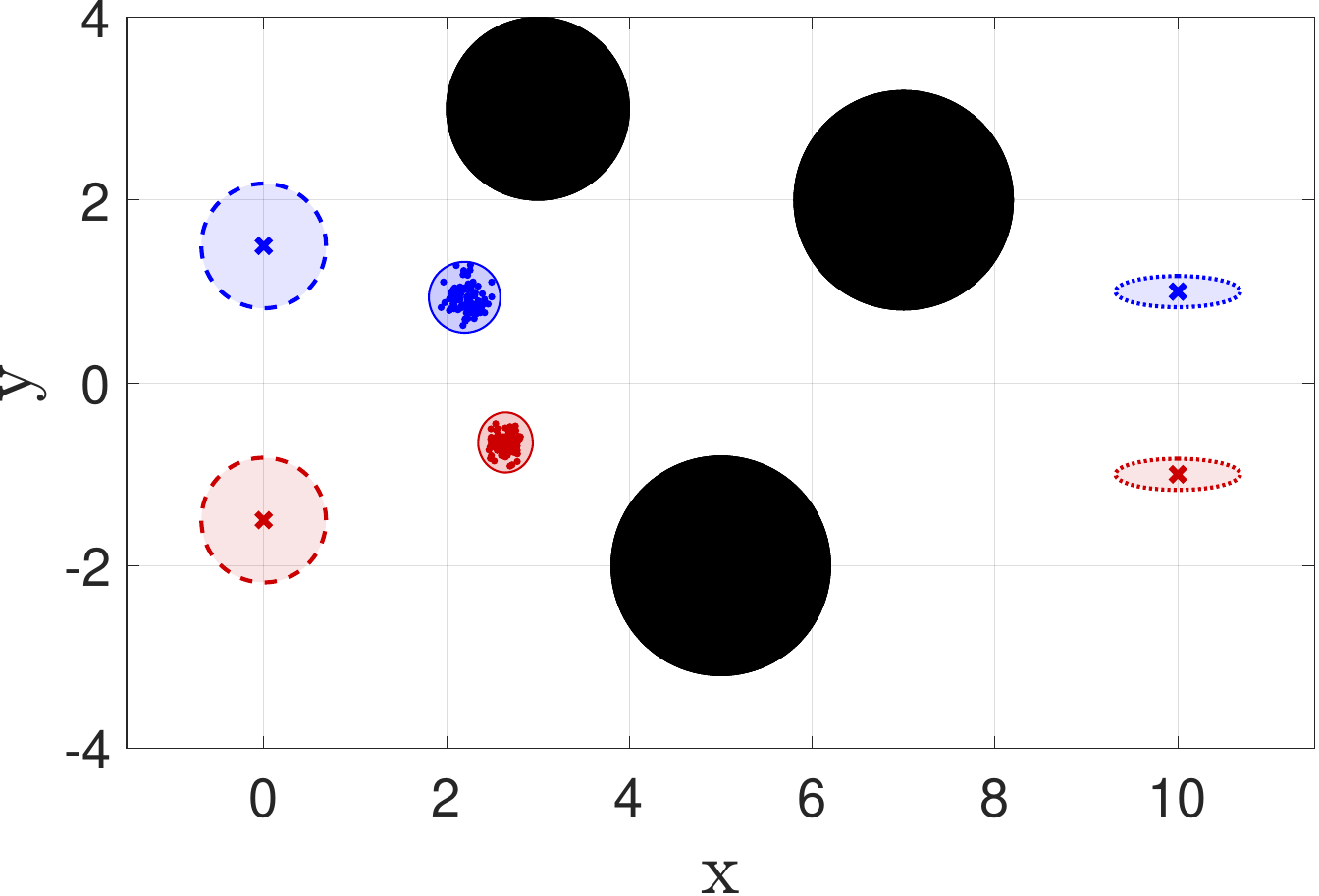}};
    \node[align=center, fill = myCyan, scale = 0.7] (c) at (3.8, 0.69) {$k=10$};
\end{tikzpicture}
}
\hfil
\subfloat{
\begin{tikzpicture}
    \node[anchor=south west,inner sep=0] at (0,0){
    \includegraphics[width=0.236\textwidth, trim={0cm 0cm 0cm 0cm},clip]{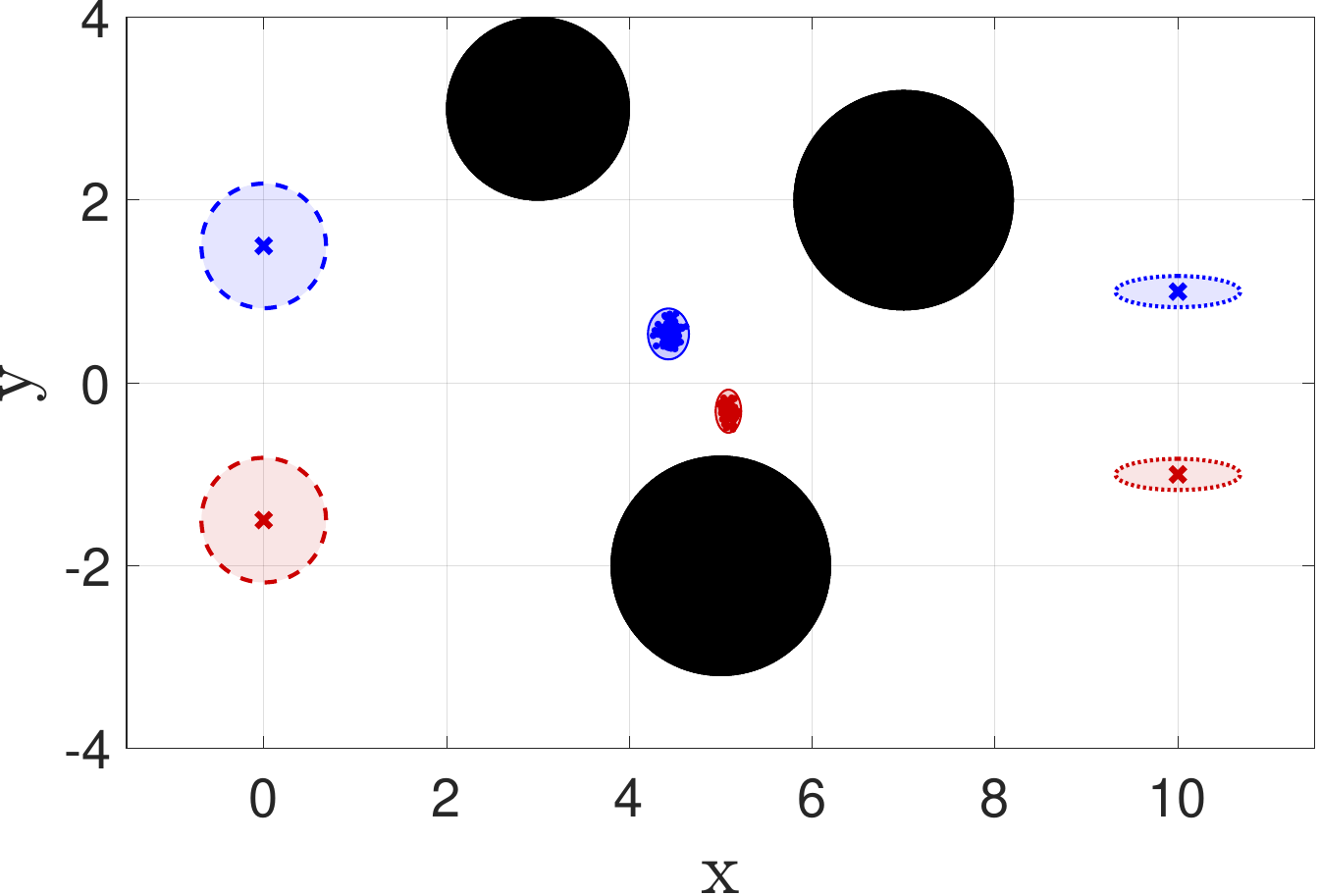}};
    \node[align=center, fill = myCyan, scale = 0.7] (c) at (3.8, 0.69) {$k=15$};
\end{tikzpicture}
}
\hfil
\subfloat{
\begin{tikzpicture}
    \node[anchor=south west,inner sep=0] at (0,0){
    \includegraphics[width=0.236\textwidth, trim={0cm 0cm 0cm 0cm},clip]{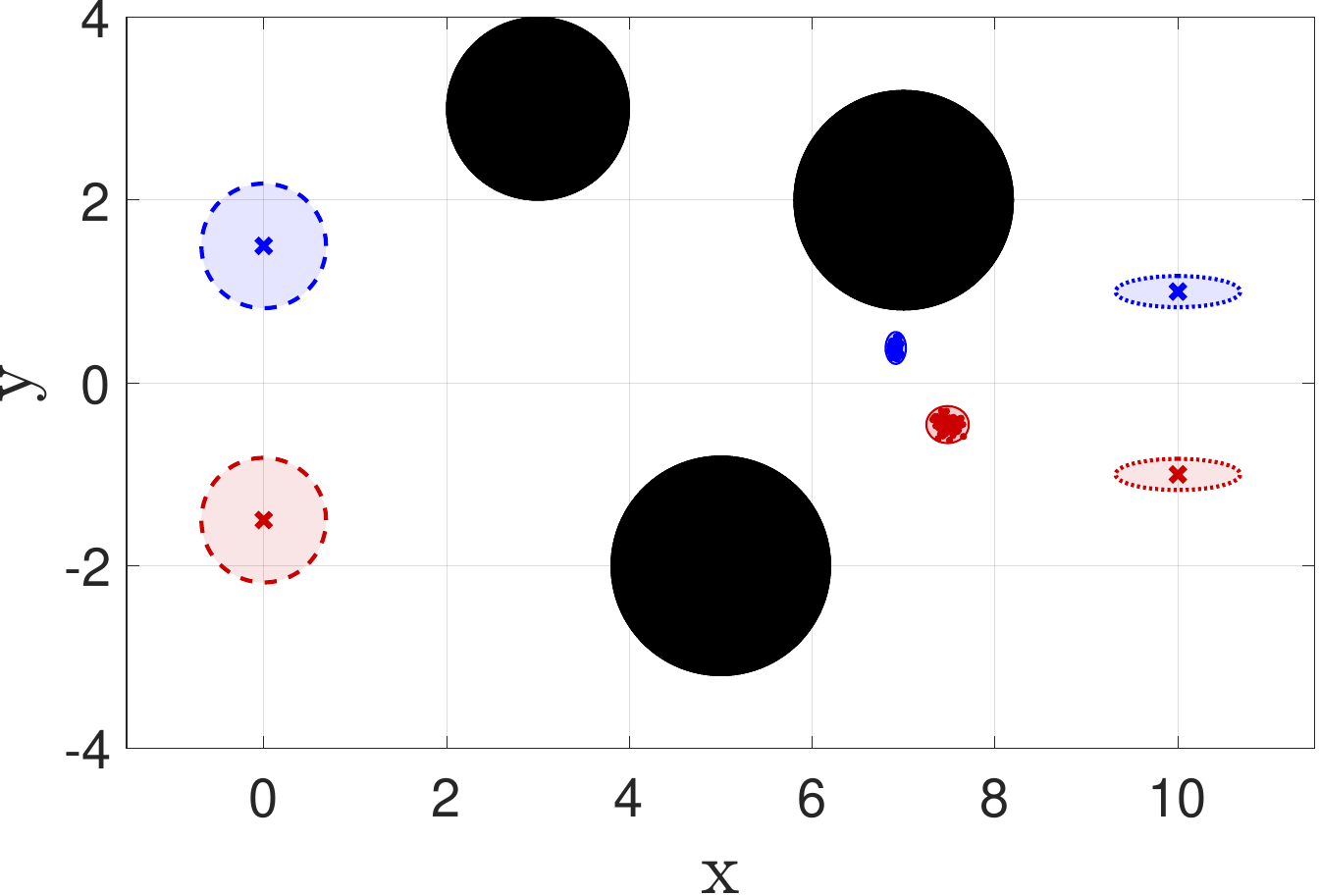}};
    \node[align=center, fill = myCyan, scale = 0.7] (c) at (3.8, 0.69) {$k=20$};
\end{tikzpicture}
}
\hfil
\\[0.1cm]
\hfil
\subfloat{
\begin{tikzpicture}
    \node[anchor=south west,inner sep=0] at (0,0){    \includegraphics[width=0.236\textwidth, trim={0cm 0cm 0cm 0cm},clip]{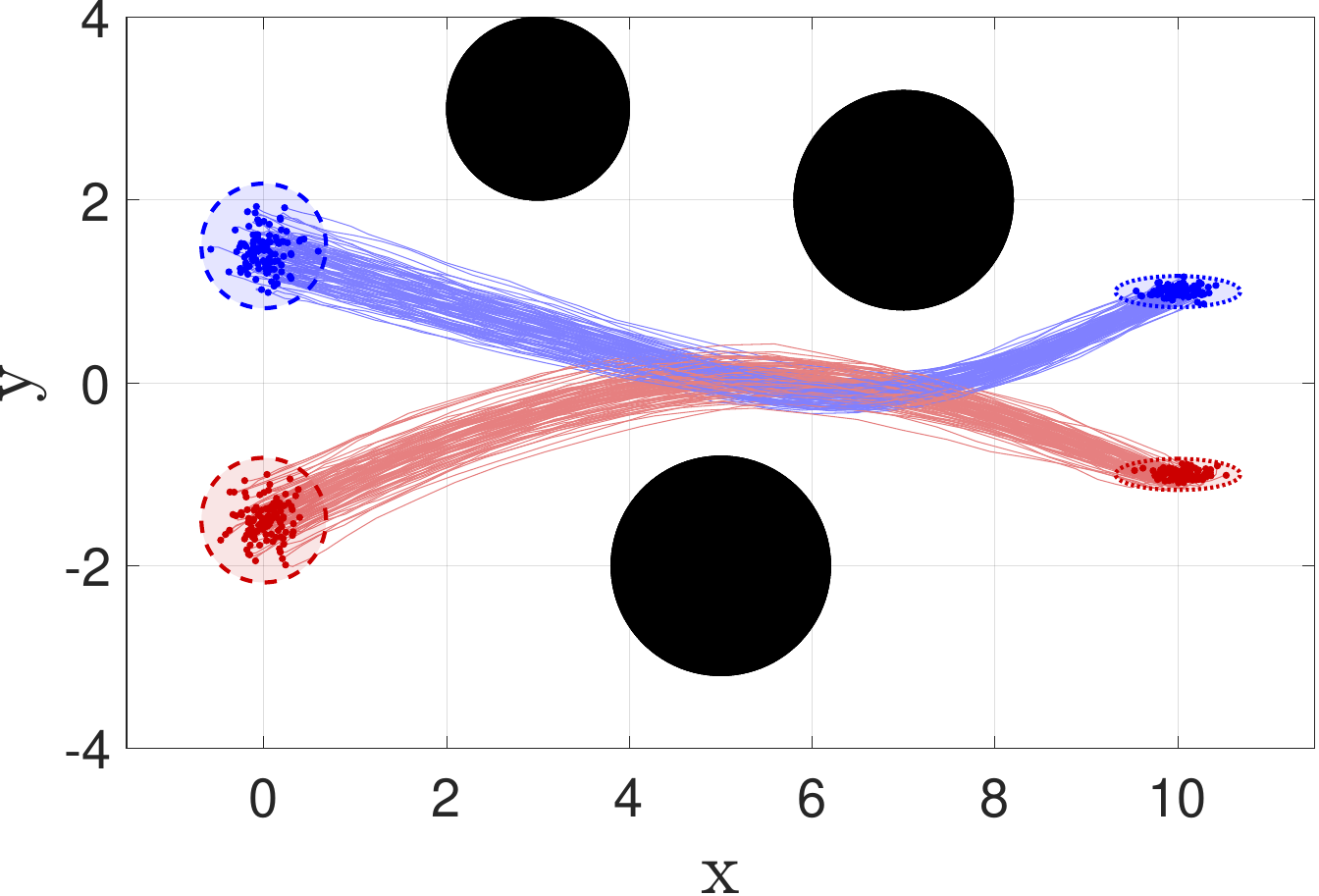}};
    \node[align=center, fill = myYellow, scale = 0.8] (c) at (1.2, 0.75) {\textbf{MC-DCS}};
\end{tikzpicture}
}
\hfil
\subfloat{
\begin{tikzpicture}
    \node[anchor=south west,inner sep=0] at (0,0){
    \includegraphics[width=0.236\textwidth, trim={0cm 0cm 0cm 0cm},clip]{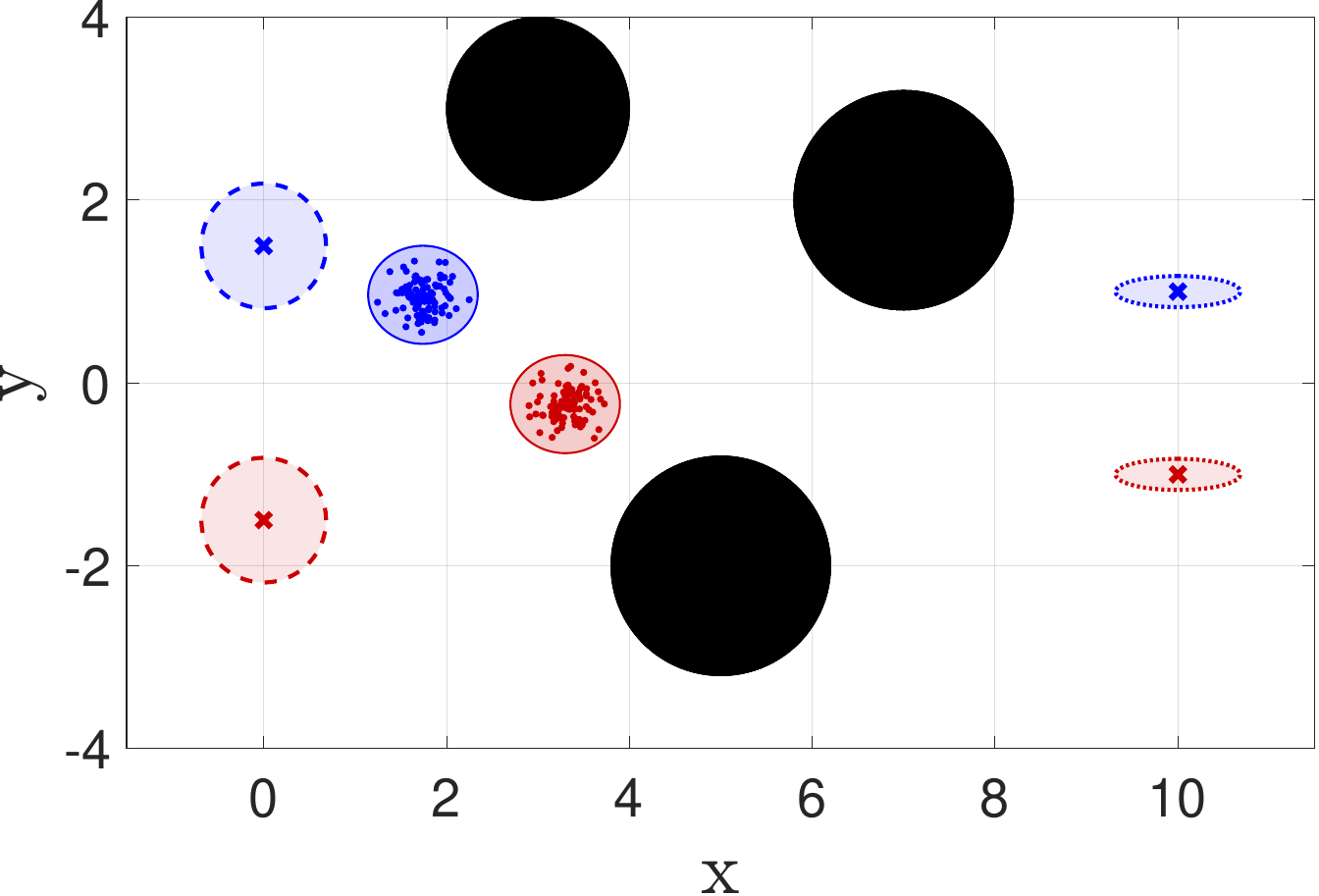}};
    \node[align=center, fill = myCyan, scale = 0.7] (c) at (3.8, 0.69) {$k=10$};
\end{tikzpicture}
}
\hfil
\subfloat{
\begin{tikzpicture}
    \node[anchor=south west,inner sep=0] at (0,0){   
    \includegraphics[width=0.236\textwidth, trim={0cm 0cm 0cm 0cm},clip]{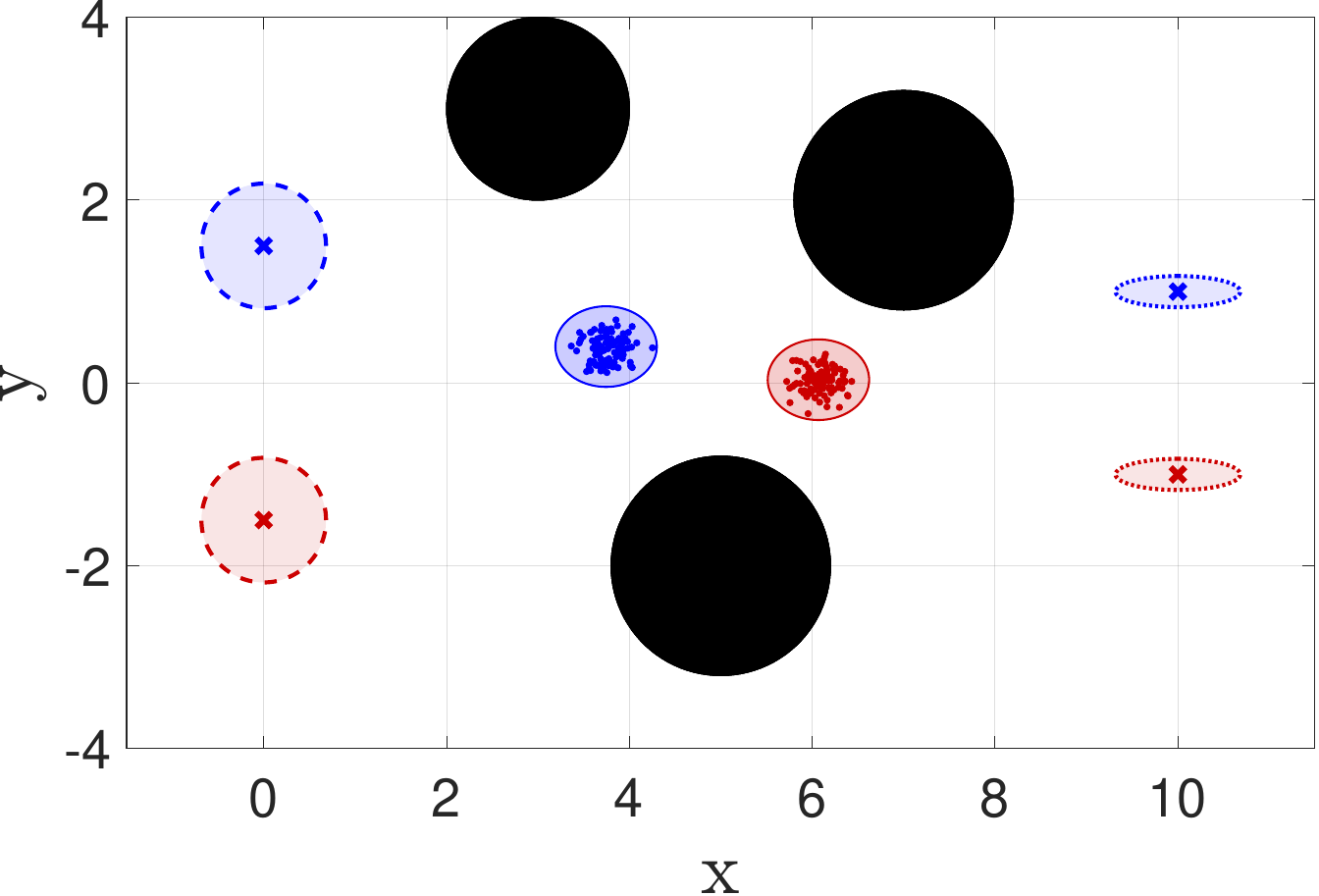}};
    \node[align=center, fill = myCyan, scale = 0.7] (c) at (3.8, 0.69) {$k=15$};
\end{tikzpicture}
}
\hfil
\subfloat{
\begin{tikzpicture}
    \node[anchor=south west,inner sep=0] at (0,0){   
    \includegraphics[width=0.236\textwidth, trim={0cm 0cm 0cm 0cm},clip]{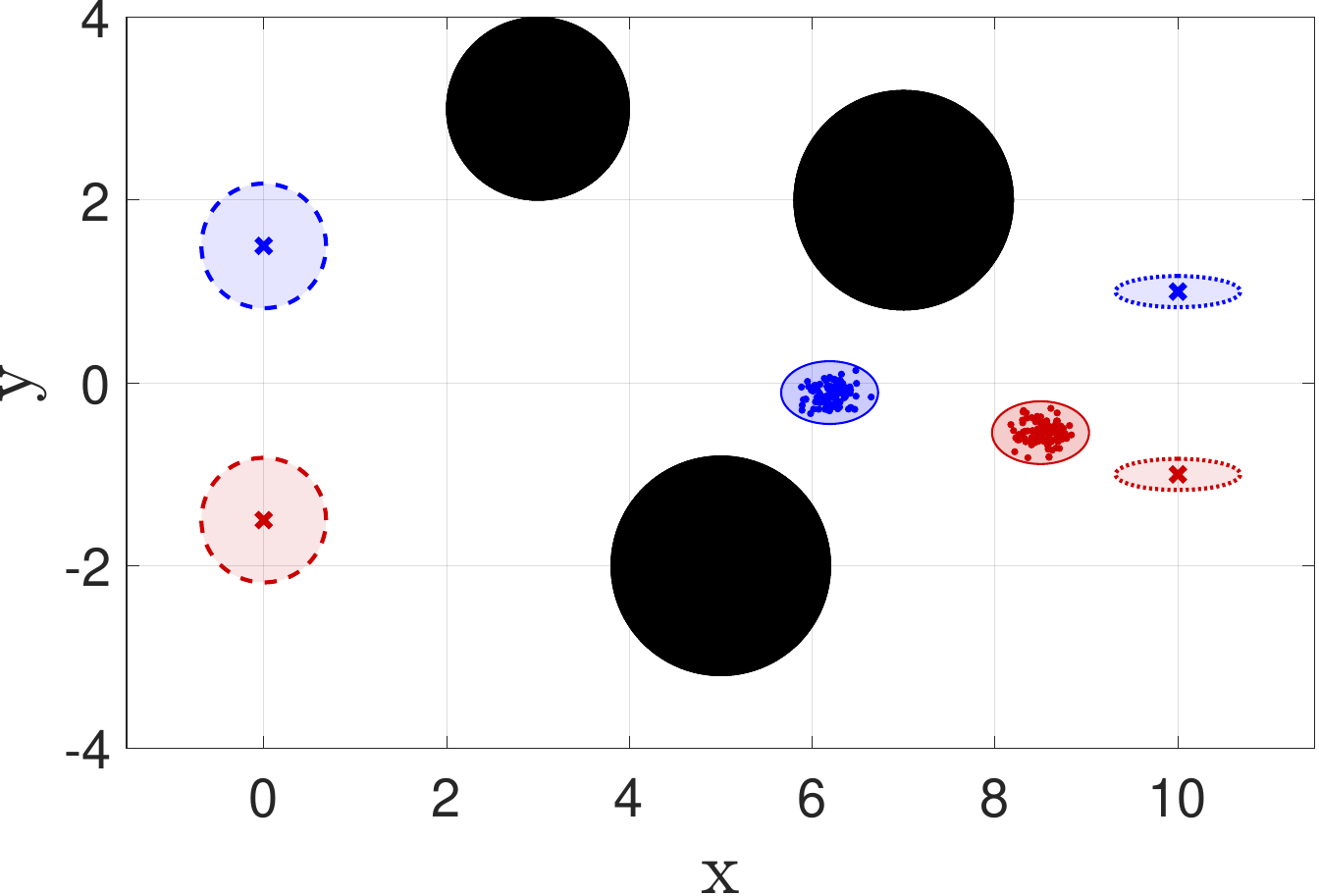}};
    \node[align=center, fill = myCyan, scale = 0.7] (c) at (3.8, 0.69) {$k=20$};
\end{tikzpicture}
}
\hfil
\vspace{-0.1cm}
\caption{\myemph{Two-agent illustrative 2D task.} The top, middle and bottom rows correspond to FCC-, PCC- and MC-DCS, respectively. The samples illustrate $100$ realizations of the distributions of the agents. The left column shows their full distribution trajectories, while the remaining subfigures on the right, show with solid ellipses the $99.7 \%$ confidence regions of their distributions at $k=10,15,20$. The dashed/dotted ellipses show their initial/target distributions. The black shapes are obstacles to be avoided.}
\label{fig: small scale}
\end{figure*}

\subsection{Convergence Guarantees}
\label{sec: convergence - main}

\refstepcounter{theorem}
\labeleditem{Theorem}{\thetheorem}{Convergence of PCC/MC-DCS}
Under Assumptions \ref{assumption: strongly convex f}-\ref{assumption: boundedness and SOOC type assumption}, the iterates $\bar{x}^\ell$, $\bar{z}^\ell$ and $\bar{y}^\ell$ converge, and any limit point of the sequence $(\bar{x}^\ell, \bar{w}^\ell, \bar{z}^\ell, \bar{y}^\ell)$ satisfies the KKT conditions \eqref{eq: KKT compact}.

\begin{proof}
Let us now consider the sum
\begin{align}
& \sum_{\ell = 0}^{\infty} c_1
\| \bar{r}^{\ell+1} \|_2^2
+ c_2
\| \bar{C} (\bar{z}^{\ell+1} - \bar{z}^\ell) \|_2^2
\leq V^0 - \lim_{\ell \rightarrow \infty} V^\ell.
\end{align}
Since the updates sequence and the stationary points of the problem lie inside a bounded set, $\lim_{\ell \rightarrow \infty} V^\ell$ must be finite. The sum of an infinite sequence of non-negative terms is finite only if that sequence converges to zero. Therefore, we obtain
\begin{equation*}
\lim_{\ell \rightarrow \infty} \| \bar{r}^{\ell+1} \| = 
\lim_{\ell \rightarrow \infty} \| \bar{C} (\bar{z}^{\ell+1} - \bar{z}^\ell) \| = 0.
\end{equation*}
Since $\bar{r}^{\ell+1}$ and $\bar{C} \bar{z}^{\ell+1}$ converge, $\bar{x}^{\ell+1} = \bar{r}^{\ell+1} + \bar{C} \bar{z}^{\ell+1}$ also converges, say to $\hat{x}$,
and $\bar{z}$ converges to $\hat{z} := (\bar{C}^\top \bar{C})^{-1} \bar{C}^\top \hat{x}$. In addition, since $\bar{y}^{\ell+1} = \bar{y}^\ell + \rho \bar{r}^{\ell+1}$ and $\bar{r}^{\ell+1}$ converges to $\hat{r}:=0$, then $\bar{y}^\ell$ also converges, say to $\hat{y}$.
%

Now that we have proved the convergence of $(\bar{x}^\ell, \bar{z}^\ell, \bar{y}^\ell)$, it remains to show that any limit point satisfies the KKT conditions \eqref{eq: KKT compact}.
Using $\bar{x}^{\ell+1} = \bar{x}^\ell = \hat{x}$, the condition \eqref{eq: KKT local - feasibility} gives $g(\hat{x}, \bar{w}^{\ell+1}) \leq 0$ and $h(\hat{x}) \leq 0$. In addition, $\hat{x} - \bar{C} \hat{z} = 0$, so the condition \eqref{eq: KKT compact - feasibility} is satisfied. The condition \eqref{eq: KKT local - optimality w} implies the satisfaction of \eqref{eq: KKT compact - optimality w}. Furthermore, since $\bar{C}^\top \bar{y}^{\ell+1} = 0$ for any $\ell$, then $\bar{C}^\top \hat{y} = 0$ which coincides with \eqref{eq: KKT compact - optimality z}.
In addition, at the limit, the optimality condition \eqref{eq: KKT local - optimality x} becomes
\begin{equation}
\nabla f(\hat{x}, \bar{w}^{\ell}) 
+ \hat{y} 
+ \nabla g(\hat{x}, \bar{w}^{\ell})^\top \sigma^{\ell}
+ \nabla h(\hat{x})^\top \nu^{\ell} = 0.
\label{eq: KKT local - limit point - optimality}
\end{equation}
The slackness conditions \eqref{eq: KKT local - slackness convex} and \eqref{eq: KKT local - slackness concave linearized} become 
\begin{subequations}
\begin{align}
& \sigma_j^{\ell} g_j(\hat{x}, \bar{w}^{\ell}) = 0, ~
\forall j \in \llbracket 1, \bar{p} \rrbracket, 
\label{eq: KKT local - limit point - slackness convex}
\\
& \nu_j^{\ell} h_j(\hat{x}) = 0, ~ 
\forall j \in \llbracket 1, \bar{q} \rrbracket.
\label{eq: KKT local - limit point - slackness concave linearized}
\end{align}
\label{eq: KKT local - limit point - slackness}%
\end{subequations}
Since $\sigma^{\ell} \geq 0$ and $\nu^{\ell} \geq 0$, then \eqref{eq: KKT local - limit point - optimality} and \eqref{eq: KKT local - limit point - slackness} coincide with the conditions \eqref{eq: KKT compact - optimality x}, \eqref{eq: KKT compact - slackness convex constraints} and \eqref{eq: KKT compact - slackness concave constraints}. Consequently, any limit point satisfies the KKT conditions \eqref{eq: KKT compact}, which proves that the algorithm reaches a stationary point of Problem \ref{problem: compact}.
\end{proof}

\begin{figure}[!t]
\centering
\hfil
\subfloat{
\includegraphics[width=0.235\textwidth, trim={0cm 0cm 0cm 0cm},clip]{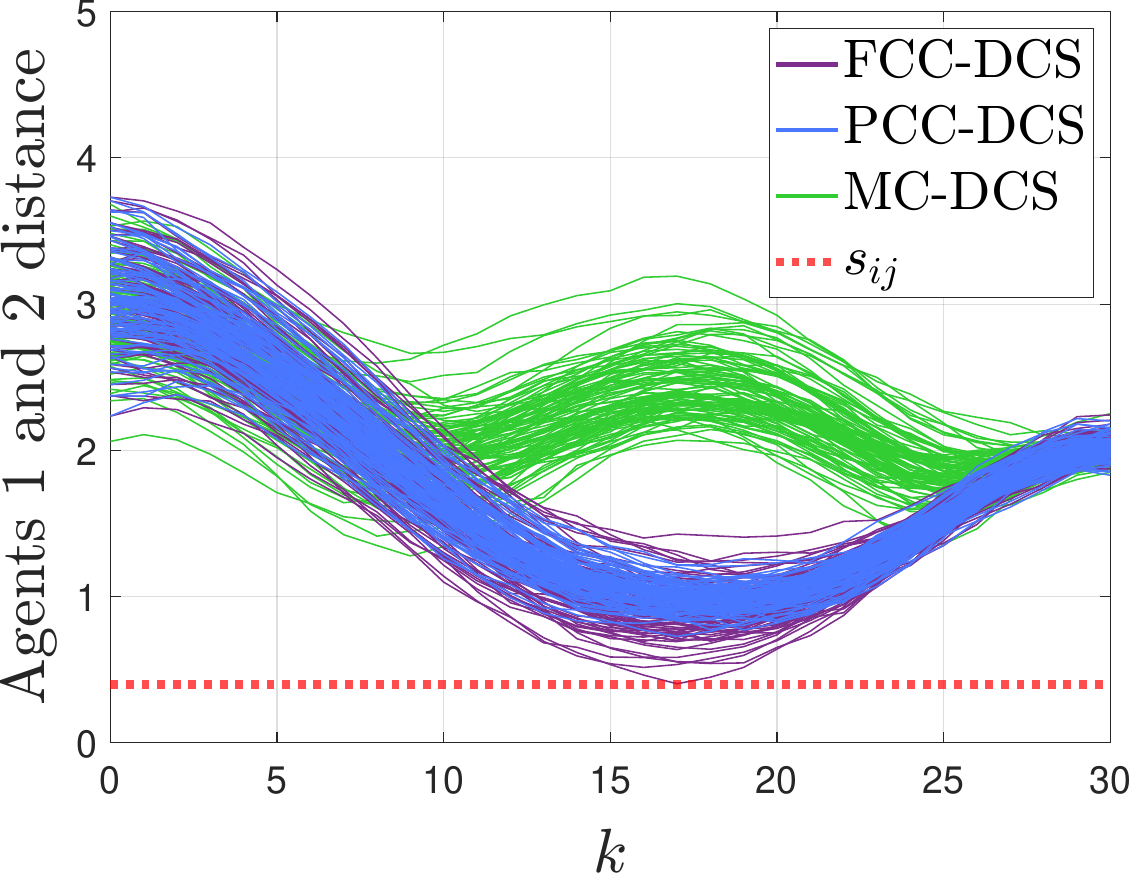}}
\hfil
\subfloat{
\includegraphics[width=0.235\textwidth, trim={0cm 0cm 0cm 0cm},clip]{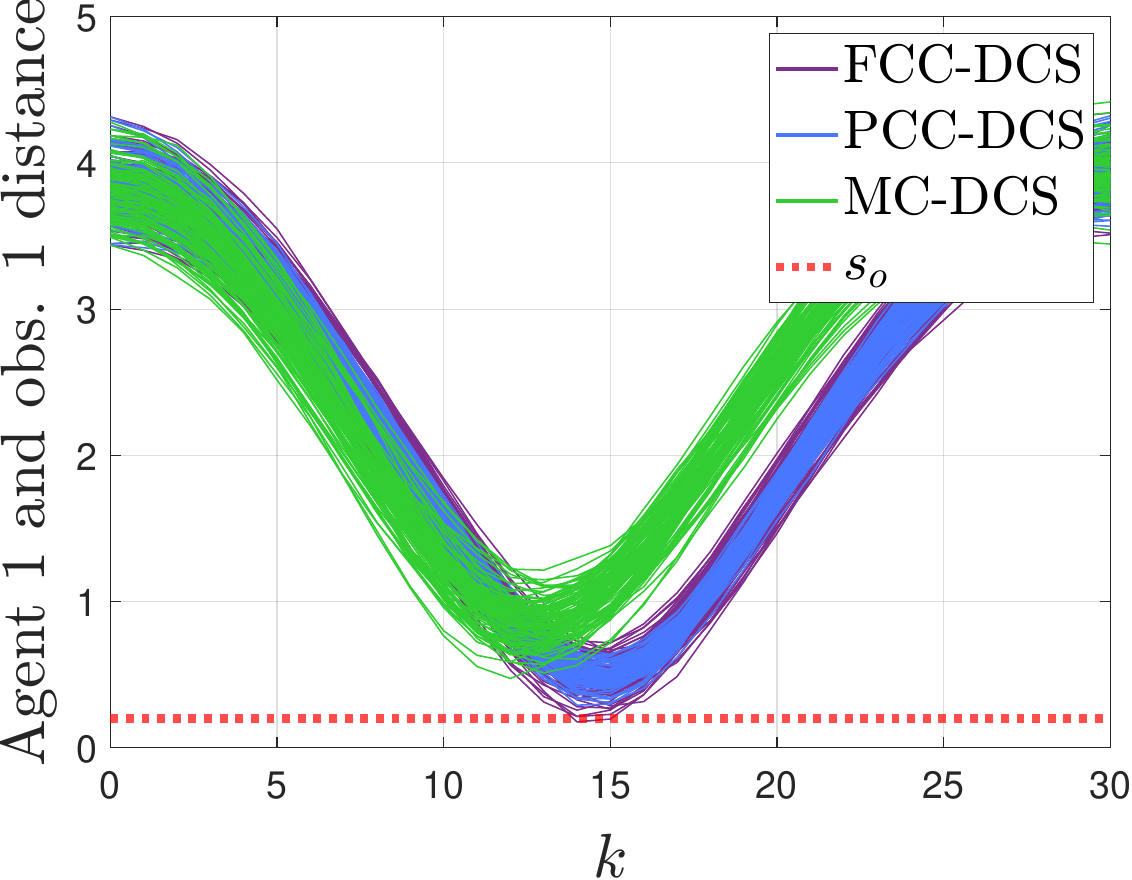}}
\hfil
\caption{\myemph{Safety distances for two-agent 2D task.} 
Left: Inter-agent distance. Right: Distance between agent 1 and obstacle 1. 
Results shown for $100$ realizations over the time horizon.}
\label{fig: small scale - distances}
\end{figure}

\begin{figure*}[!t]
\centering
\hfil
\subfloat{
\includegraphics[width=0.322\textwidth, trim={1.67cm 1cm 1.7cm 1cm},clip]{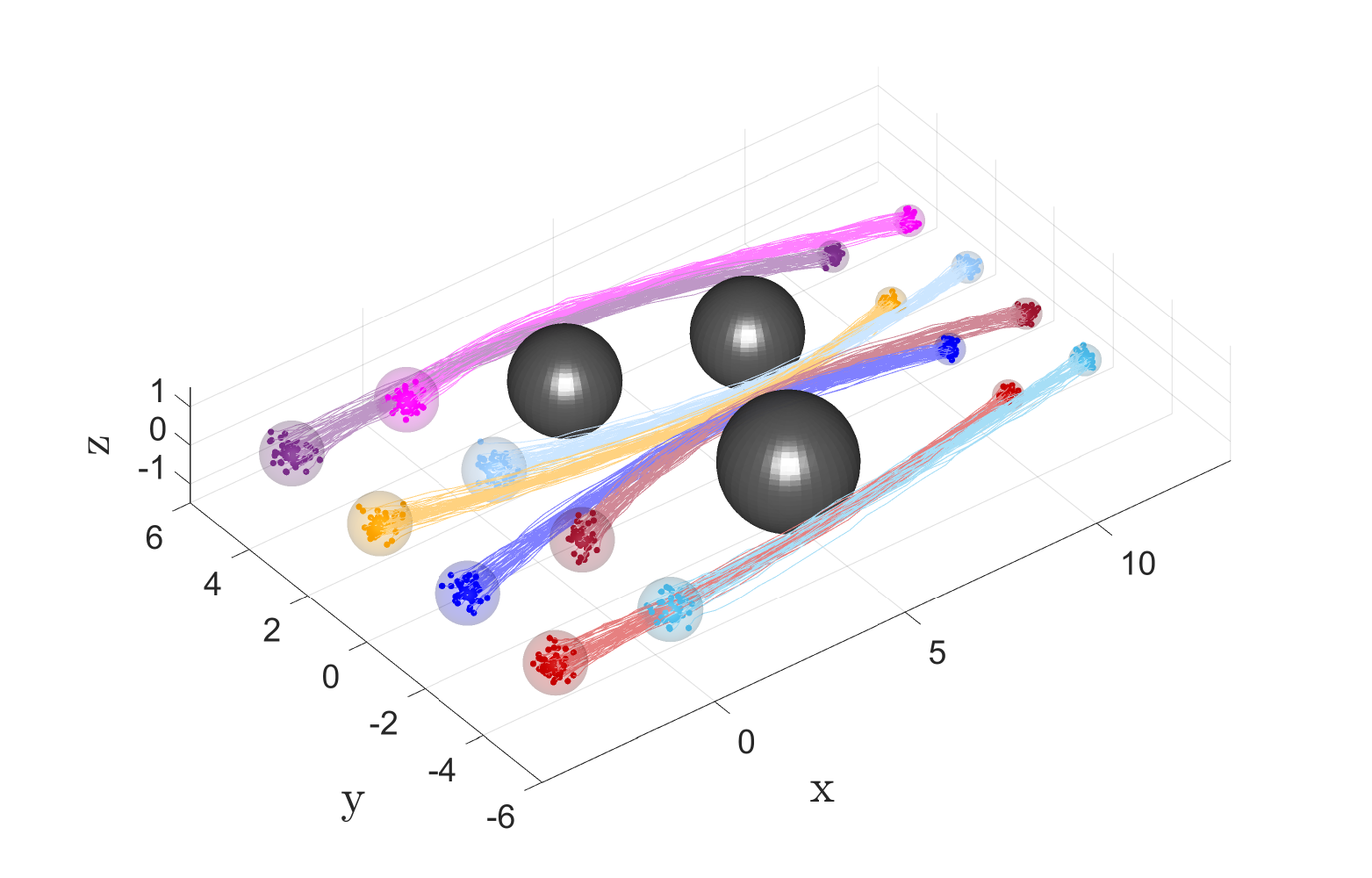}}
\hfil
\subfloat{
\begin{tikzpicture}
    \node[anchor=south west,inner sep=0] at (0,0){   
    \includegraphics[width=0.322\textwidth, trim={1.67cm 1cm 1.7cm 1cm},clip]{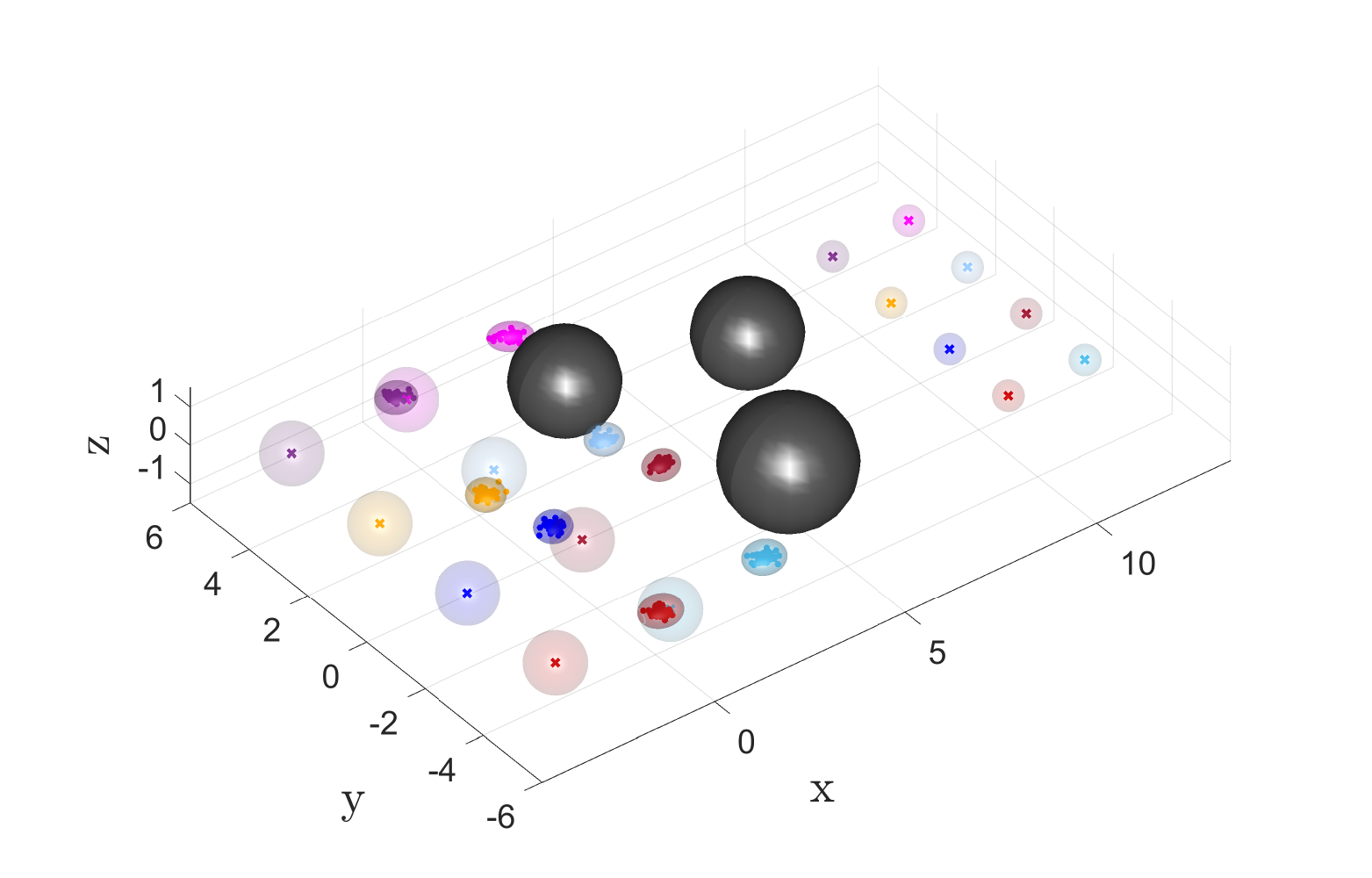}};
    \node[align=center, fill = myCyan, scale = 0.7] (c) at (5.1, 0.4) {$k=10$};
\end{tikzpicture}
}
\hfil
\subfloat{
\begin{tikzpicture}
    \node[anchor=south west,inner sep=0] at (0,0){   
    \includegraphics[width=0.322\textwidth, trim={1.67cm 1cm 1.7cm 1cm},clip]{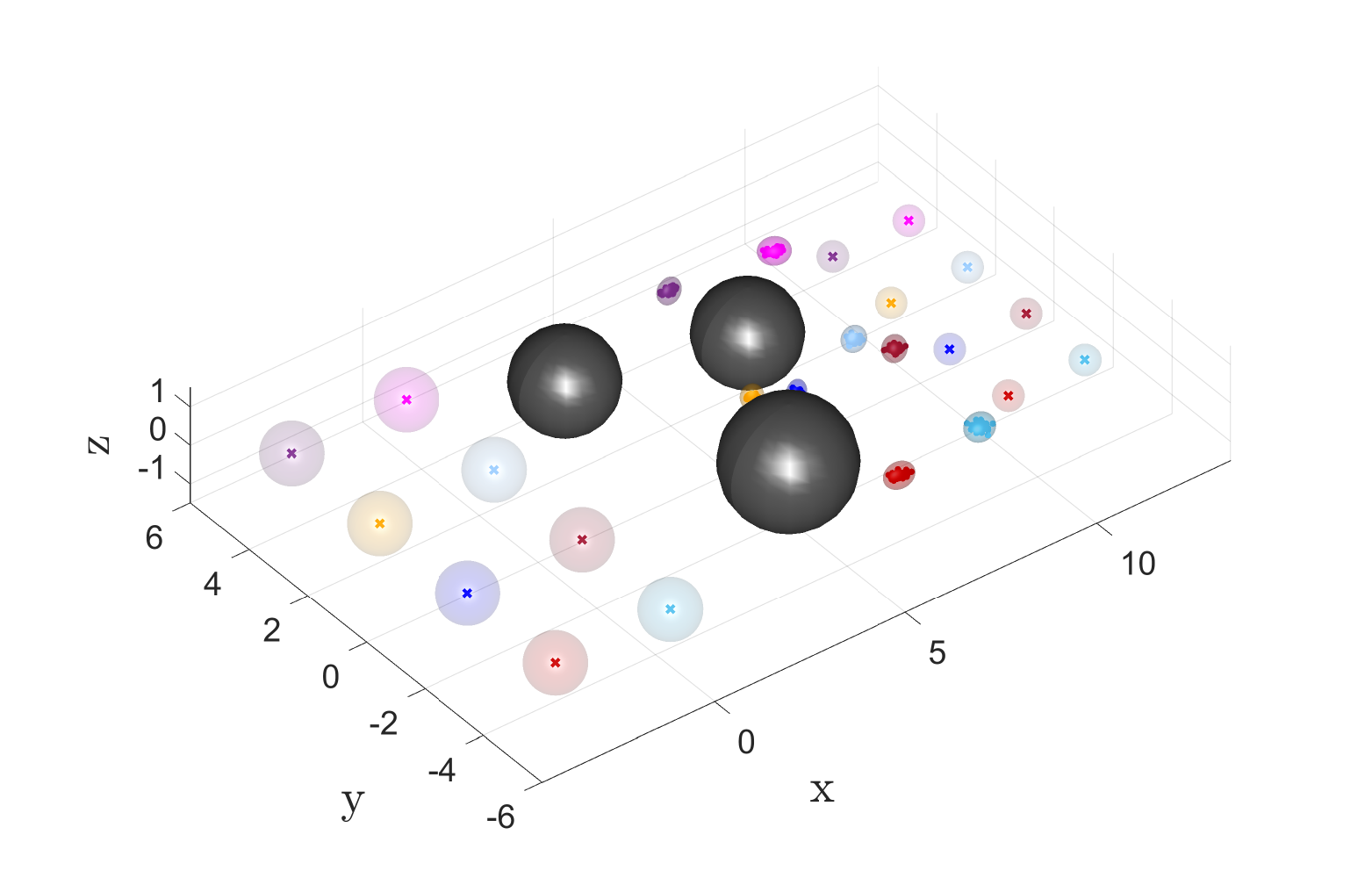}};
    \node[align=center, fill = myCyan, scale = 0.7] (c) at (5.1, 0.4) {$k=20$};
\end{tikzpicture}
}
\hfil
\caption{\myemph{Multi-drone 3D task with FCC-DCS.} The three subfigures show the $99.7 \%$ confidence ellipsoids of the initial and target distributions of the agents (faded colors) and of their current distributions (solid colors). The dark gray shapes are obstacles.}
\label{fig: drones}
\end{figure*}

\begin{figure*}[!t]
\centering
\hfil
\subfloat{
\begin{tikzpicture}
\node[anchor=south west,inner sep=0] at (0,0){   
\includegraphics[width=0.315\textwidth, trim={0cm 0cm 0cm 0cm},clip]{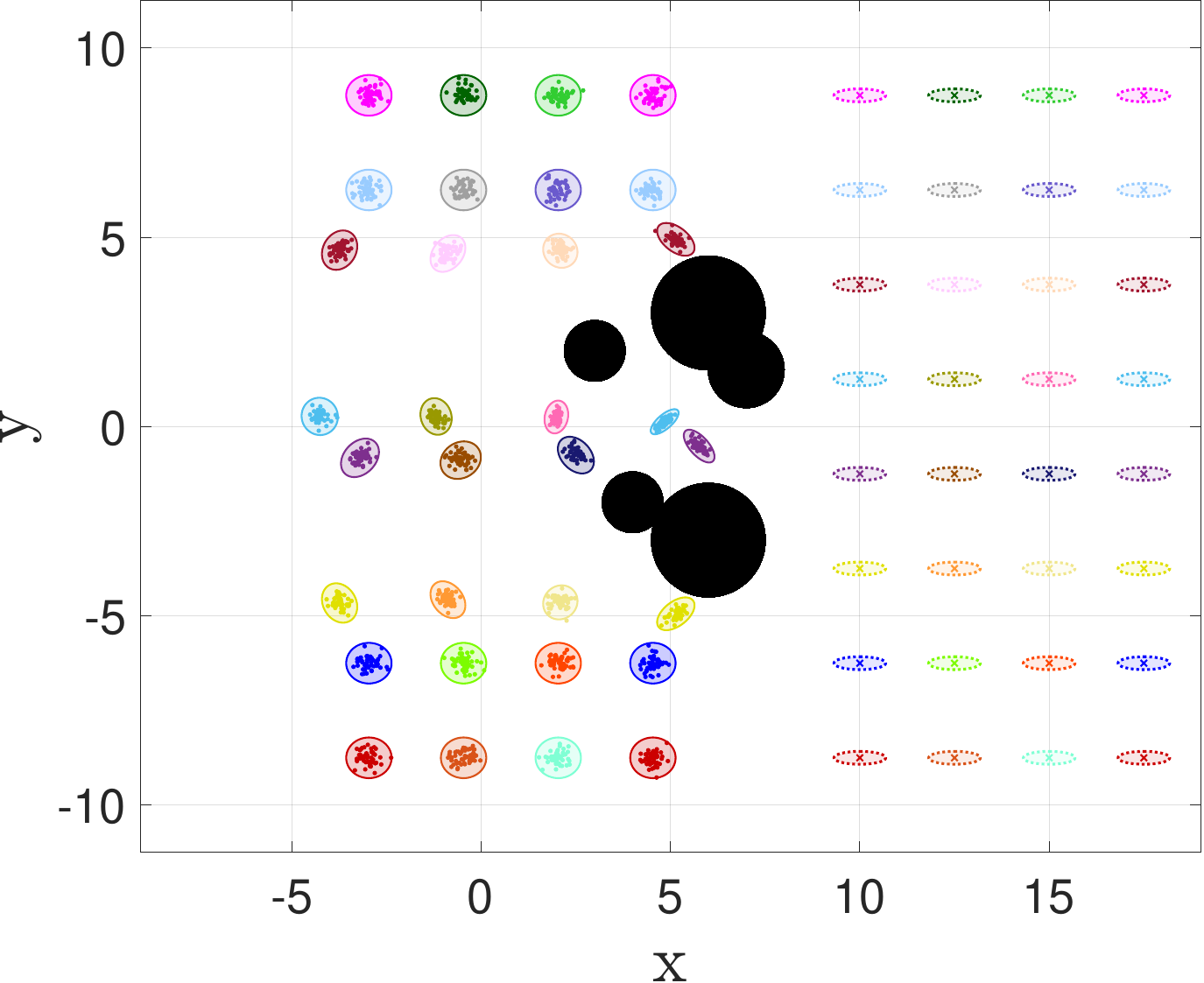}};
\node[align=center, fill = myCyan, scale = 0.7] (c) at (5.23, 0.83) {$k=10$};
\end{tikzpicture}
}
\hfil
\subfloat{
\begin{tikzpicture}
\node[anchor=south west,inner sep=0] at (0,0){   
\includegraphics[width=0.315\textwidth, trim={0cm 0cm 0cm 0cm},clip]{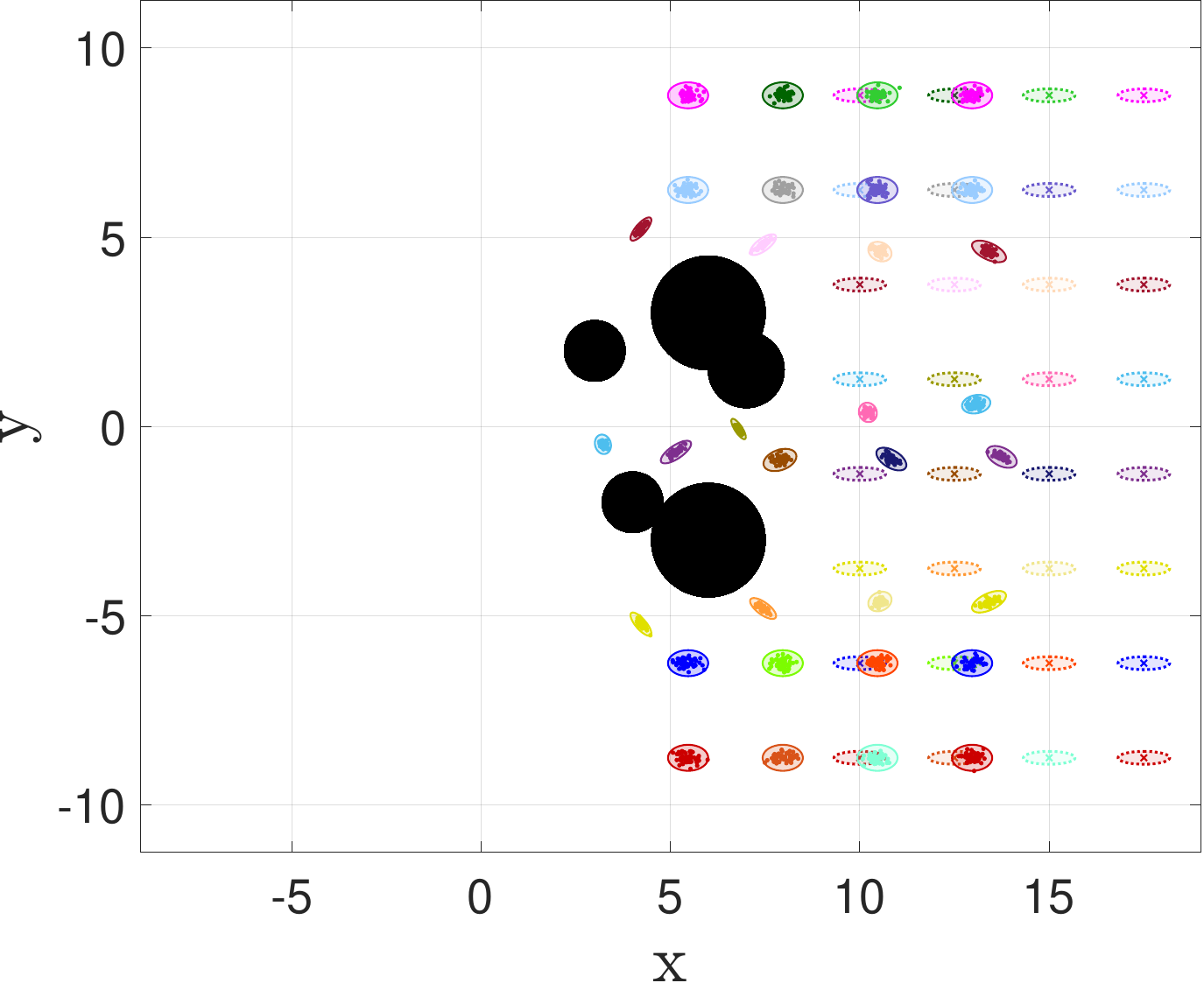}};
\node[align=center, fill = myCyan, scale = 0.7] (c) at (5.23, 0.83) {$k=20$};
\end{tikzpicture}
}
\hfil
\subfloat{
\begin{tikzpicture}
\node[anchor=south west,inner sep=0] at (0,0){   
\includegraphics[width=0.315\textwidth, trim={0cm 0cm 0cm 0cm},clip]{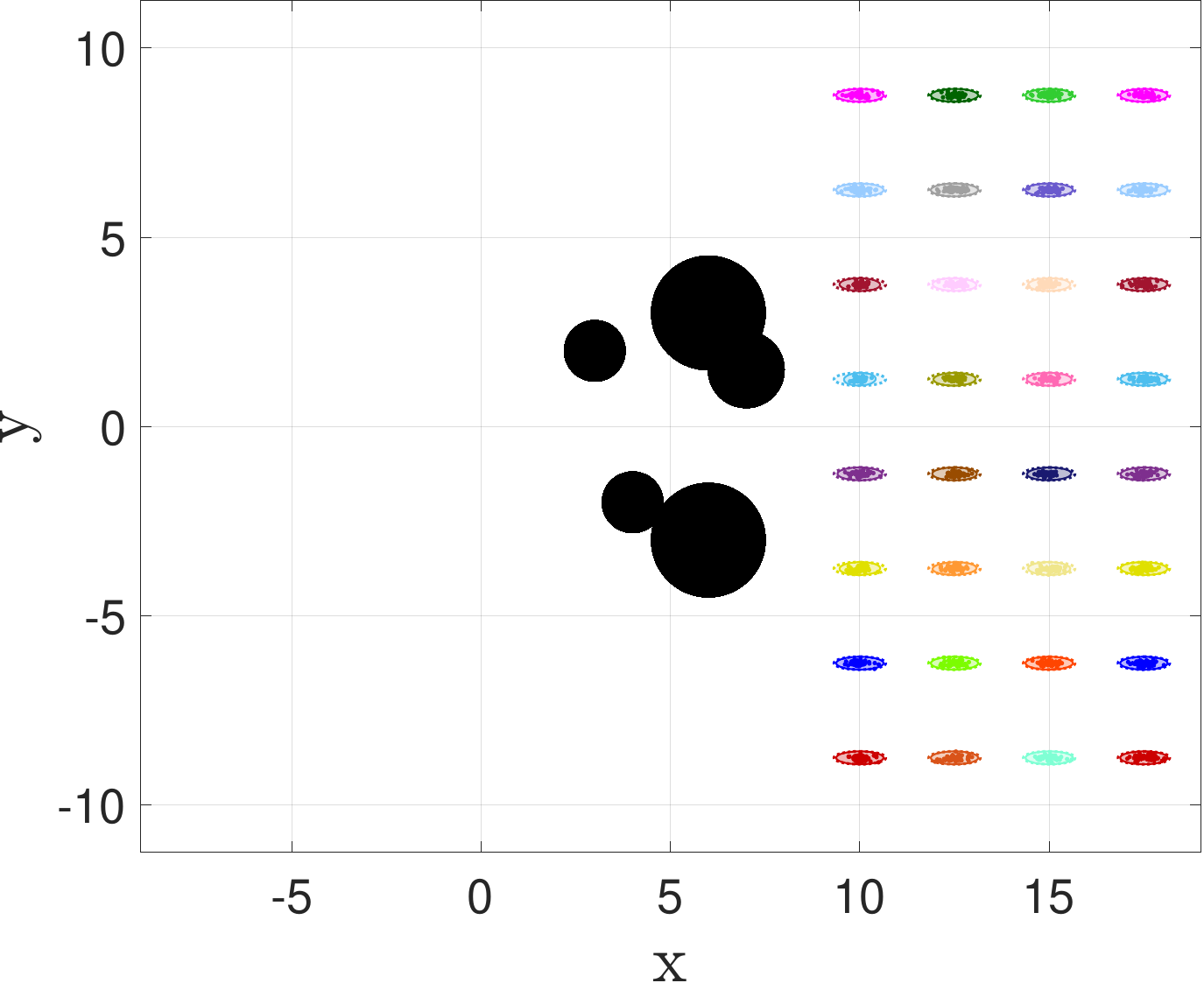}};
\node[align=center, fill = myCyan, scale = 0.7] (c) at (5.23, 0.83) {$k=30$};
\end{tikzpicture}
}
\hfil
\vspace{-0.2cm}
\caption{\myemph{Multi-agent 2D task with 32 agents via FCC-DCS.} The three subfigures correspond to time instants $k=10,20,30$.}
\label{fig: fcc dcs 16 agents}
\end{figure*}
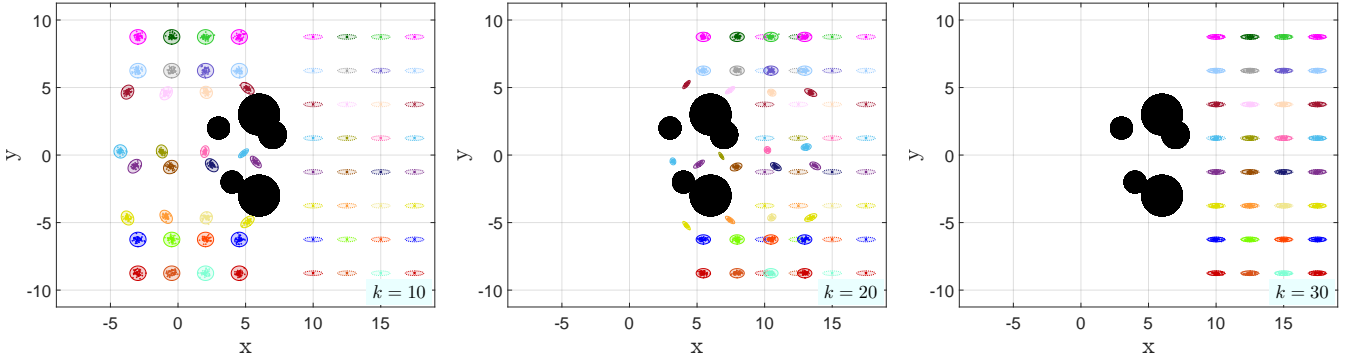


\refstepcounter{remark}
\labeleditem{Remark}{\theremark}{Novelty of Convergence Analysis}
Previous analyses for non-convex ADMM have focused either on non-convex objectives \cite{li2015global, guo2017convergence, hong2016convergence} or addressing non-convex constraints through complex schemes that reduce computational efficiency \cite{sun2023two, sun2021two, tang2022fast}.
In contrast, we present a novel analysis for distributed ADMM with iterative linearization of the non-convex constraints, guaranteeing convergence to a KKT point.

\refstepcounter{remark}
\labeleditem{Remark}{\theremark}{Discussion on the Convergence of FCC-DCS}
Studying the convergence of FCC-DCS includes an additional layer of complexity beyond non-convexity, since the original chance constraints lack a closed form and the iterative linearization takes place inside the arguments of the chance constraints (see Remark \ref{remark: successive convexification}). We provide the following statement for guaranteeing the convergence of the algorithm to the optimum of the (convex) fixed Problem \ref{problem: multi-agent cs - fcc version - consensus}. 

\refstepcounter{proposition}
\labeleditem{Proposition}{\theproposition}{Convergence of FCC-DCS}
%
The iterates of FCC-DCS converge to the optimum of Problem \ref{problem: multi-agent cs - fcc version - consensus}.

\begin{proof}
By introducing a similar notation as in Table \ref{tab: compact notation convergence}, Problem \ref{problem: multi-agent cs - fcc version - consensus} is rewritten in the form of Problem \ref{problem: compact}, where
$\bar{x}_i = [ \tilde{v}_i; \vect(\tilde{K}_i)]$, 
$\bar{z} = [z; Z]$, 
$f_i(\bar{x}_i) = \calJ_i(v_i, K_i)$, 
the convex constraints $g_i(\bar{x})_i \leq 0$ encompass 
$\cala_i(v_i) = 0$, 
$\calB_i(K_i) \succeq 0$, 
$\calc_i^{\text{FCC}}(v_i, K_i) \leq 0$,  
$\tilde{\cald}_i^{\text{FCC}}(\tilde{v}_i, \tilde{K}_i) \leq 0$,
and the non-convex ones $h_i(\bar{x}_i) \leq 0$ are empty. 
Note that Assumptions \ref{assumption: strongly convex f}, \ref{assumption: convex g} and \ref{assumption: C full column rank} are met.  
Thus, since $\bar{C}$ has full column rank, then, by standard convergence of ADMM in convex optimization \cite{deng2016global}, the iterates converge to the optimum of Problem \ref{problem: multi-agent cs - fcc version - consensus}.
\end{proof}

\section{Simulation Results}
\label{sec: simulation results}

This section presents simulations that verify the safety capabilities and scalability of the proposed methods.
Section \ref{subsec: sims - section a} illustrates a two-agent 2D example, showing the main differences between the algorithms. Section \ref{subsec: sims - section b} presents a more complex 3D multi-drone scenario. Section \ref{subsec: sims - section c} demonstrates the scalability of the methods to large-scale systems. All simulations were performed in Matlab with an Intel(R) Core(TM) i9-13900K and 64GB RAM. The MOSEK solver \cite{mosek} was used  for SDP and OSQP \cite{stellato2020osqp} for QP problems. For completeness, a \href{https://youtu.be/4tFcab6_NWg}{supplementary video} is provided including the full-motion animations of the multi-agent trajectories.


\begin{table}
\centering
\caption{Performance metrics for two-agent 2D task (Section \ref{subsec: sims - section a}) and eight-agent 3D task (Section \ref{subsec: sims - section b}).}
\vspace{-0.1cm}
\label{table}
\setlength{\tabcolsep}{3pt}
\renewcommand{\arraystretch}{1.2} 
\begin{tabular}{|c|c|c|c|c|}
\hline
\textbf{Task} & \textbf{Metrics}
& \textbf{FCC-DCS}
& \textbf{PCC-DCS}
& \textbf{MC-DCS}
\\
\hline
2D Task & Average Cost 
& 174.6
& 193.9
& 205.9
\\
(Sec. \ref{subsec: sims - section a}) & Safety viol. rate $(\%)$ 
& 0.02\%
& 0.00\%
& 0.00\%
\\[0.05cm]
\hline
3D Task & Average Cost 
& 1145.9
& 1408.3
& 1638.3
\\
(Sec. \ref{subsec: sims - section b}) & Safety viol. rate $(\%)$ 
& 0.01\%
& 0.00\%
& 0.00\%
\\[0.05cm]
\hline
\end{tabular}
\label{tab: task metrics}
\end{table}

\subsection{Two-Agent Illustrative 2D Task}
\label{subsec: sims - section a}

We consider a two-agent scenario with 2D double integrator dynamics. Each agent $i$ has a state $x_i = [p_\rx^i, p_\ry^i, v_\rx^i, v_\ry^i]$ and control $u_i = [a_\rx^i, a_\ry^i]$, where $(p_\rx^i, p_\ry^i)$, $(v_\rx^i, v_\ry^i)$ and $(a_\rx^i, a_\ry^i)$  are the 2D position, velocity and acceleration coordinates, respectively. 
The continuous-time dynamics are given by $A_{\text{cont}} = [0_{2 \times 2}, I_2; 0_{2 \times 4}]$ and $B_{\text{cont}} = [0_{2 \times 2}; I_2]$, and the discretization step and time horizon are $\Delta t = 0.05$s and $T = 30$. 
The initial mean states are 
$\mu_\rs^1 = [0; -1.5; 0; 0]$ and 
$\mu_\rs^2 = [0; 1.5; 0; 0]$, 
while the target ones are 
$\mu_\rf^1 = [10; -1; 0; 0]$ 
and $\mu_\rf^2 = [10; 1; 0; 0]$. 
The initial and target covariances are $\Sigma_\rs^i = \mathrm{bdiag}(0.04 I_2, 0.25 I_2)$ and $\Sigma_\rf^i = \mathrm{bdiag}(0.04, 0.0025, 0.25 I_2)$, while the noise covariance is $W_k^i = \mathrm{bdiag}(0.02 I_2, 0.2 I_2)$ $\forall k \in \llbracket 0,T-1 \rrbracket$, for all agents. 
We choose to penalize only the control effort, so we set $R_i = 0.01  I_2$ and $Q_i = 0_{4 \times 4}$ for each agent. For the safety constraints, we set $s_o = 0.2$, $s_{ij} = 0.4$ and $\epsilon = 3 \cdot 10^{-3}$. For MC-DCS, we choose $\hat{r}_i = 0.65$ for all agents. The penalty parameters are selected as $\rho_v = \rho_K = 1$ and $\rho_r = 10$, and each algorithm is ran for $30$ ADMM rounds.

Figure \ref{fig: small scale} illustrates the $99.7\%$ confidence regions of the distribution trajectories of the agents, along with $100$ sampled realizations for each algorithm. For the same sampled trajectories, Fig. \ref{fig: small scale - distances} shows the inter-agent distance and the distance between agent 1 and obstacle 1 during the entire time horizon. Table \ref{tab: task metrics} provides the average cost and safety violation rate for all methods.
All three methods successfully steer the two agents to their target distributions, while avoiding collisions with each other and the obstacles. Further, they all achieve a safety violation rate below the prescribed threshold $\epsilon = 0.3\%$.
The FCC-DCS method yields the most cost-efficient and least conservative trajectories, as illustrated in Figs. \ref{fig: small scale} and \ref{fig: small scale - distances}, as well as in Table \ref{tab: task metrics}, followed by PCC-DCS, and then MC-DCS.

\begin{figure*}[!t]
\centering
\hfil
\subfloat{
\begin{tikzpicture}
\node[anchor=south west,inner sep=0] at (0,0){   
\includegraphics[width=0.315\textwidth, trim={2.5cm 4.5cm 2.0cm 4cm},clip]{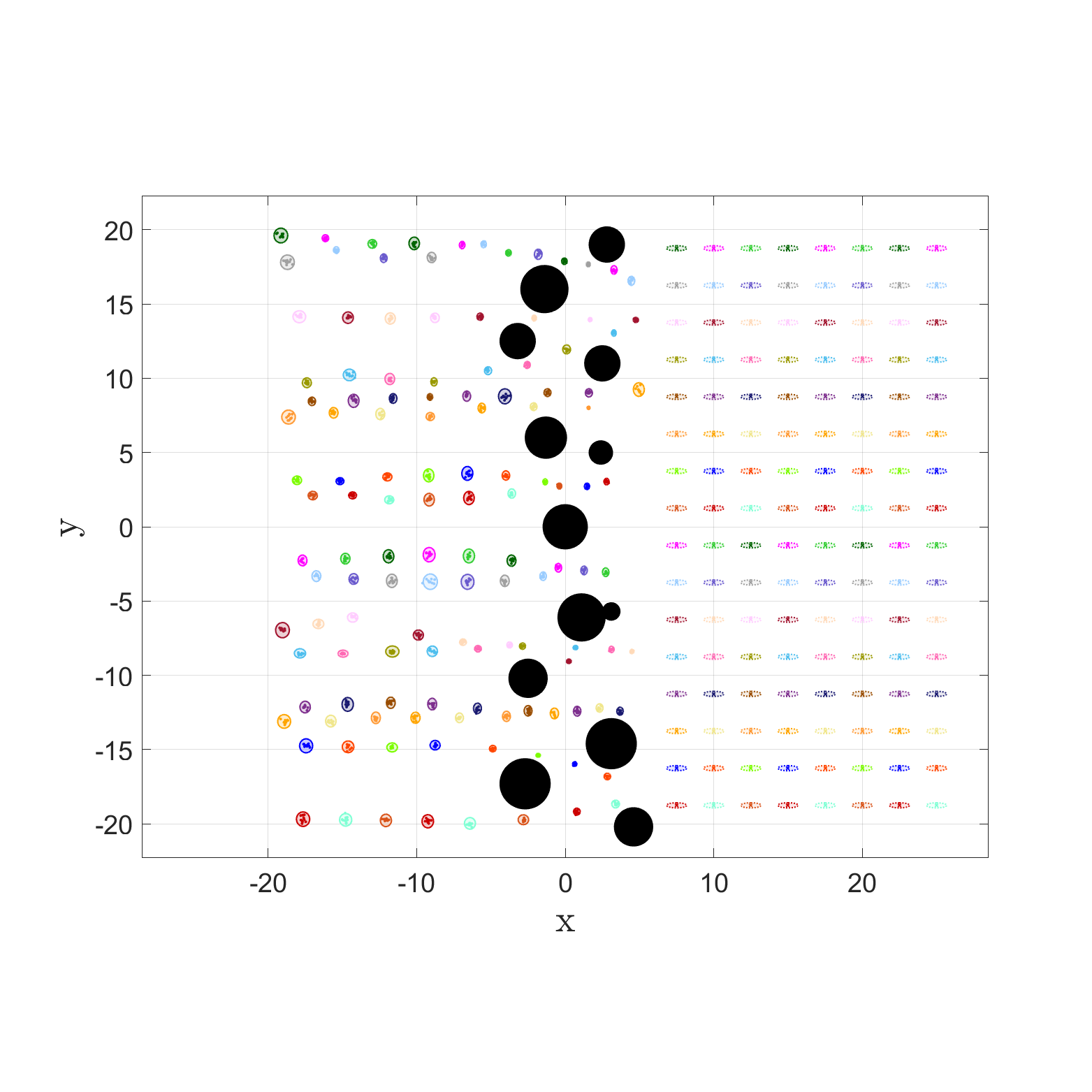}};
\node[align=center, fill = myCyan, scale = 0.7] (c) at (0.73, 0.52) {$k=10$};
\end{tikzpicture}
}
\hfil
\subfloat{
\begin{tikzpicture}
\node[anchor=south west,inner sep=0] at (0,0){   
\includegraphics[width=0.315\textwidth, trim={2.5cm 4.5cm 2.0cm 4cm},clip]{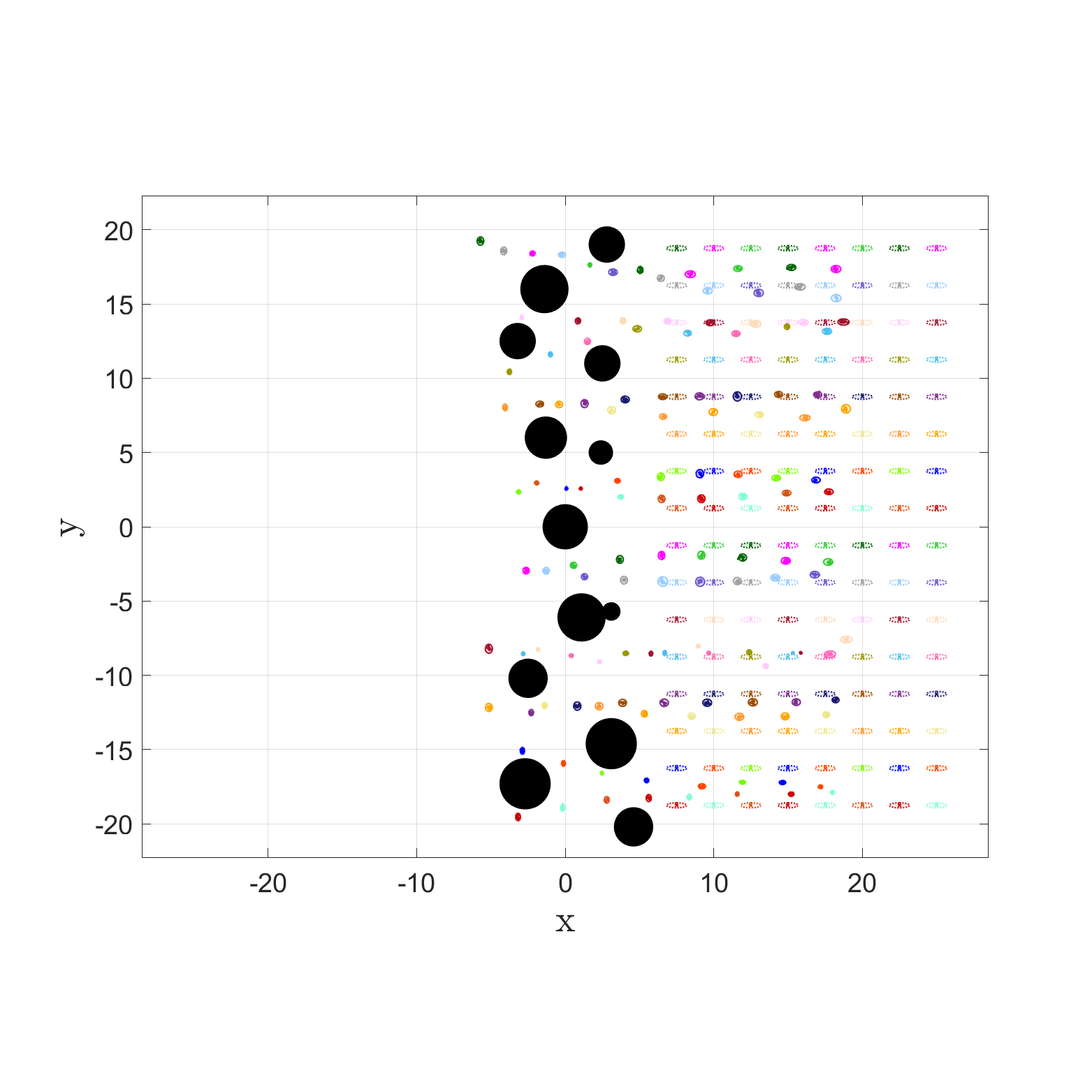}};
\node[align=center, fill = myCyan, scale = 0.7] (c) at (0.73, 0.52) {$k=20$};
\end{tikzpicture}
}
\hfil
\subfloat{
\begin{tikzpicture}
\node[anchor=south west,inner sep=0] at (0,0){   
\includegraphics[width=0.315\textwidth, trim={2.5cm 4.5cm 2.0cm 4cm},clip]{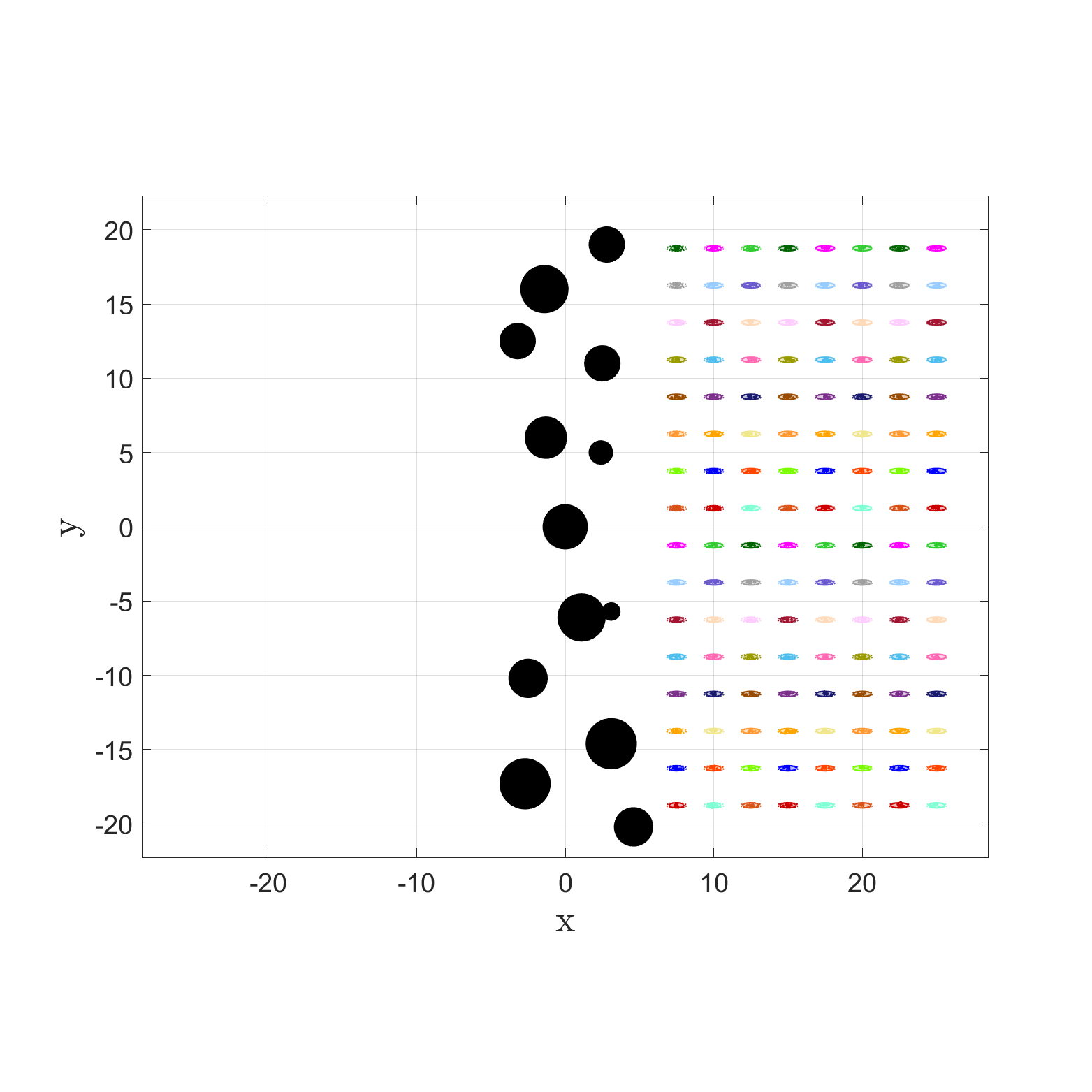}};
\node[align=center, fill = myCyan, scale = 0.7] (c) at (0.73, 0.52) {$k=30$};
\end{tikzpicture}
}
\hfil
\caption{\myemph{Large-scale 2D task with 128 agents via PCC-DCS.} The two subfigures correspond to time instants $k=10,20,30$.}
\label{fig: pcc dcs 128 agents}
\end{figure*}
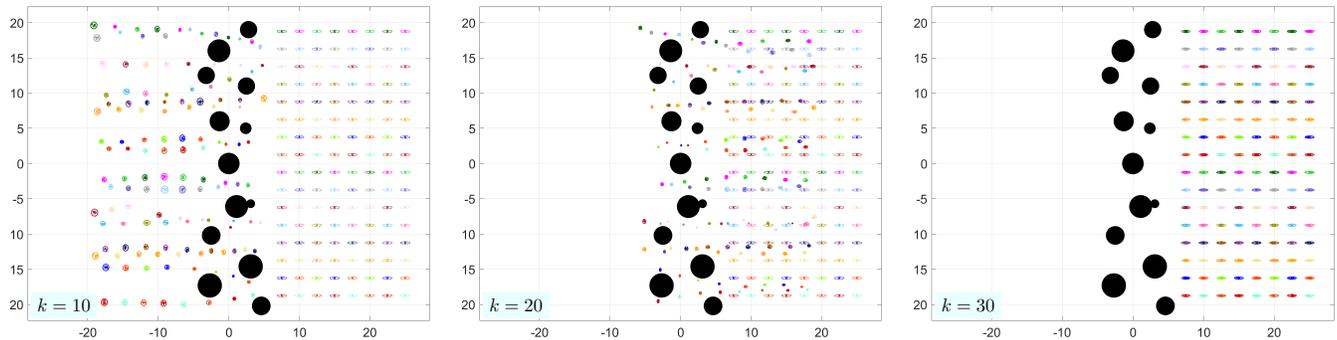

\subsection{Multi-Drone 3D Task}
\label{subsec: sims - section b}

Next, we consider a more complex 3D multi-agent task with linearized drone dynamics \cite{zhang2025gcbf+}. Each agent has a state $x_i = [p_\rx^i, p_\ry^i, p_\rz^i, v_\rx^i, v_\ry^i, v_\rz^i]$, where $(p_\rx^i, p_\ry^i, p_\rz^i)$ and $(v_\rx^i, v_\ry^i, v_\rz^i)$ correspond to the 3D position and velocity coordinates, and a control $u_i = [a_\rx^i, a_\ry^i, a_\rz^i]$, including the 3D acceleration coordinates. The continuous-time dynamics are given by $A_{\text{cont}} = [0_{3 \times 3}, I_3; 0_{3 \times 3}, \diag(-1.1, -1.1, -6)]$ and $B_{\text{cont}} = [0_{3 \times 3}; \diag(1.1, 1.1, 6)]$. The initial and target covariances are $\Sigma_\rs^i = \mathrm{bdiag}(0.04 I_3, 0.25 I_3)$ and $\Sigma_\rf^i = \mathrm{bdiag}(0.04 I_3, 0.25 I_3)$, while the noise covariance is $W_k^i = \mathrm{bdiag}(0.02 I_3, 0.2 I_3)$ for all agents. The rest of the parameters are set as in Section \ref{subsec: sims - section a}. Figure \ref{fig: drones} demonstrates all drones guided safely to their target distributions with the FCC-DCS method. Table \ref{tab: task metrics} presents again the average cost and safety violation rate for each method, verifying that FCC-DCS provides the least conservative solution followed by PCC-DCS and MC-DCS.

\subsection{Scalability on Large-Scale Multi-Agent Systems}
\label{subsec: sims - section c}

\begin{figure}[!t]
\centering
\includegraphics[width=0.49\textwidth, trim={1.5cm 7cm 2cm 7cm},clip]{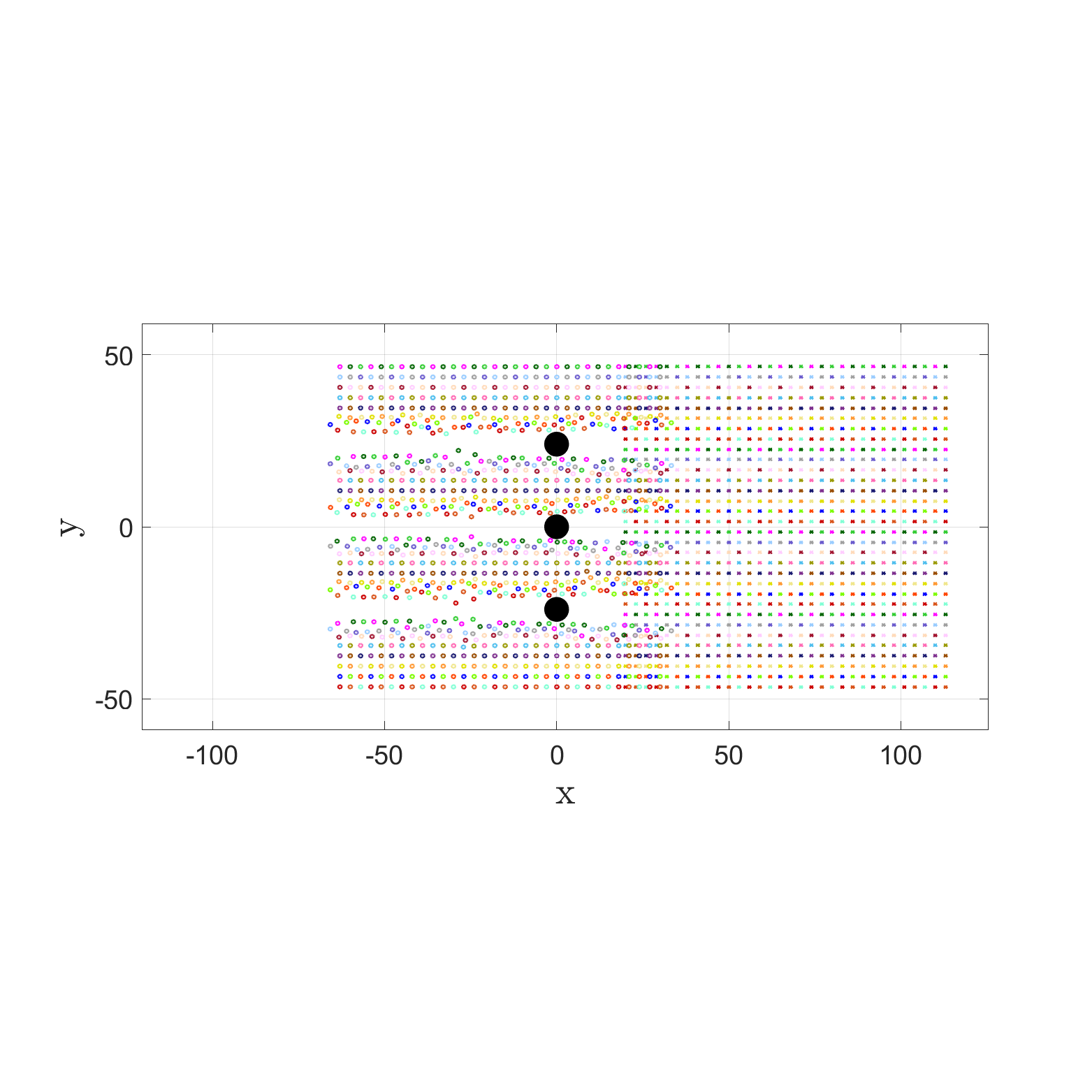}
\caption{\myemph{Large-scale 2D task with 1024 agents via MC-DCS.} Snapshot at time instant $k=20$.}
\label{fig: mc dcs 1024 agents}
\end{figure}

\begin{figure}[!t]
\centering
\includegraphics[width=0.49\textwidth, trim={0cm 0cm 0cm 0cm},clip]{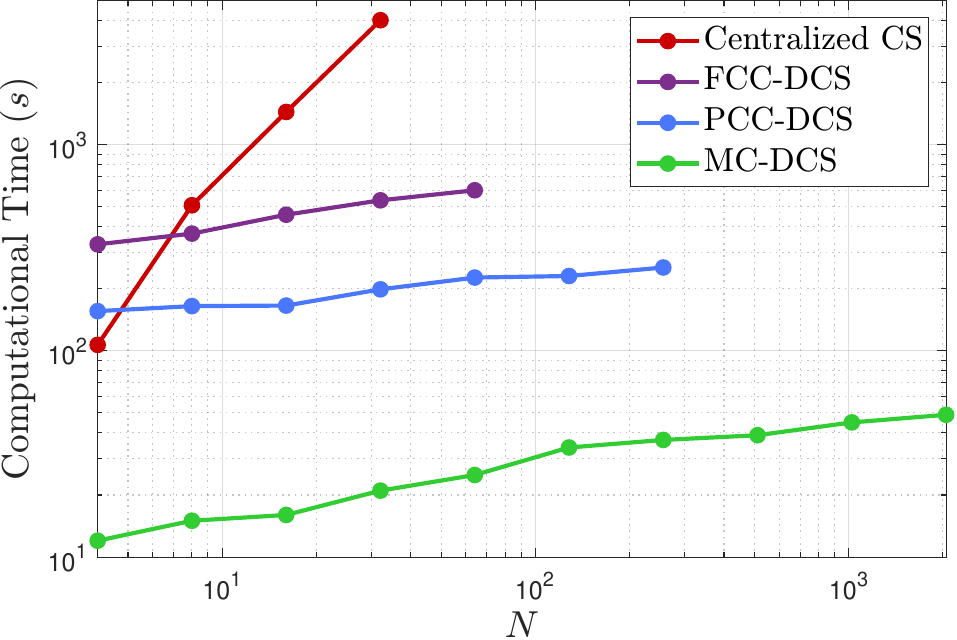}
\caption{\myemph{Scalability comparison.} Computational times of  Centralized CS and FCC-, PCC- and MC-DCS for an increasing number of agents $N$.}
\label{fig: times}
\end{figure}

Finally, we illustrate the scalability of the proposed methods to large-scale multi-agent systems. Figure \ref{fig: fcc dcs 16 agents} shows a system of $32$ agents safely steered to their target distributions with FCC-DCS. In Fig. \ref{fig: pcc dcs 128 agents}, PCC-DCS is demonstrated on a team of $128$ agents, while Fig. \ref{fig: mc dcs 1024 agents} shows MC-DCS with $1024$ agents.
Figure \ref{fig: times} shows the total computational times of each method for an increasing number of agents $N$. The MC-DCS algorithm preserves the best scalability, followed by PCC-DCS and FCC-DCS. Solving these high-dimensional problems with centralized CS becomes intractable beyond $8$ agents.


\section{Conclusion}
\label{sec: conclusion}

This article introduces a family of distributed methods for the multi-agent covariance steering problem. The proposed ADMM-based algorithms enforce different levels of consensus, i.e., full covariance, partial covariance, and mean, thereby providing a trade-off between conservatism and computational burden. Furthermore, convergence is established via a novel analysis of distributed ADMM with iteratively linearized non-convex constraints. Numerical results demonstrate the safety capabilities and scalability of the methods to systems with up to thousands of agents. Future work includes extensions to nonlinear dynamics \cite{ridderhof2019nonlinear}, GMM-based distributions \cite{balci2024density}, and learning-aided distributed optimization architectures \cite{saravanos2025deep}.

\appendices

\section*{Acknowledgment}

The authors thank Arkadi Nemirovski for valuable discussions on the convergence and complexity of the methods presented in this article.


\section*{References}

\bibliographystyle{IEEEtran}

\bibliography{references}

\begin{IEEEbiography}[{\includegraphics[width=1in,height=1.25in, trim={1.5cm 0cm 1.0cm 0cm}, clip,keepaspectratio]{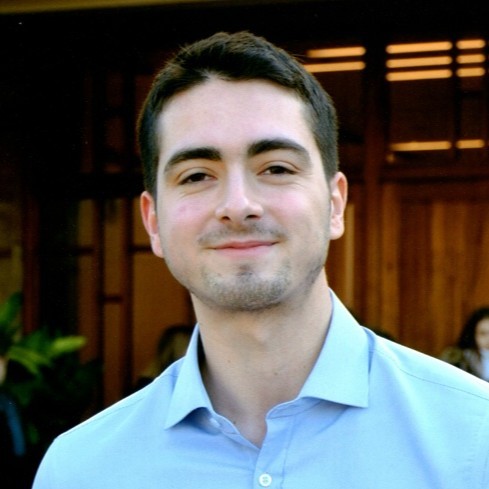}}]{Augustinos D. Saravanos} (Graduate Student Member, IEEE) received his Diploma in Electrical and Computer Engineering with highest honors from the University of Patras, Greece in 2019 and his M.S. in Aerospace Engineering and Ph.D. in Machine Learning from the Georgia Institute of Technology in 2024 and 2025. 
He is currently a Postdoctoral Associate at the Department of of Aeronautics and Astronautics at the Massachusetts Institute of Technology. His research interests lie at the intersection of optimization, control and machine learning for large-scale systems.
\end{IEEEbiography}

\begin{IEEEbiography}[{\includegraphics[width=1in,height=1.25in, trim={0cm 0.6cm 0cm 0.3cm}, clip,keepaspectratio]{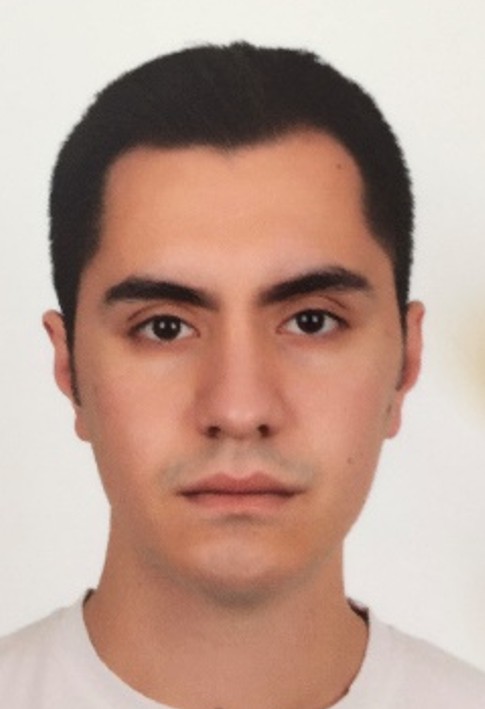}}]{Isin M. Balci} received a B.Sc. in Mechanical Engineering from the Bogazici University, Istanbul, Turkey, in 2018 and the M.S. and Ph.D. degrees in Aerospace Engineering from the University of Texas at Austin, Austin, TX, USA, in 2020 and 2024, respectively. He is currently a software engineer at Applied Intuition. His research is mainly focused on control of uncertain and stochastic systems and optimization-based control.
\end{IEEEbiography}

\begin{IEEEbiography}[{\includegraphics[width=1in,height=1.25in, trim={0cm 0cm 0cm 0cm}, clip,keepaspectratio]{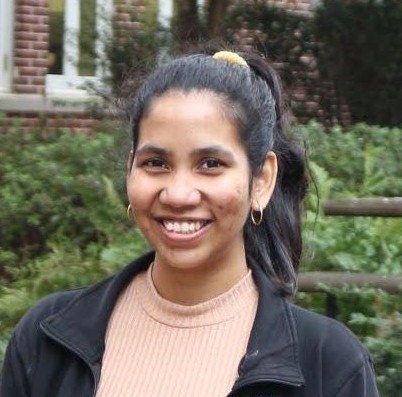}}]{Arshiya Taj Abdul} (Student Member, IEEE) received a B.Tech in Electrical and Electronics Engineering from the National Institute of Technology, Warangal, India. She is currently pursuing a Ph.D. in Electrical and Computer Engineering at Georgia Institute of Technology, Atlanta, USA. Her research interests lie in distributed and robust optimization, focusing on developing safe and scalable frameworks to address uncertainty.
\end{IEEEbiography}

\begin{IEEEbiography}[{\includegraphics[width=1in,height=1.25in,clip,keepaspectratio]{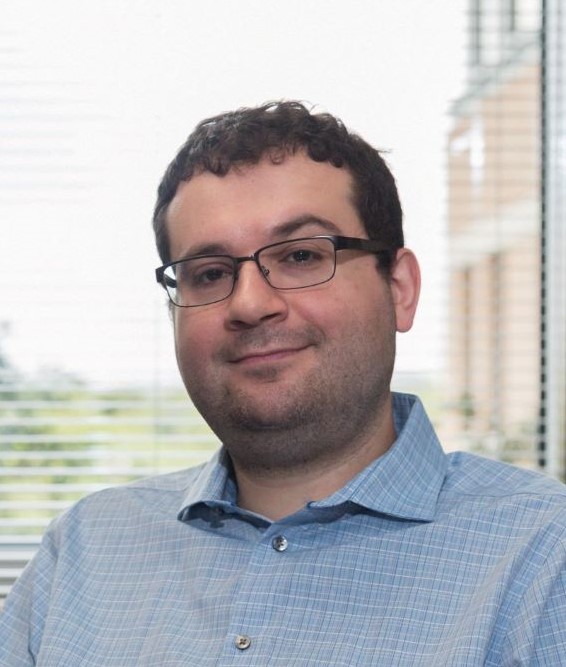}}]{Efstathios Bakolas} (Senior Member, IEEE) received the Diploma degree in Mechanical Engineering with highest honors from the National Technical University of Athens, Athens, Greece, in 2004 and the M.S. and Ph.D. degrees in Aerospace Engineering from the Georgia Institute of Technology, Atlanta, GA, USA, in 2007 and 2011, respectively. He is currently an Associate Professor with the Department of Aerospace Engineering and Engineering Mechanics, University of Texas at
Austin, Austin, TX, USA. His research is mainly focused on control of uncertain and stochastic systems, data-driven control of complex systems, optimal control theory, decision making and control of autonomous agents and multi-agent networks and differential games.
\end{IEEEbiography}

\begin{IEEEbiography}[{\includegraphics[width=1in,height=1.25in,clip,keepaspectratio]{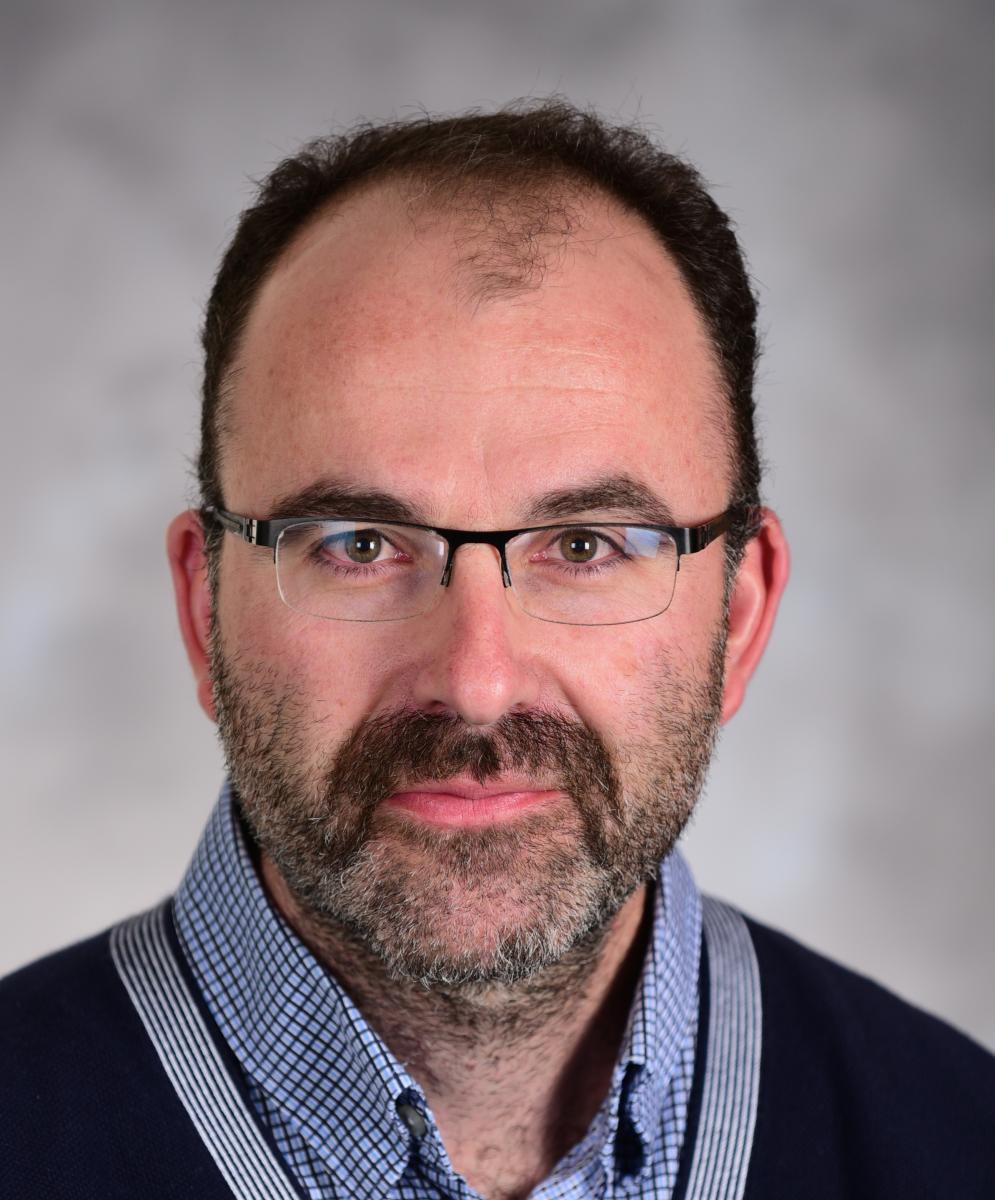}}]{Evangelos A. Theodorou} (Member, IEEE) is an Associate Professor with the Daniel Guggenheim School of Aerospace Engineering at Georgia Institute of Technology. He is also the director of the Autonomous Control and Decision Systems Laboratory, and he is affiliated with the Institute of Robotics and Intelligent Machines and the Center for Machine Learning Research at Georgia Institute of Technology. Dr. Theodorou holds a BS in Electrical Engineering,  from the Technical University of Crete (TUC), Greece in 2001. He also holds three MSc degrees in Production Engineering from TUC in 2003,  Computer Science and Engineering from University of Minnesota in 2007, and Electrical Engineering  from the University of Southern California (USC) in 2010. In 2011, he graduated with his PhD in Computer Science from USC. From 2011 to 2013, he was a Postdoctoral Research Fellow with the department of Computer Science and Engineering, University of Washington. Dr. Theodorou is the recipient of the King-Sun Fu best paper award of the IEEE Transactions on Robotics in 2012 and recipient of several best paper awards and nominations  in machine learning and robotics conferences. His theoretical research spans the areas of stochastic optimal control theory, machine learning, statistical physics and optimization.
\end{IEEEbiography}

\end{document}